\documentclass{aa}  
\usepackage{adjustbox}
\usepackage{caption}
\usepackage{float}
\usepackage{graphicx}
\usepackage{xcolor}
\usepackage{txfonts}
\usepackage[]{hyperref}
\hypersetup{
    colorlinks=true,
    linkcolor=blue,
    citecolor=blue,
    filecolor=magenta,      
    urlcolor=magenta,
    }
\usepackage{morefloats}
\extrafloats{100}
\usepackage{soul}
\usepackage{subcaption}

\usepackage{xurl} 
\usepackage{booktabs}
\usepackage{array}
\renewcommand{\arraystretch}{1.2}
\usepackage{orcidlink} 
\usepackage{longtable}

\usepackage{cuted}
\usepackage{capt-of}

\begin{document}
   \title{A multi-epoch VLBI study of compact symmetric objects as contaminants in the search for milli-lenses}
   \titlerunning{A multi-epoch VLBI study of CSOs as contaminants in the search for milli-lenses}
   \subtitle{}

   \authorrunning{D. \'Alvarez-Ortega et al.}
   \author{D. \'Alvarez-Ortega\inst{1,2}\orcidlink{0000-0002-9998-5238}
          \and
          F. M. P{\"o}tzl\inst{1,2}\orcidlink{0000-0002-6579-8311}
          \and
          C. Casadio\inst{1,2}\orcidlink{0000-0003-1117-2863}
          }
   \institute{
           Institute of Astrophysics, Foundation for Research and Technology – Hellas, N. Plastira 100, Voutes GR-70013, Heraklion, Greece
           \email{dalvarez@physics.uoc.gr}
       \and
           Department of Physics, University of Crete, 70013 Heraklion, Greece
   }
   \date{Received date /
Accepted date }
\abstract
  {The $\Lambda$ Cold Dark Matter ($\Lambda$CDM) model predicts a high abundance of low-mass dark matter halos, which can be probed through milliarcsecond-scale gravitational lensing (milli-lensing) of background radio-loud active galactic nuclei (AGNs). The Search for MIlli-LEnses (SMILE) project aims to find these rare systems using high-resolution Very Long Baseline Interferometry (VLBI). However, intrinsically compact radio sources, particularly compact symmetric objects (CSOs), can closely resemble milli-lenses in VLBI images and therefore constitute the predominant contaminant in such searches.}
  {We aim to determine the physical nature of the nine remaining milli-lens candidates identified in the SMILE pilot project. Recognizing that CSOs will constitute the main contaminant in the upcoming main SMILE survey, we also aim to thoroughly characterize their intrinsic physical properties to establish a robust classification framework.}
  {We analyzed an extensive multi-epoch, multi-frequency dataset comprising all available archival data, new dedicated High Sensitivity Array (HSA) and European VLBI Network (EVN) observations, as well as single-dish and connected-element interferometer measurements. We evaluated the viability of the lensing hypothesis by studying the morphology and spectral properties of the sources. At the same time, we characterized their intrinsic AGN physics by analyzing kinematics, spatially resolved spectral indices, equipartition magnetic fields, and integrated spectra using homogeneous and inhomogeneous synchrotron self-absorption (SSA) and free-free absorption (FFA) models.}
  {Our multi-wavelength analysis identifies eight of the nine candidates as CSOs rather than milli-lenses, including four sources that had not previously been classified as such in the literature. We discover severe spectral index asymmetries between components and previously undetected emission connecting them, all of which rule out the lensing hypothesis. We observe low-frequency spectral turnovers potentially produced by ionized absorbing media. For the sources with known redshifts, we also derive equipartition magnetic fields and lower limits on the kinematic ages. Only one source (J0923-3435) remains ambiguous and will require deeper observations to confirm its nature.}
   {This work establishes a comprehensive framework for the follow-up and evaluation of milli-lens candidates and classifies eight of the nine remaining sources in the SMILE pilot sample. For the CSOs with known redshifts, the lobes have equipartition magnetic fields of $\sim$20--320\,mG and projected separations of $\sim$60--240\,pc, and show no measurable expansion, implying kinematic ages of at least $\sim$50--950\,yr. The integrated spectra of J0210-2213 and J2347-1856 favor an inhomogeneous external free-free absorber over both homogeneous and inhomogeneous SSA models. Since CSOs will be the main contaminant in the main SMILE survey, applying this framework to its candidates will provide the same magnetic-field, age, and absorption constraints for a substantially larger sample of compact radio-loud AGNs.}


   \maketitle

\section{Introduction}\label{sec:introduction}

The $\Lambda$ Cold Dark Matter ($\Lambda$CDM) model stands as the widely accepted paradigm of cosmology. While highly successful at describing the Universe on large scales, it faces unresolved challenges at the sub-galactic level \citep{Bullock2017, Perivolaropoulos2022}. At these scales, alternative dark matter (DM) models, such as warm, self-interacting or fuzzy DM, make distinct predictions from $\Lambda$CDM regarding both the abundance and the internal structure of low-mass DM halos ($\lesssim 10^{10}\,M_\odot$), as shown by recent cosmological simulations \citep[e.g.,][]{Bose2016, May2023, Despali2025}. Constraining the number density of these small-scale structures is therefore essential for discriminating between competing DM models.

Strong gravitational lensing provides a direct way to probe these sub-galactic structures, because lensing is sensitive to mass regardless of whether it emits light. In galaxy-scale lenses, low-mass halos can be detected through two complementary approaches. Flux-ratio anomalies in multiply imaged quasars can constrain the abundance of halos down to $\sim$10$^7\,M_\odot$ and the free-streaming length of dark matter \citep[e.g.,][]{Nierenberg2024, Keeley2024, Gilman2026}, while images of extended lensed arcs can be used to detect sub-halos by analyzing the distortions they imprint on the arcs. Combined with the angular resolution of Very Long Baseline Interferometry (VLBI), the latter technique has recently reached the million-solar-mass regime: \citet{Powell2025} detected an object of $\sim$10$^6\,M_\odot$ through the  milliarcsecond-scale perturbation of the arc in JVAS B1938+666 \citep[see also][]{McKean2025}, and VLBI observations of MG J0751+2716 have been used to set new constraints on fuzzy dark matter \citep{Powell2023}. However, these methods detect low-mass halos only as perturbers of a much more massive galaxy-scale lens, and are therefore limited to the small number of known systems with suitable lensed sources.

A complementary approach is to search for strong lensing in which the low-mass object is itself the lens. When a compact object with a mass in the range of $10^6 - 10^9~M_\odot$ intersects the line of sight to a distant, compact background source, it can produce multiple lensed images separated by milliarcseconds (mas), an effect referred to as milli-lensing. Detecting such systems, or their absence, using a large sample of background sources constrains the number density of low-mass DM halos along the line of sight \citep{Loudas2022}, as well as primordial black holes and supermassive black holes hosted by intervening galaxies \citep{Banik2019, Dike2023}. The extreme angular resolution of VLBI at radio frequencies makes it the ideal tool for identifying these mas-scale image separations.

The main aim of the Search for MIlli-LEnses (SMILE) project \citep{Casadio2021} is to search for mas-scale gravitational lens systems in a large sample of radio-loud active galaxies, in order to constrain the density of such compact objects and to improve upon previous results \citep{Wilkinson2001}. The project targets a VLBI subsample of 4\,968 flat-spectrum ($\alpha \geq -0.5$; 1.4\,-\,5\,GHz) radio sources with $S_{8\,\mathrm{GHz}} \geq 50$ mJy, drawn from the Cosmic Lens All Sky Survey (CLASS) survey \citep{Myers2003}. As a precursor to the main SMILE search, a pilot project defined in \citet{Casadio2021} identified 40 initial milli-lens candidates by visually inspecting multi-frequency VLBI data on a different sample of 13\,828 sources taken from the Astrogeo VLBI image database\footnote{\url{http://astrogeo.org/vlbi_images/}} \citep{Petrov2025}. Recently, \citet{Potzl2025} incorporated new VLBI data to study these 40 candidates in detail, reducing the initial pool to nine robust candidates by evaluating them against a strict set of defined lensing criteria. Specifically, we required the putative lensed images to preserve flux ratios, surface brightness, and spectral indices, alongside showing compatible morphologies and no proper motion between emitting regions.

A major challenge in the search for milli-lenses is the presence of intrinsic active galactic nuclei (AGN) structures that can mimic the morphology of a lensed system. Core-jet structures and supermassive black hole binaries (SMBHBs) can sometimes appear as multiple compact components. VLBI-resolved SMBHBs are, however, very rare. A systematic search for double flat-spectrum cores in archival VLBI data of 3\,114 radio-loud AGNs found only one resolved binary, corresponding to a detection rate of $\sim$0.03\% \citep{BurkeSpolaor2011}. This binary was 0402+379, with a projected separation of \(7.3\,\mathrm{pc}\) \citep{Rodriguez2006} and remains, to date, the only SMBHB resolved with VLBI. Far more common, and therefore the most prominent contaminants, are compact symmetric objects (CSOs). 

CSOs were originally identified as compact radio sources with emission on both sides of the center of activity and a total extent below 1\,kpc \citep{Readhead1993, Wilkinson1994, Readhead1996}. Because a symmetric morphology alone can be mimicked by relativistically beamed sources, \citet{Kiehlmann2024} redefined bona fide CSOs by additionally requiring low variability ($\leq$20\%\,yr$^{-1}$) and the absence of superluminal motion ($v_\mathrm{app} < 2.5c$), and identified 79 such objects; a number that has since grown to 144 \citep{Sheldahl2025}. Many CSOs have peaked radio spectra and form part of the broader population of peaked-spectrum (PS) and compact steep-spectrum (CSS) sources \citep[see the review by][]{ODea2021}. The nature of CSOs remains debated. Where hot spot motions have been measured, their separation speeds are typically a few tenths of the speed of light, implying kinematic ages of hundreds to thousands of years \citep[e.g.,][]{Polatidis2003, An2012}. This led to the interpretation of CSOs as young radio galaxies that may evolve into large-scale Fanaroff-Riley sources \citep[e.g.,][]{ODea1998}, while others proposed that their growth is frustrated by a dense ambient medium \citep[e.g.,][]{Bicknell1997}. More recently, \citet{Kiehlmann2024} and \citet{Readhead2024} showed that CSOs form a distinct population of jetted AGN, most of which are short-lived rather than young progenitors of larger sources, and proposed that the high-luminosity CSOs may be powered by single accretion episodes such as tidal disruption events \citep[see also][]{Sullivan2024}. CSOs are divided into two morphological classes analogous to Fanaroff-Riley I and II radio galaxies \citep{Tremblay2016, Kiehlmann2024}: CSO 1s are edge-darkened and dominated by diffuse jets, while CSO 2s are edge-brightened and feature compact hot spots at their outermost regions, with the CSO 2s further divided into subclasses that may trace an evolutionary sequence \citep{Readhead2024}. It is precisely the CSO 2 morphology, consisting of two distinct, compact, and remarkably symmetric emission regions, that can most easily be misclassified as lensed images of a single background core. CSOs make up 6.8--8.5\% of complete flux-density-limited AGN samples \citep{Kiehlmann2024}. The flat-spectrum selection between 1.4 and 5 GHz that characterizes the SMILE sample should exclude CSOs whose spectra peak outside this frequency range, while the remaining ones may be identified as milli-lens candidates.

The primary aim of this work is to determine the physical nature of the nine remaining milli-lens candidates presented in \citet{Potzl2025}. Milli-lens systems are expected to be rare: \citet{Wilkinson2001} found no lensed systems among 300 compact radio sources, and the predictions of \citet{Loudas2022} for the full SMILE sample of $\sim$5000 sources range from zero to a few lensed systems in the most optimistic dark matter scenarios. The identification of even a single milli-lens would therefore place significant constraints on the abundance of compact objects in the $10^6$--$10^9\,M_\odot$ range \citep{Loudas2022}. In contrast, 36 of the 40 candidates examined in the pilot study showed CSO-like properties \citep{Potzl2025}, suggesting that CSOs will constitute the main contaminant in the full SMILE sample, and making it likely that many, if not all, of the nine remaining candidates belong to this class. Efficiently rejecting these contaminants is essential to keep the follow-up of a survey of this size feasible.

A secondary but equally important goal of this study is to characterize the physical properties of these candidates, which also allows us to test the two classic scenarios for the compactness of CSOs. If CSOs are young, freely expanding sources, their hot spots should separate at a few tenths of the speed of light \citep{Polatidis2003, An2012}, and their spectral turnovers should arise from synchrotron self-absorption in hot spots close to equipartition, as found for several young radio sources \citep{Orienti2008, Orienti2012}. If, instead, their growth is confined by a dense and clumpy circumnuclear medium \citep{Bicknell1997}, their low-frequency spectra should be shaped by free-free absorption in an inhomogeneous ionized screen, as inferred for a number of compact radio sources \citep{Kameno2000, Marr2001, Callingham2015, Tingay2015}. We test these predictions by combining limits on the expansion speeds, equipartition and SSA magnetic field estimates, and SSA and FFA models of the integrated spectra. Because their double-component morphologies mimic milli-lens images, these sources also provide a benchmark to refine the selection criteria and follow-up strategy of the full SMILE survey.

This paper is organized as follows: Section~\ref{sec:sample} presents the data sample used in this study. Section~\ref{sec:analysis} describes the analysis methodology. Section~\ref{sec:results} presents the results: we first test the milli-lens hypothesis (Sect.~\ref{sec:resmillilens}), then present the global properties of the sample as CSOs and peaked-spectrum sources (Sect.~\ref{sec:rescso}), and finally review each source individually (Sect.~\ref{sec:ressources}). Section~\ref{sec:discussion} discusses the implications of our results, and Sect.~\ref{sec:conclusions} summarizes our main conclusions. Throughout this paper, we adopt a $\Lambda$CDM cosmology with $H_0 = 67.8$ km s$^{-1}$ Mpc$^{-1}$, $\Omega_M = 0.3089$, and $\Omega_\Lambda = 0.6911$ \citep{Planck2016}.

\section{Data sample}\label{sec:sample}

The sample used in this work consists of both archival and new VLBI observations of the nine milli-lens candidates, alongside single-dish and connected-element interferometer data. A brief description of the sources is given in Table~\ref{tab:redshifts}.

\begin{table*}[t]
\caption{Source properties. Positions are J2000. Redshift type: s = spectroscopic, p = photometric.
$S_{\rm C}$ and $S_{\rm X}$ are the total flux densities at C-band and X-band from the Radio Fundamental Catalogue (RFC 2024a)\protect\tablefootmark{a}.
References: (1) \citealt{deVries1995}; (2) \citealt{deVries2007}; (3) \citealt{DESIDR8};
(4) \citealt{SDSSDR18}; (5) \citealt{Labiano2007}; (6) \citealt{Sheldahl2025};
(7) \citealt{DAbrusco2019}; (8) \citealt{Taylor2003}; (9) \citealt{Kiehlmann2024}.}
\label{tab:redshifts}
\centering
\begin{tabular}{lcccccc}
\hline\hline
Source & RA (h:m:s) & Dec (d:m:s) & $z$ & Classification & $S_{\rm C}$ [mJy] & $S_{\rm X}$ [mJy] \\ \hline
J0024-4202 & 00:24:42.99 & $-$42:02:03.95 & $0.9364\pm0.0010^{\,\rm s,1}$ & GPS$^{5}$    & --          & 514 \\
J0210-2213 & 02:10:10.06 & $-$22:13:36.90 & $1.491\pm0.03^{\,\rm s,2}$    & CSO$^{6}$    & 410         & 188 \\
J0923-3435 & 09:23:53.88 & $-$34:35:26.10 & --                            & --           & 284         & 167 \\
J1143+1834 & 11:43:26.07 & $+$18:34:38.36 & $1.13\pm0.22^{\,\rm p,3}$     & CSO$^{6}$    & 345         & 230 \\
J1218-2159 & 12:18:58.82 & $-$21:59:45.40 & --                            & --           & --          &  92 \\
J1632+3547 & 16:32:31.25 & $+$35:47:37.74 & $1.08\pm0.08^{\,\rm p,3}$     & CSO$^{6}$    & 239         &  -- \\
J1653+3503 & 16:53:53.16 & $+$35:03:27.03 & $0.81\pm0.10^{\,\rm p,4}$     & --           & 101         &  -- \\
J1805-0438 & 18:05:31.12 & $-$04:38:09.69 & --                            & BL Lac$^{7}$ & --          &  66 \\
J2347-1856 & 23:47:08.63 & $-$18:56:18.86 & $0.22\pm0.13^{\,\rm p,4}$     & CSO$^{8,9}$  & --          & 288 \\ \hline
\end{tabular}
\tablefoot{\tablefoottext{a}{\url{https://astrogeo.org/sol/rfc/rfc_2024a/rfc_2024a_cat.txt}}}
\end{table*}

\subsection{VLBI observations}

\citet{Potzl2025} based their analysis on the Astrogeo VLBI image database and dedicated European VLBI Network (EVN) follow-up observations (EC071; PI: Casadio). Here we extend that dataset with all available archival Very Long Baseline Array (VLBA) data for the nine candidates and with new High Sensitivity Array (HSA) observations at 5 and 8 GHz (BA189; PI: \'Alvarez-Ortega) and EVN observations at 22 GHz (EA075; PI: \'Alvarez-Ortega). The new HSA and EVN observations were conducted at a recording rate of 4 Gbps, with a total bandwidth of 512 MHz per polarization. The data were correlated with a spectral channel width of 500 kHz and an integration time of 2 s. The HSA observations included the VLBA antennas and the phased Karl G. Jansky Very Large Array (VLA), while the EVN observations included Jb, Ef, Mc, O6, Mh, Ys, Hh, Sr, Kt, and Ku. All datasets used in this work, including their exposure times, synthesized clean beams, and achieved image sensitivities, are summarized in Table~\ref{table:usedvlbidata}.

\subsection{Connected-element and single-dish observations}\label{sec:integrateddata}

We also collected single-dish and connected-element interferometer measurements to investigate the integrated properties of the sources. Data were taken from the LOFAR Two-metre Sky Survey \citep{LOFARDR2}, the GMRT 150 MHz all-sky radio survey (TGSS, \citealt{TGSSDR1}), the GaLactic and Extragalactic All-sky Murchison Widefield Array eXtended survey (GLEAM-X, \citealt{GLEAMXDR2}), the Texas survey of radio sources \citep{TEXASDR1}, the Rapid ASKAP Continuum Survey (RACS, \citealt{RACSDR1}), the Faint Images of the Radio Sky at Twenty cm survey (FIRST, \citealt{FIRSTDR}), the NRAO VLA Sky Survey (NVSS, \citealt{NVSSDR}), the Very Large Array Sky Survey (VLASS, \citealt{VLASSQuickLook}), the Green Bank 6-cm survey (GB6, \citealt{GB6Catalog}), the  Combined Radio All-Sky Targeted Eight GHz Survey (CRATES, \citealt{CRATESDR}), and the Cosmic Lens All-Sky Survey (CLASS, \citealt{Myers2003}). The corresponding measurements and their errors are compiled in Table~\ref{table:usedsingledishdata}.

In addition, we obtained new connected-element flux densities of J0024-4202 and J2347-1856 between 4.7 and 10.1\,GHz from the VLA data recorded during the HSA observations (BA189). The data were calibrated with the standard VLA calibration pipeline in \texttt{CASA} \citep{CASA2022}, using 3C48 as the flux-density calibrator on the \citet{PerleyButler2017} scale. Each of the eight sub-bands, four at C-band (centered at 4.74, 5.25, 5.72, and 6.25\,GHz) and four at X-band (centered at 8.54, 9.05, 9.54, and 10.05\,GHz), was then imaged separately in \texttt{difmap} \citep{Sheperd1997}, and the integrated flux density of each source was measured as the total flux density of the CLEAN model. We adopted an uncertainty of 5\% to account for the absolute flux-density calibration.


\section{Analysis methods}\label{sec:analysis}

This section describes the various methods used to distinguish between the different physical scenarios associated with the milli-lens candidates.

\subsection{VLBI maps}

To determine the precise morphology of each candidate and accurately measure the flux densities and positions of their individual components, we first generated continuum maps for all available VLBI data. The EVN pilot observations (EC071) were manually calibrated and imaged by \citet{Potzl2025} using the \texttt{AIPS} package \citep{Greisen2003}. The archival datasets, as well as the new HSA (BA189) and EVN (EA075) observations, were calibrated using the \texttt{VIPCALs} pipeline \citep{AlvarezOrtega2025}. Data obtained from the Astrogeo VLBI image database were not recalibrated, as fully calibrated uvfits files are already provided.

All datasets were imaged in \texttt{difmap} using the CLEAN algorithm \citep{Hogbom1974}. We applied an iterative self-calibration procedure, performing phase-only calibration before introducing amplitude corrections. The solution interval was initially set to the entire observation time and progressively shortened down to the visibility integration time over successive iterations. Final images were produced using natural weighting (\texttt{uvweight 0,-1} in \texttt{difmap}). Lastly, we performed a least-squares fitting of the visibility data with a set of circular Gaussian components using the \texttt{modelfit} procedure. To determine the appropriate number of components, we compared models with different numbers of Gaussian components and added a new component only when the improvement in the reduced $\chi^2$ satisfied the statistical significance criterion described in Section 2.2 of \citet{Schinzel2012}.


\subsection{Lens properties}\label{sec:lensproperties}

Gravitational lensing imposes specific theoretical constraints on the properties of multiple images produced by a lens. To test the milli-lens hypothesis and distinguish lensing candidates from any intrinsic AGN properties, we considered the following five criteria defined in \cite{Potzl2025}:

\begin{enumerate}
    \item The sources should exhibit two or more compact radio components at all observing frequencies and epochs. Any extended emission or elongation should be oriented tangentially to the Einstein radius.

    \item The surface brightness ratio (SBR) between the components should satisfy $0.25 < \mathrm{SBR} < 4$ within the uncertainties. Although a ratio close to unity is expected in gravitational lensing, these conservative limits account for unmodelled systematic uncertainties in the imaging and modelling.

    \item At any given observing frequency, the flux ratio between the components should not vary between observing epochs by more than the combined $2\sigma$ uncertainty. For the small image separations considered here, the expected lensing time delays are short compared with the VLBI observing times, and propagation effects are neglected.
    
    \item The spectral index $\alpha$ (defined as $S_{\nu} \propto \nu^{\alpha}$) between the components should be comparable ($\Delta\alpha < 0.46$).

    \item No significant radial proper motion should be observed between the components  ($<\!2\text{-}\sigma$ deviation considering the combined error between epochs).
\end{enumerate}

To assess these criteria, we examined the radio maps, the flux ratio between components, the surface brightness ratio, the distance between components, and, where possible, the spectral index of the components. The analysis and error estimation are identical to those conducted in \cite{Potzl2025}.


\subsection{VLBI synchrotron self-absorption spectra}
\label{sec:vlbissaspectra}

To understand the intrinsic emission and absorption mechanisms of the individual components, we fitted the spectrum of each VLBI component with both a simple power law and a synchrotron self-absorption (SSA) model, and adopted the SSA model only where the data require it (see below). In cases where a primary component was resolved into multiple sub-components during model fitting (e.g., A0, A1, and A2 in J1632+3547; see Fig.~\ref{fig:vlbimaps}), their individual flux densities were summed to represent the total spectrum for that region. We followed the parameterization defined in \cite{Karamanavis2016}:
\begin{equation}
    S_\nu = S_m \left(\frac{\nu}{\nu_m}\right)^{\alpha_t}\frac{1-\text{e}^{-\tau_m(\nu/\nu_m)^{\alpha-\alpha_t}}}{1-\text{e}^{-\tau_m}},
    \label{eq:karamanavis}
\end{equation}
where $\nu_m$ is the turnover frequency, $S_m$ is the turnover flux, $\alpha_t$ is the spectral index of the optically thick part of the emission, $\alpha$ is the optically thin spectral index, and $\tau_m$ is the optical depth at the turnover frequency, which can be obtained by solving:
\begin{equation}
    \text{e}^{\tau_m} = 1 + \left(1 - \frac{2\alpha}{5}\right)\tau_m.
    \label{eq:opticaldepth}
\end{equation}

We fixed $\alpha_t$ to the theoretical value of 5/2, as expected for a homogeneous synchrotron self-absorbed emitting region, and fitted the values of $S_m$, $\nu_m$, and $\alpha$ using the \texttt{curve\_fit} function from the \texttt{scipy} library, which employs the Levenberg–Marquardt algorithm for nonlinear least-squares fitting. The errors on the fitted parameters were derived as the square root of the sum of the squared diagonal elements in the covariance matrix. 

To plot the best-fit spectra alongside their $1\sigma$ confidence regions, we employed Monte Carlo error propagation. For each component, $10^4$ parameter realizations were drawn from a multivariate normal distribution defined by the best-fit values and their associated covariance matrix. This generated an ensemble of spectra consistent with the parameter uncertainties. From this ensemble, the $1\sigma$ confidence values were computed at each frequency, producing a $1\sigma$ uncertainty envelope around the best-fit spectrum.

For each component, we also fitted a simple power law, $S_\nu = S_{\text{1GHz}}\,\nu^{\alpha}$, where $S_{\text{1GHz}}$ is the flux density at 1\,GHz. To compare the two models, both were fitted in linear flux-density space with identical uncertainties, and we computed the Bayesian information criterion, $\mathrm{BIC} = \chi^2 + k\ln N$, where $k$ is the number of free parameters (two for the power law and three for the SSA model) and $N$ the number of data points. We adopted the SSA model only when it was strongly preferred ($\Delta\mathrm{BIC} = \mathrm{BIC}_\mathrm{PL} - \mathrm{BIC}_\mathrm{SSA} > 10$, \citealt{Kass1995}) and the fitted turnover frequency lay within the observed frequency range. Otherwise, the spectrum was described by the power law, whose parameters and uncertainties were obtained from a weighted linear least-squares fit in logarithmic space.


\subsection{Spectral index maps}

We generated maps to study the spatial distribution of the spectral index. This is crucial for distinguishing between flat-spectrum AGN cores and the steep-spectrum lobes which typically characterize the two-sided CSO morphology. Spectral index maps for each source were produced using \texttt{VCAT} \citep{Baczko2025}. For this analysis, we selected the pair of maps at different frequencies with the highest sensitivity per source, ensuring they were always observed within 24 hours of each other. The uv-ranges of the data were matched across frequencies, and the higher-frequency maps were convolved with the lower-frequency beam. The maps were then aligned using a per-pixel cross-correlation approach without masking, since in most cases the emission is dominated by the optically thin radio lobes, whose positions are expected to be independent of the observing frequency.   

\subsection{Magnetic field}\label{sec:magmethods}
To probe the physical conditions of the emitting plasma and determine if the component properties are consistent with typical AGN hot spots, we estimated the magnetic field strength. In a milli-lensing scenario, the observed components correspond to multiple lensed images of a compact AGN core. Previous studies show that compact core regions should, in principle, exhibit significantly larger magnetic fields \citep{Pushkarev2012}. Conversely, if the components represent expanding radio hot spots at the ends of a jet, these are expected to feature much weaker magnetic fields. Furthermore, comparing different magnetic field estimates allows us to test whether the plasma in these regions is in a state of equipartition, i.e. the energy density stored in the magnetic field is comparable to that carried by the relativistic particles \citep{Pacholczyk1970}. While magnetic field measurements could serve as a diagnostic against the lensing hypothesis, they simultaneously allow us to physically characterize the emission regions in the highly probable event that they are expanding CSO hot spots.

\subsubsection{Synchrotron self-absorption magnetic field}
We first estimated the magnetic field strength under the assumption that the spectral turnover is caused by SSA in a single homogeneous emitting region, using the formula described in \cite{Pushkarev2019}:
\begin{equation}
    B_\mathrm{SSA} = 10^{-5} b(\alpha) \theta^4S^{\prime\hspace{2pt}-2}_m\nu_m^5\left(\frac{\delta}{1+z}\right),
    \label{eq:pushkarevB}
\end{equation}
where $B_\mathrm{SSA}$ is the magnetic field strength in Gauss, $\theta$ is the source size in mas, $S^{\prime}_m$ is the intrinsic flux at the turnover frequency in Jy, $\nu_m$ is the turnover frequency in GHz, $\delta$ is the Doppler factor, $z$ is the redshift, and $b(\alpha)$ is a dimensionless parameter dependent on the spectral index $\alpha$. Assuming a spherical emitting region, we calculate the component size $\theta$ by multiplying the full width at half maximum of the fitted Gaussian by a factor of 1.8 \citep{Marscher1983}.

This approach follows the one introduced in \cite{Marscher1983}, but instead of using tabulated values for the $b(\alpha)$ parameter, it employs an analytical expression:
\begin{equation}
    b(\alpha) = 5.5\times10^{62}\left[\frac{\pi}{6}\frac{c_5(\alpha)}{c_6(\alpha)}(e^{\tau_m}-1)\right]^2(2c_1)^{-5},
\end{equation}
where $c_1$, $c_5(\alpha)$, and $c_6(\alpha)$ are constants and functions defined in Appendix A of \cite{Pushkarev2019}. Equation~\ref{eq:pushkarevB} also uses the absorption-corrected flux density at the turnover frequency $S^{\prime}_m$, defined as:
\begin{equation}
    S^{\prime}_m = S_\nu|_{\nu=\nu_m} \text{e}^{\tau_m},
\end{equation}
where the optical depth $\tau_m$ is obtained by solving Eq.~\ref{eq:opticaldepth}.

\subsubsection{Equipartition magnetic field}
To complement the SSA calculations, we also derived the equipartition magnetic field, $B_\mathrm{eq}$. The magnetic field strength is computed using the formula in \cite{Perucho2002}:
\begin{equation}
    B_\mathrm{eq} = \left(\frac{6\pi AL_{hs}}{V_{hs}}\right)^{2/7},
\end{equation}
where $V_{hs}$ is the volume of the emitting region (assumed spherical),  $L_{hs}$ refers to the intrinsic total radio luminosity of the hot spot, and $A$ is a function that depends on the spectral index and the optically thin frequency range. The radio luminosity can be computed by assuming that synchrotron emission follows a power law in the optically thin regime, so that the total luminosity is equal to:
\begin{equation}
    L_{hs} = S_{\nu_0}4\pi D_L^2 \nu_0^{-\alpha}\frac{\nu_{\text{max}}^{1+\alpha} - \nu_{\text{min}}^{1+\alpha}}{1+\alpha}\hspace{10pt}(\alpha\neq-1)
    \label{eq:luminosity}
\end{equation}
\begin{equation}
    L_{hs} = S_{\nu_0}4\pi D_L^2 \nu_0^{-\alpha}\ln\left(\frac{\nu_{\text{max}}}{\nu_{\text{min}}}\right)\hspace{10pt}(\alpha=-1),
\end{equation}
where 
\begin{equation}
    D_L = (1+z)^2\cdot D_A
\end{equation}
is the luminosity distance, $D_A$ the angular diameter distance, and $S_{\nu_0}$ is the flux at any reference frequency in the optically thin regime, $\nu_0$. The minimum and maximum frequencies, $\nu_\text{min}$ and $\nu_\text{max}$, correspond to 10$^7$ and 10$^{11}$ Hz respectively \citep{Perucho2002}. Finally, $A$ is defined as:
\begin{equation}
    A = \frac{C_1^{1/2}}{C_3} \frac{2+2\alpha}{1+2\alpha}\frac{\nu_{\text{max}}^{1/2+\alpha} - \nu_{\text{min}}^{1/2+\alpha}}{\nu_{\text{max}}^{1+\alpha} - \nu_{\text{min}}^{1+\alpha}},
\end{equation}
where $C_1$ and $C_3$ are two constants ($C_1 \simeq 6.3\cdot10^{18}$ and $C_3 \simeq 2.40\cdot10^{-3}$ in cgs units). 

Uncertainties in both magnetic field estimates were propagated with Monte Carlo simulations. The SSA fit of each component (Sect.~\ref{sec:vlbissaspectra}) provides a single set of turnover parameters, whereas the component size is measured separately in each VLBI dataset (i.e. each epoch and frequency). For $B_\mathrm{SSA}$, we therefore used only the datasets observed close to the turnover frequency, where the measured size best represents the self-absorbed region. For each of these datasets, we drew $10^4$ realizations of the turnover flux density, turnover frequency, and optically thin spectral index from the multivariate normal distribution defined by the best-fit SSA parameters and their covariance matrix, together with the component size and the redshift, each drawn from a normal distribution with its measured uncertainty. For $B_\mathrm{eq}$, we used all datasets and, for each, drew $10^4$ realizations of the flux density, component size, redshift, and spectral index. Because $B_\mathrm{SSA} \propto \theta^4\nu_m^5$, the resulting distributions are strongly skewed; we therefore characterized each distribution by its median and its 16th and 84th percentiles, rather than by its mean and standard deviation.

The magnetic field of each component is finally given as a range, from the lowest 16th to the highest 84th percentile among all its datasets. Before combining the datasets, we excluded outliers, defined as datasets whose median field deviates by more than $5\sigma$ from the median of all datasets of that component. These outliers mostly corresponded to datasets in which the component was unresolved or its size was poorly constrained.


\subsection{Integrated emission spectra}\label{sec:singledishmodels}

To complement the high-resolution VLBI data, we fitted different models to the single-dish and connected-element interferometer data. Because these measurements capture the integrated emission, they are ideal for probing the global absorption mechanisms at play. If these candidates are indeed CSOs, determining the physical origin of the turnover frequency in their spectrum can provide vital clues about their surrounding environment. For instance, detecting a free-free absorption (FFA) signature acts as a probe of a dense, ionized gas covering the radio emitter. Characterizing this surrounding medium allows us to test hypotheses regarding the frustrated expansion of AGN jets, turning these potential lensing contaminants into valuable laboratories to study CSO evolution.

Since the models have different numbers of free parameters $k$ and are not all nested, we compared them with the Bayesian information criterion ($\mathrm{BIC}$), as described in Sect.~\ref{sec:vlbissaspectra}.

\subsubsection{Internal FFA}
We first tested free-free absorption by an internal absorber (i.e., where the thermal absorbing gas is cospatial with the non-thermal synchrotron electron population). The formulation, originally from \cite{Kellermann1966} and parameterized by \cite{Shao2022}, is given by:
\begin{equation}
    S_\nu = S_\mathrm{I}\cdot\nu^{\alpha}\left(\frac{1-e^{-\tau_{\nu}}}{\tau_{\nu}}\right),
\end{equation}
where the optical depth is described by
\begin{equation}
    \tau_{\nu} = \left(\frac{\nu}{\nu_m}\right)^{-2.1},
    \label{eq:opticaldepthffa}
\end{equation}
$S_\mathrm{I}$ is a normalization parameter, and $\alpha$ is the spectral index of the intrinsic synchrotron emission.
\subsubsection{External FFA}
We also modeled scenarios in which the free-free absorbing screen is external to the synchrotron-emitting region. Within this framework, we distinguish two physical cases. First, we considered a homogeneous ionized medium, where the spectrum is characterized by:
\begin{equation}
    S_\nu = S_{\mathrm{EH}}\cdot\nu^{\alpha}e^{-\tau_{\nu}},
\end{equation}
where $S_{\mathrm{EH}}$ is a normalization constant and the optical depth $\tau_\nu$ is also described by Eq.~\ref{eq:opticaldepthffa}.

Second, we studied the case of an inhomogeneous external absorbing screen. As the expanding radio lobes of a CSO propagate outwards, they drive shocks into the surrounding host galaxy medium, potentially rendering the ionized gas thermally unstable. In such a scenario, the absorbing medium can be thought of as a clumpy ensemble of clouds with varying optical depths \citep{Bicknell1997}. Assuming that the distribution of the absorption parameter $a$ of these clouds follows a power law of the form $a^p$, the resulting radio spectrum is described by:
\begin{equation}
    S_\nu = S_{\mathrm{EI}}(p+1)\left(\frac{\nu}{\nu_m}\right)^{2.1(p+1) + \alpha}\gamma\left[p+1, \left(\frac{\nu}{\nu_m}\right)^{-2.1}\right],
\end{equation}
where $S_{\mathrm{EI}}$ is a normalization constant, $\gamma$ is the lower incomplete gamma function of order $p + 1$, and $p$ describes the distribution of the clouds. The fitting of the spectra and the subsequent error analysis follows the same approach described in Sect.~\ref{sec:vlbissaspectra}.

\subsubsection{Inhomogeneous SSA}
Finally, we tested whether a flattened optically thick spectrum could instead arise from an inhomogeneous synchrotron source, in which components with different turnover frequencies contribute to the integrated emission \citep{TingayDeKool2003}. Following \citet{Callingham2015}, we approximated such a source by the sum of two independent homogeneous SSA components:
\begin{equation}
    S_\nu = \sum_{i=1}^{2} S_{m,i} \left(\frac{\nu}{\nu_{m,i}}\right)^{\alpha_t}\frac{1-\text{e}^{-\tau_{m,i}(\nu/\nu_{m,i})^{\alpha_i-\alpha_t}}}{1-\text{e}^{-\tau_{m,i}}},
    \label{eq:doublessa}
\end{equation}
where each component $i$ has its own turnover flux density $S_{m,i}$, turnover frequency $\nu_{m,i}$, optically thin spectral index $\alpha_i$, and optical depth $\tau_{m,i}$; the latter follows from Eq.~\ref{eq:opticaldepth} with $\alpha = \alpha_i$. As in Eq.~\ref{eq:karamanavis}, $\alpha_t$ is fixed to 5/2. Because the uncertainties of this six-parameter model are poorly described by the covariance matrix, they were derived from a parametric bootstrap, refitting 500 synthetic data sets obtained by perturbing the measured flux densities within their uncertainties.


\section{Results}\label{sec:results}

This section presents the observational results, spectral fits, and derived physical parameters for the nine milli-lens candidates. We first test the milli-lens hypothesis against the observational lensing criteria (Sect.~\ref{sec:resmillilens}), then present the global properties of the sample as CSOs and peaked-spectrum sources (Sect.~\ref{sec:rescso}), and finally review the evidence for each source individually (Sect.~\ref{sec:ressources}). To illustrate the methodology and data quality, a representative example is shown in the main text for each analysis step. The complete set of figures for all candidates is provided in Appendices~\ref{app:vlbimaps}--\ref{app:vlbilightcurves}.


\subsection{Milli-lens hypothesis test}\label{sec:resmillilens}

In this subsection, we test whether the nine candidates satisfy the observational criteria for gravitational lensing defined in Sect.~\ref{sec:lensproperties}, using the full multi-epoch, multi-frequency VLBI dataset. 


\subsubsection{Morphology}

\begin{figure}[h]
    \centering
    \includegraphics[width=0.49\textwidth]{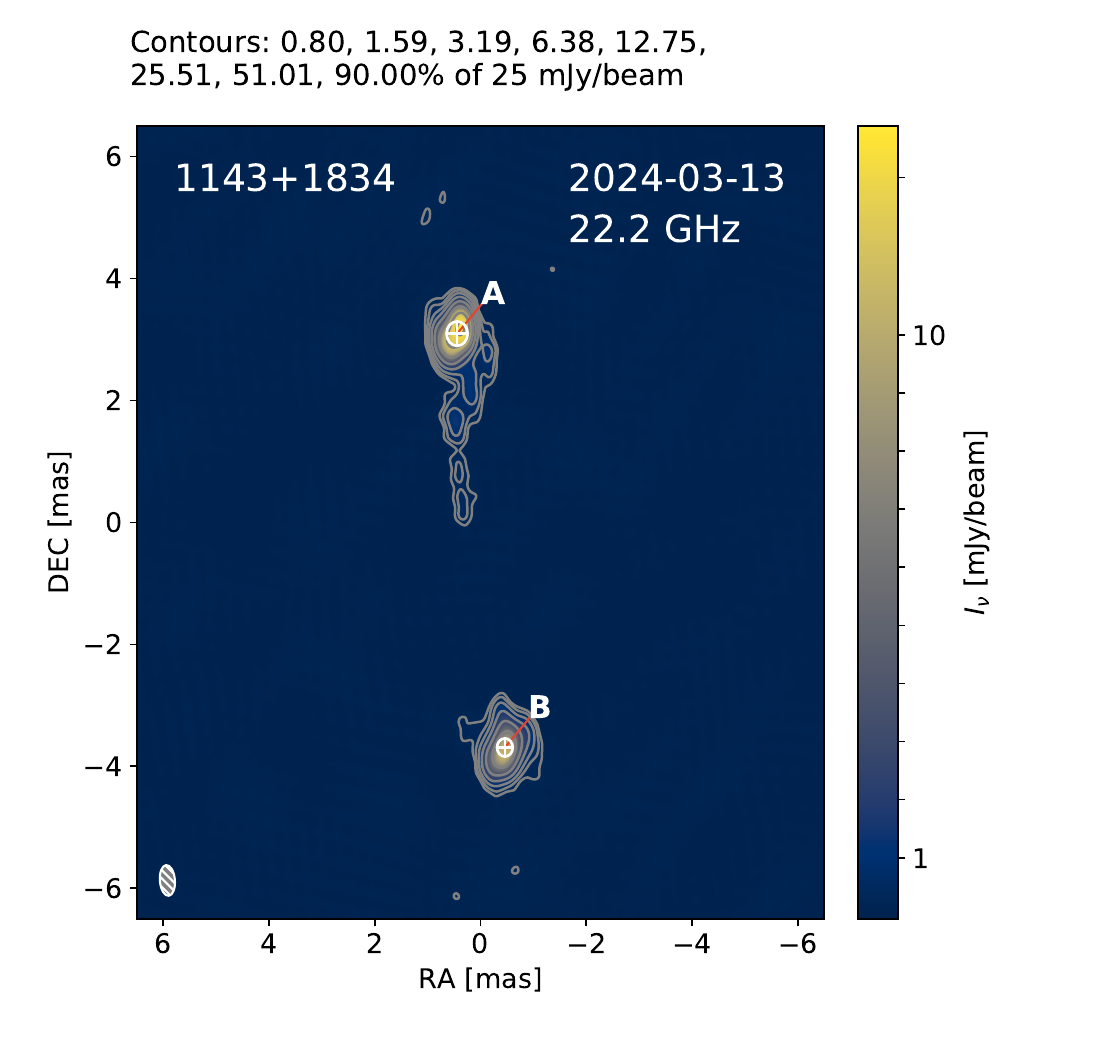}
\caption{Representative VLBI intensity map of J1143+1834. Lowest contours are drawn at four times the rms of the residuals from the CLEAN model. Intensity maps of the rest of the candidates can be found in Fig.~\ref{fig:vlbimaps}.}
    \label{fig:vlbimapsexample}
\end{figure}

Figure~\ref{fig:vlbimapsexample} presents a representative total intensity map (J1143+1834), along with the fitted Gaussian components. The lowest contour corresponds to four times the rms of the residuals from the CLEAN model. Based purely on the observed morphology in these maps, four sources (J0210-2213, J1143+1834, J1218-2159, J2347-1856) exhibit extended, connecting emission or hint at a central component. These bridging structures firmly point toward compact radio lobes and are sufficient to immediately rule them out as viable milli-lens candidates.
In contrast, the remaining sources display morphologies that leave the milli-lens hypothesis open. For instance, J1632+3547 exhibits extended emission that is still geometrically compatible with lensing distortion, since gravitational lensing can introduce tangential shear along the lens potential critical curves. Meanwhile, J0024-4202, J0923-3435, and J1653+3503 are resolved into two compact components. Finally, J1805-0438 exhibits three main components in all epochs, with a fourth, fainter component required in the most sensitive observations. A simple lens model, i.e. a point mass or a singular isothermal sphere, produces two images of the background source, so an extended or multi-component background source can appear as more than two VLBI components \citep[e.g.,][]{Spingola2018}. The morphology of J1805-0438 is therefore also compatible with lensing, with two of its components being the lensed images of the same compact emitting region.


\subsubsection{Flux ratio, surface brightness, and proper motion}\label{result:lensparameters}

To test the lensing criteria defined in Sect.~\ref{sec:lensproperties}, we derived, for every VLBI dataset (i.e. each source, epoch, and observing frequency), the following quantities from the model-fitted components: the flux density and observed brightness temperature of each component, and the flux ratio (FR), surface brightness ratio (SBR), and angular separation between components. Where simultaneous observations at two frequencies are available, we also computed the components' spectral indices. All values and their uncertainties are tabulated for the nine sources and are available on Zenodo\footnote{\url{https://doi.org/10.5281/zenodo.23862271}}. For J1805-0438, three main components (A, B, and C) are detected in all epochs. In the most sensitive observations, a fourth, fainter component (D) is additionally required to describe the brightness distribution (Fig.~\ref{fig:vlbimaps}), but it is not detected consistently enough to be included in the analysis. Of the three main components, only A and C are analyzed because, based on the spectral data (see Sect.~\ref{result:speccons}), these two are the most compatible with a multiple-image lensing scenario. With a few exceptions attributed to modeling uncertainties, all sources fulfill the flux ratio and surface brightness conservation constraints required for gravitational lensing \citep{Potzl2025}. The remaining criterion, the absence of significant proper motion between the components, is also fulfilled by all sources: none of the fitted separation speeds deviates from zero by more than $2\sigma$. These measurements are presented in Sect.~\ref{result:kinematics}, where they are discussed together with the projected sizes and kinematic ages of the sources.


\subsubsection{Spectral index conservation}\label{result:speccons}

Since gravitational lensing is achromatic, lensed images of the same background source must share the same spectrum. We test this criterion using the VLBI spectra of the individual components, whose fits are described in Sect.~\ref{result:ssaspectra} (see Fig.~\ref{fig:vlbispectraexample} for an example), together with the spectral index maps.

For most sources, the modeled spectra of the individual components are compatible with one another within the uncertainties. However, it is worth noting that the data used for these fits were collected from epochs spanning up to 30 years, and as a consequence, intrinsic source variability is not accounted for and could systematically affect the fits. In two cases (J0923-3435 and J1632+3547), the lack of low-frequency observations in the optically thick regime prevents any conclusions based solely on the VLBI spectra. Conversely, for J1653+3503, the compelling difference between the optically thin spectral indices ($\alpha_A = -1.07 \pm 0.17$, $\alpha_B = -3.03 \pm 0.35$) definitively rules out the possibility that these components are lensed images of the same background object. For J1805-0438, components A and C do not exhibit a spectral turnover and are well described by simple power laws, whereas component B is well-fitted by an SSA spectrum. This discrepancy leaves components A and C as the only potentially viable lensed images in this system.

Spectral index maps for each source were produced using \texttt{VCAT}. A representative map is shown in Fig.~\ref{fig:spixmapsexample}, with the full set provided in Appendix~\ref{app:spixmaps}. In each map, pixels are displayed only where the intensity is at least five times greater than the rms of each intensity map. The overlaid contours correspond to the intensity from the lower-frequency map.

\begin{figure}[h]
    \centering
    \includegraphics[width=0.49\textwidth]{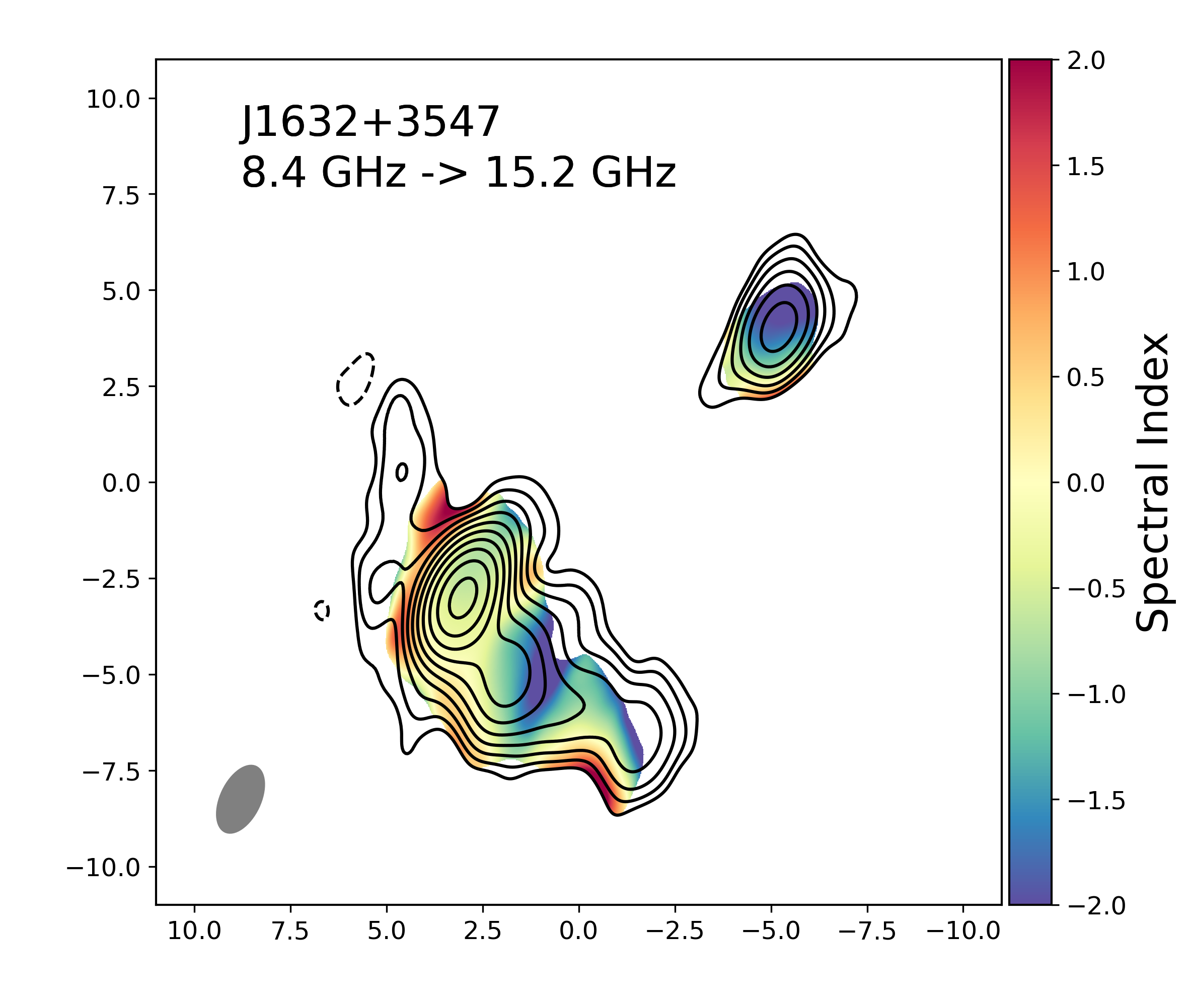}
\caption{Spectral index map of J1632+3547 produced with \texttt{VCAT}. Spectral index maps for the other sources are shown in Fig.~\ref{fig:spixmaps}.}
    \label{fig:spixmapsexample}
\end{figure}

The diagnostic power of the spectral index maps is highlighted in Fig.~\ref{fig:spixmapsexample}, showcasing J1632+3547, where the southeastern component (A) shows a notably flat spectrum in its central region between the X- and U-bands, in sharp contrast to the steep-spectrum emission of the northwestern component (B). Because the VLBI spectrum merges the flux from multiple sub-components (see Sect.~\ref{sec:vlbissaspectra}), it effectively masks this internal spatial gradient. While the majority of candidates exhibit spatially uniform, steep-spectrum components consistent with typical radio lobes, a few deviate from this trend. J0024-4202 displays a highly complex spectral index distribution that prevents a simple morphological interpretation. Additionally, J1805-0438 appears to host a flat central region between the S- and X-bands. However, the VLBI spectrum (Fig.~\ref{fig:vlbispectra}) is still well described by a steep power law, and the spectral index map alone does not provide convincing evidence for the presence of a flat-spectrum radio core.


\subsection{Compact symmetric object properties of the sample}\label{sec:rescso}

Having tested the lensing hypothesis, we now characterize the sample as a population of compact radio sources, focusing on the properties commonly used to identify and study CSOs and peaked-spectrum sources. We consider their projected sizes and kinematics, the spectra, variability, and magnetic fields of the individual VLBI components, as well as the integrated radio spectra and different possible absorption mechanisms.


\subsubsection{Sizes and kinematics}\label{result:kinematics}

The distances between components were computed across all available frequencies and epochs, with the average values for the most densely sampled band presented in Table~\ref{tab:distances}. We also derived the corresponding projected physical distances and performed linear fits to the component positions to estimate their angular and linear separation speeds. For candidates lacking redshift measurements, we established an upper limit on their physical size by evaluating the angular diameter distance, $D_A(z)$, at its cosmological maximum (which occurs at $z \approx 1.6$ in the cosmology adopted in this work). For all sources, the projected physical separations are strictly sub-kpc (see Table~\ref{tab:distances}), matching the defining spatial constraints of CSOs \citep{Kiehlmann2024}. Kinematically, none of the sources shows a statistically significant ($\geq 3\sigma$) proper motion: all fitted separation speeds are consistent with zero within $2\sigma$. The $3\sigma$ upper limits on the separation speeds range from 0.011 to 0.26\,mas\,yr$^{-1}$, corresponding to 0.41--14.7\,$c$ for the sources with redshifts. For these sources, we derived lower limits on their kinematic ages by dividing the projected component separation by the $3\sigma$ upper limit on the linear separation speed, assuming constant expansion. These lower estimates range from 54 to 951\,yr (Table~\ref{tab:distances}). Although these values are broadly consistent with those of previous CSO studies \citep{Taylor2000}, all kinematic estimates presented here are poorly constrained and should be interpreted with caution.

\begin{table*}[h]
\caption{Kinematic parameters of each of the sources.}
\vspace*{1mm}
\label{tab:distances}
\centering
\begin{tabular}{ccccccc}
\hline\hline
Source     & Band & Angular speed {[}mas yr$^{-1}${]} & Linear speed {[}$c${]} & \begin{tabular}[c]{@{}c@{}}Mean angular \\ distance {[}mas{]}\end{tabular} & \begin{tabular}[c]{@{}c@{}}Mean projected \\ distance {[}pc{]}\end{tabular} & \begin{tabular}[c]{@{}c@{}} Age lower limit \\ (3$\sigma$) {[}yr{]} \end{tabular}\\ \hline
J0024-4202 & X    & $0.073 \pm 0.063$            & $3.74 \pm 3.24$   & 28.20                                                                      & 228.20                                                                     & 55       \\ 
J0210-2213 & C    & $0.011 \pm 0.066$            & $0.74 \pm 4.65$   & 27.82                                                                      & 241.41                                                                     & 54       \\ 
J0923-3435 & C    & $-0.031 \pm 0.070$    & -                    & 17.39                                                                      & $< 151.07$                                                                          & -            \\
J1143+1834 & K    & $0.011 \pm 0.011$     & $0.66 \pm 0.66$   & 6.82                                                                       & 57.54                                                                      & 71        \\ 
J1218-2159 & S    & $0.023 \pm 0.012$& -                    & 4.41                                                                       & $<38.31$                                                                         & -            \\ 
J1632+3547 & C    & $0.002 \pm 0.003$     & $0.12 \pm 0.18$   & 10.82                                                                      & 90.49                                                                      & 430       \\ 
J1653+3503 & C    & $0.004 \pm 0.009$     & $0.19 \pm 0.42$   & 10.00                                                                      & 77.57                                                                      & 176       \\ 
J1805-0438 & S    & $-0.038 \pm 0.029$    & -                    & 9.61                                                                           & $<83.48$                                                                         & -            \\ 
J2347-1856 & X    & $0.004 \pm 0.008$     & $0.05 \pm 0.12$   & 33.37                                                                      & 123.17                                                                     & 951      \\ \hline
\end{tabular}
\end{table*}


\subsubsection{VLBI component spectra}\label{result:ssaspectra}

CSOs are generally classified as peaked-spectrum radio sources, with spectral turnovers typically occurring at frequencies of a few GHz in the observer's frame \citep{Orienti2016, ODea2021}. Their convex spectra are commonly interpreted as the result of SSA, although free-free absorption could also contribute in some sources. The observed inverse correlation between turnover frequency and projected linear size strongly supports SSA as the dominant absorption mechanism in the youngest and most compact radio sources \citep{ODea2021}. In the optically thin regime, CSOs also tend to exhibit steeper spectra than large-scale radio galaxies, with typical spectral indices of $\alpha \sim -0.7$ to $-1.2$, likely reflecting strong radiative losses in the compact lobes and hot spots, although intrinsically steep electron energy distributions may also contribute \citep{Orienti2016, ODea2021}. The comparison between SSA and power-law fits shows very strong evidence for a spectral turnover ($\Delta\mathrm{BIC} > 10$, with $\nu_m$ within the observed frequency range) only for the components of J0024-4202, J1143+1834, J1218-2159, and J2347-1856, and for component B of J1805-0438. All other components are adequately described by a power law and are modelled as such in the following. A representative example is shown in Fig.~\ref{fig:vlbispectraexample}, with the complete set of spectra available in Appendix~\ref{app:ssaspectra}. The best-fit SSA spectrum is overplotted alongside the 1-$\sigma$ confidence envelope, computed as described in Sect.~\ref{sec:vlbissaspectra}.

Finally, none of the modeled spectra show evidence of a high-frequency cutoff driven by the radiative aging of the electron population, a feature  observed in other GHz-peaked compact sources \citep{Murgia2002, Orienti2007}. The absence of this cutoff prevents us from deriving a spectral age, not allowing any direct comparison with the kinematic ages estimated in Sect.~\ref{result:kinematics}.

\begin{figure}[h]
    \centering
    \includegraphics[width=0.49\textwidth]{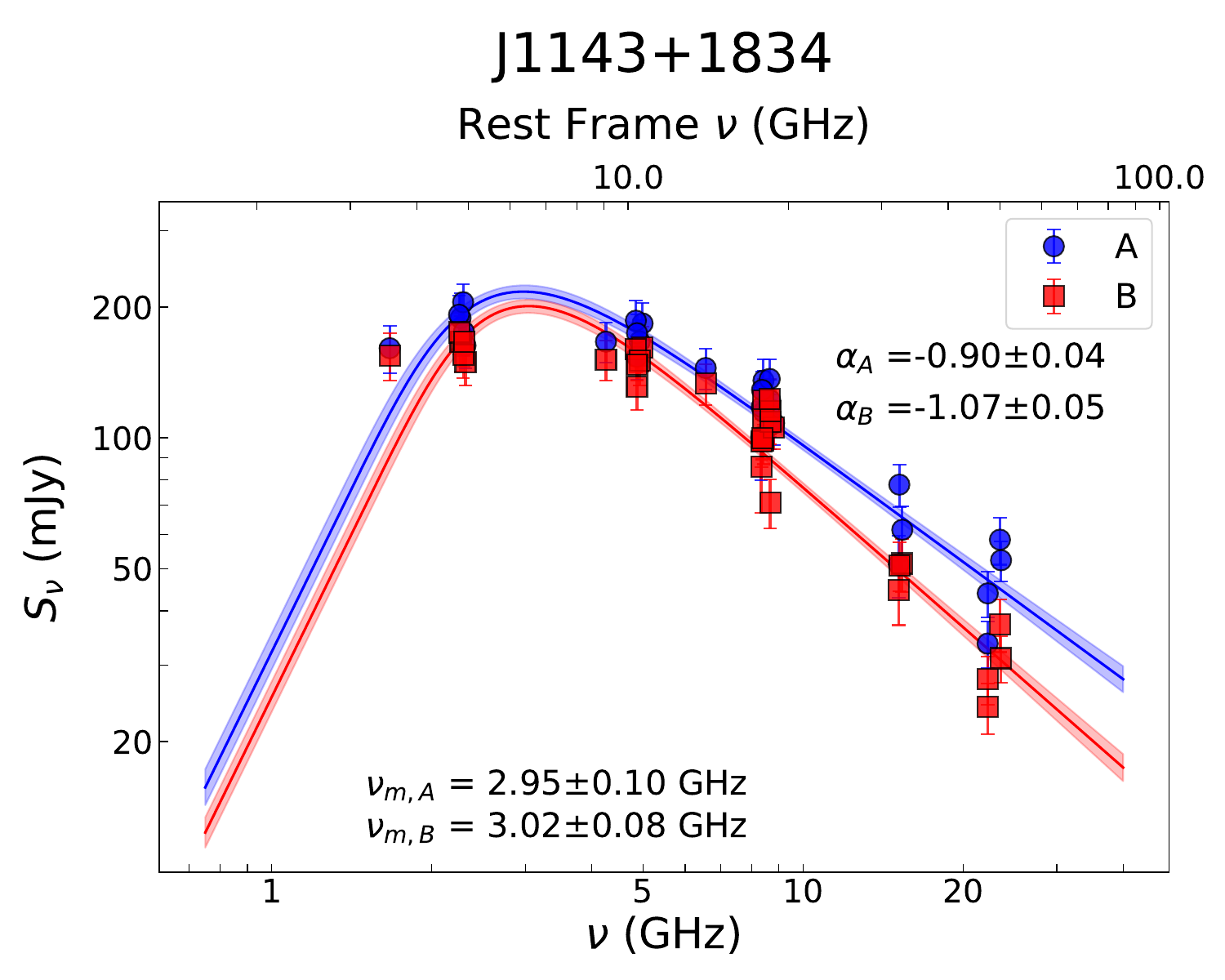}
\caption{VLBI spectra of the different components of J1143+1834. Overplotted are the synchrotron self-absorption spectra along with their 1-$\sigma$ confidence envelope. Best fit values of the turnover frequency and the spectral index and their errors are indicated on the plot. VLBI spectra of the rest of the candidates can be found in Fig.~\ref{fig:vlbispectra}.}
    \label{fig:vlbispectraexample}
\end{figure}


\subsubsection{Integrated flux spectra}\label{sec:intfluxspectra}
The integrated flux spectra capture the broad-band radio emission of each source. Two representative examples, J0024-4202 and J2347-1856, are shown in Fig.~\ref{fig:integratedspectraexample} (see Appendix~\ref{app:sdspectra} for all sources), with all fitted models and their 1-$\sigma$ confidence bands overplotted. For comparison, the sum of the individual VLBI component flux densities is also shown. In several cases (J0210-2213, J1218-2159, J1632+3547, and J2347-1856), the total integrated flux significantly exceeds the sum of the VLBI components at certain frequencies. This indicates either the presence of extended, large-scale emission that is resolved out by the VLBI observations, or simply intrinsic variability of the source not reflected due to the limited time sampling. Since strong intrinsic variability is atypical for CSOs \citep{Kiehlmann2024}, resolved-out emission is favored as an explanation under the CSO hypothesis.

In most cases, the homogeneous SSA model yields a poor fit to the integrated flux spectra, reflected by high $\chi^2_{\text{red}}$ values. This is driven by several factors: a limited number of data points (e.g., J1218-2159), unrealistic measurement errors (e.g., J1653+3503), or the presence of additional physical absorption mechanisms (free-free absorption). An ionized medium with thermal electrons surrounding the emitting region can alter the optically thick spectral slope, deviating from the theoretical SSA value of 5/2. This difference of the optically thick slope appears to be present in sources like J0210-2213, J1218-2159, and especially J2347-1856. To investigate this, we compared the SSA fits of all sources with the alternative absorption models described in Sect.~\ref{sec:singledishmodels}. The best-fit parameters are listed in Tables~\ref{tab:FFAtable} and~\ref{tab:SSA2table}, and the fitted spectra are shown in Fig.~\ref{fig:integratedspectra}.

\begin{figure}[h]
    \centering
    \includegraphics[width=0.49\textwidth]{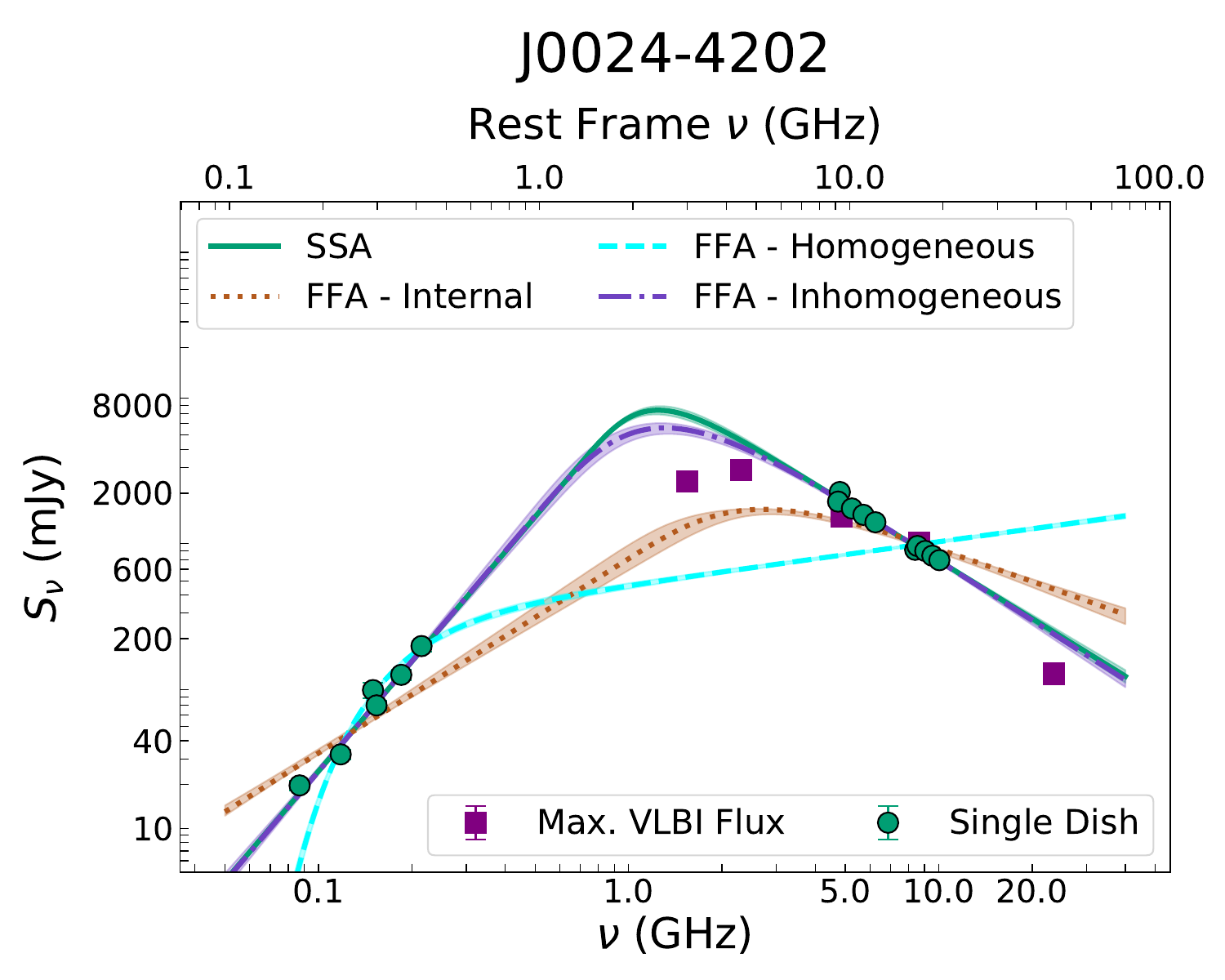}\\
    \includegraphics[width=0.49\textwidth]{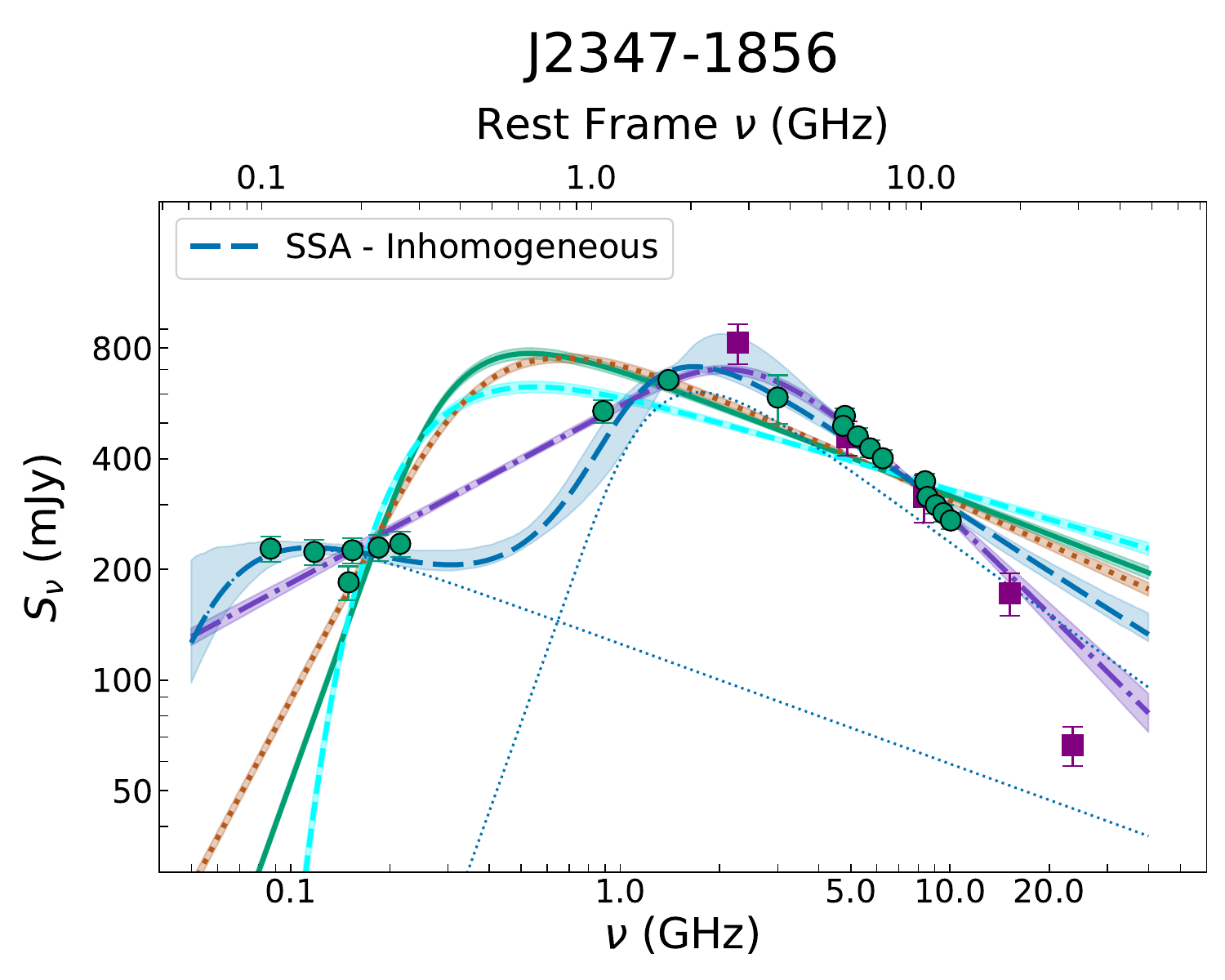}
\caption{Integrated radio spectra of J0024-4202 (top) and J2347-1856 (bottom). Green circles show the integrated flux densities, and purple squares the largest sum of the VLBI component flux densities. The lines show the best-fit SSA, internal FFA, homogeneous external FFA, and inhomogeneous external FFA models, and, for J2347-1856, the inhomogeneous SSA model with their 1-$\sigma$ confidence bands. The spectra of all sources are shown in Fig.~\ref{fig:integratedspectra}.}
    \label{fig:integratedspectraexample}
\end{figure}

The model comparison is summarized by the $\Delta\mathrm{BIC}$ values relative to the best model of each source. For four sources (J0210-2213, J0923-3435, J1632+3547, and J2347-1856), all FFA models except the homogeneous external one fit significantly better than homogeneous SSA ($\Delta\mathrm{BIC} = 72$--291). For J0210-2213 and J2347-1856, the inhomogeneous external absorber is strongly preferred over the internal FFA model ($\Delta\mathrm{BIC} = 46$ and 191), whereas for J0923-3435 the preference is only marginal ($\Delta\mathrm{BIC} = 3.4$) and for J1632+3547 the two models are statistically equivalent ($\Delta\mathrm{BIC} = 0.5$). The success of the inhomogeneous external screen model is consistent with findings for other compact radio sources \citep{Kameno2000, Marr2001, Tingay2015, Callingham2015}. In contrast, the spectrum of J0024-4202 is best described by a single SSA component, and the models are statistically indistinguishable for J1143+1834 and J1653+3503 ($\Delta\mathrm{BIC} < 3$). For J1218-2159 and J1805-0438, the four available data points do not constrain the four-parameter inhomogeneous FFA model, and the three-parameter fits are only marginally constrained; the same applies to the inhomogeneous FFA model for J1143+1834. In none of the sources does the homogeneous external absorber fit better than SSA. Denser frequency coverage below $\sim$1\,GHz is essential to distinguish conclusively between the absorption models.

We also tested whether the spectra can be explained by an inhomogeneous synchrotron source instead of an external absorber, by fitting the sum of two independent SSA components. Since this model has six free parameters, we only applied it to the sources with more than six data points (J0024-4202, J0210-2213, J1632+3547, J1653+3503, and J2347-1856). For J0024-4202, the fit converges to a single dominant SSA component, and a second component is not required (F-test $p = 0.92$). For J1632+3547 and J1653+3503, the fits did not yield physically meaningful solutions. We therefore report the two-component results only for J0210-2213 and J2347-1856 (Table~\ref{tab:SSA2table}). For these two sources, the two-component model improves on a single SSA component, but it remains strongly disfavored with respect to the inhomogeneous external FFA model ($\Delta\mathrm{BIC} = 25$ and 16, respectively). The low-frequency flattening of these spectra is therefore better explained by an inhomogeneous ionized screen than by a superposition of synchrotron components with different turnover frequencies. This mirrors the result obtained for the GPS source PKS B0008-421, for which the inhomogeneous FFA model was also statistically favored over a double-component SSA model \citep{Callingham2015}.

While these spectral fits suggest that an ionized medium drives the low-frequency turnover via free-free absorption in some of the sources, independent multi-wavelength diagnostics are ultimately required to definitively confirm the presence and properties of this absorbing gas. The neutral and ionized phases of the surrounding medium may coexist in a multiphase interstellar environment, with warm neutral gas potentially present alongside the warm ionized medium \citep{Wolfire2003}. Therefore, high-resolution H\,{\sc i} 21-cm absorption observations could directly trace the neutral atomic phase and provide complementary information on the gaseous environment interacting with the radio lobes. Furthermore, X-ray observations provide a powerful independent metric by measuring the intrinsic equivalent hydrogen column density ($N_H$). If the radio components are indeed embedded within a dense gaseous environment causing severe FFA, this should manifest as heavily absorbed X-ray emission \citep{Siemiginowska2016, Sobolewska2019, Gan2025}.

\begin{table*}[h]
\caption{Comparison between FFA and SSA}
\vspace*{1mm}
\label{tab:FFAtable}
\centering
\begin{tabular}{cccccccccccc}
\hline\hline
Source & Model    & $S_m$ [Jy] & $\Delta S_m$ [Jy]& $\nu_m$ [GHz]& $\Delta \nu_m$ [GHz]& $\alpha$ & $\Delta \alpha$ & $p$ & $\Delta p$ & $\chi^2_{red}$ & $\Delta\mathrm{BIC}$ \\ \hline
J0024-4202      & SSA               &  7.42          &    0.50              &    1.23       &     0.02            &    -1.32        &   0.06                & -          & -                &   1.48     & 0.0    \\
                & Internal FFA      &    1.49        &    0.09              &    2.51        &   0.50               &   -0.76         &  0.09                 & -          & -                &   18.58    & 222.2     \\
                & Hom. FFA   &   0.10       &      \textless{0.01}             &     0.16       & \textless{0.01}     &     0.29       &   0.02                & -          & -                &    51.94   & 655.9    \\
                & Inhom. FFA &   5.56         &  0.45                &   1.38         &      0.10            &   -1.37         &    0.06
                &  0.85       &    0.06           &    1.63   & 3.0     \\ \hline
J0210-2213      & SSA               & 1.05          & 0.03                & 0.77          & 0.01                & -0.59          & 0.03                 & -          & -                & 24.16    & 170.6     \\
                & Internal FFA      & 0.84          & 0.02                & 1.49          & 0.15                & -0.79          & 0.06                 & -          & -                & 8.60     & 46.1     \\
                & Hom. FFA   & 0.68          & 0.03                & 0.36          & \textless{}0.01     & -0.55          & 0.03                 & -          & -                & 78.84    & 608.1     \\
                & Inhom. FFA & 1.05          & 0.06                & 2.33          & 0.21                & -1.42          & 0.14                 & 0.28       & 0.06             & 2.90     & 0.0     \\ \hline
J0923-3435      & SSA               & 0.59          & 0.02                & 0.64          & 0.04                & -0.57          & 0.03                 & -          & -                & 40.81    & 75.3     \\
                & Internal FFA      & 0.51          & 0.02                & 2.00          & 0.19                & -1.14          & 0.07                 & -          & -                & 4.87     & 3.4     \\
                & Hom. FFA   & 0.44          & 0.03                & 0.27          & 0.01                & -0.57          & 0.03                 & -          & -                & 40.78    & 75.2     \\
                & Inhom. FFA & 0.54          & 0.04                & 2.62          & 0.26                & -1.52          & 0.17                 & 0.18       & 0.08             & 4.73     & 0.0     \\ \hline
J1143+1834      & SSA               &  0.39          &    0.03              &    2.30        &    0.05              &   -0.49         &   0.10                & -          & -                &    0.29  & 0.0      \\
                & Internal FFA      & 0.35             & 0.03                   & 1.91            & 0.28                   & -0.56              & 0.12                    & -          & -                & 0.30   & 0.1     \\
                & Hom. FFA   &   0.25         &      0.02            &    1.24        &     0.14             &    -0.54        &    0.11               & -          & -                &   0.30   & 0.0      \\
                & Inhom. FFA & -             & -                   & -             & -                   & -              & -                    & -          & -                & -        & -     \\ \hline
J1218-2159      & SSA               & 0.25          & 0.02                & 2.41          & 0.06                & -0.89          & 0.11                 & -          & -                & 5.29     & 0.2     \\
                & Internal FFA      & -             & -                   & -             & -                   & -              & -                    & -          & -                & -        & - \\
                & Hom. FFA   & 0.21          & 0.02                & 1.50          & 0.06                & -0.96          & 0.11                 & -          & -                & 5.05     & 0.0     \\
                & Inhom. FFA & -             & -                   & -             & -                   & -              & -                    & -          & -                & -        & -     \\ \hline
J1632+3547      & SSA               & 0.74          & 0.01                & 0.53          & \textless{}0.01                & -0.45          & 0.02                 & -          & -                & 46.14    & 207.6     \\
                & Internal FFA      & 0.63          & \textless{}0.01                & 0.69          & 0.02                & -0.66          & 0.03                 & -          & -                & 4.72     & 0.5     \\
                & Hom. FFA   & 0.44          & 0.01                & 0.23          & \textless{}0.01                & -0.41          & 0.02                 & -          & -                & 61.15    & 282.6     \\
                & Inhom. FFA & 0.61          & 0.02                & 0.75          & 0.05                & -0.66          & 0.03                 & -0.06      & 0.04             & 5.26     & 0.0     \\ \hline
J1653+3503      & SSA               & 0.42          & 0.01                & 0.49          & 0.02                & -0.60          & 0.03                 & -          & -                & 8.90     & 2.5     \\
                & Internal FFA      & 0.32          & 0.01                & 0.63          & 0.04                & -0.65          & 0.03                 & -          & -                & 8.26     & 0.0     \\
                & Hom. FFA   & 0.18          & 0.09                & 0.75          & 0.38                & -0.75          & 0.17                 & -          & -                & 35.88    & 110.5     \\
                & Inhom. FFA & 0.26          & 0.02                & 1.30          & 0.62                & -0.79          & 0.18                 & -0.23      & 0.03             & 10.48    & 0.3     \\ \hline
J1805-0438      & SSA               & 0.79          & 0.03                & 0.41          & 0.02                & -0.55          & 0.05                 & -          & -                & 20.46    & 11.3     \\
                & Internal FFA      & 0.79          & 0.03                & 0.58          & 0.09                & -0.94          & 0.15                 & -          & -                & 9.20     & 0.0     \\
                & Hom. FFA   & 0.54          & 0.03                & 0.22          & 0.01                & -0.58          & 0.05                 & -          & -                & 18.37    & 9.2     \\
                & Inhom. FFA & -             & -                   & -             & -                   & -              & -                    & -          & -                & -        & -     \\ \hline
J2347-1856      & SSA               & 0.77          & 0.03                & 0.53          & 0.01                & -0.35           & 0.02                 & -          & -                & 19.82   & 290.6     \\
                & Internal FFA      & 0.69          & 0.02                & 0.44          & 0.02                & -0.40          & 0.02                 & -          & -                & 13.59    & 190.7     \\
                & Hom. FFA   & 0.34          & 0.01                & 0.20          & 0.01                & -0.27           & 0.02                 & -          & -                & 28.11    & 423.1     \\
                & Inhom. FFA & 0.68          & 0.02                & 2.55          & 0.21                & -0.90          & 0.08                 & -0.34      & 0.04             & 1.58     & 0.0     \\ \hline
\end{tabular}
\tablefoot{$\Delta\mathrm{BIC}$ is given relative to the best-fitting model of each source among all tested models; $\Delta\mathrm{BIC} > 10$ indicates very strong evidence against a model.}
\end{table*}

\begin{table*}
\caption{Inhomogeneous SSA fits (sum of two SSA components) to the integrated spectra.}
\label{tab:SSA2table}
\centering
\begin{tabular}{lcccccc}
\hline\hline
Source & Comp. & $S_m$ [Jy] & $\nu_m$ [GHz] & $\alpha$ & $\chi^2_\mathrm{red}$ & $\Delta\mathrm{BIC}$ \\
\hline
J0210$-$2213 & 1 & $0.067 \pm 0.005$ & $0.24 \pm 0.03$ & $-0.33 \pm 0.11$ & 8.10 & 25.0 \\
             & 2 & $0.94 \pm 0.03$   & $1.38 \pm 0.07$ & $-0.83 \pm 0.05$ &      &      \\
\hline
J2347$-$1856 & 1 & $0.227 \pm 0.011$ & $0.11 \pm 0.03$ & $-0.33 \pm 0.12$ & 2.59 & 15.9 \\
             & 2 & $0.61 \pm 0.06$   & $1.71 \pm 0.22$ & $-0.66 \pm 0.33$ &      &      \\
\hline
\end{tabular}
\tablefoot{Component 1 (2) is the component with the lower (higher) turnover frequency. Uncertainties are derived from a parametric bootstrap. $\Delta\mathrm{BIC}$ is given relative to the inhomogeneous external FFA model, the best-fitting model for both sources (Table~\ref{tab:FFAtable}).}
\end{table*}


\subsubsection{Variability}\label{results:variability}

We analyzed the multi-frequency VLBI light curves of all sources to assess their temporal variability. A representative light curve is shown in Fig.~\ref{fig:timevarexample}, while the full set for all sources is available in Appendix~\ref{app:vlbilightcurves}. Whenever possible, we selected observing bands with more than three epochs to ensure a meaningful trend analysis. For sources lacking this cadence (J0923-3435 and J1653+3503), we instead used the band with the largest number of available epochs.

An important caveat applies to all light curves: in most sources, the flux densities of the individual components vary in a strongly correlated way from epoch to epoch (e.g., Fig.~\ref{fig:timevarexample}). Such correlated fluctuations are the expected signature of uncertainties in the absolute amplitude calibration of the VLBI data, which scale the flux densities of all components in a given epoch by the same factor, rather than of intrinsic variability. Indeed, when considering the mean flux densities and their associated $1\sigma$ standard deviations, none of the individual components exhibits statistically significant variability: all measurements remain consistent with a constant flux density at the $2\sigma$ level. The variability estimates derived below should therefore be regarded as upper limits on the intrinsic variability.

To quantify the variability of each individual component, we computed the maximum fractional variability per year, $V$, following the definition of \citet{Kiehlmann2024}:

\begin{equation} 
    V = \frac{S_\mathrm{max}-S_\mathrm{min}}{S_\mathrm{min}} \cdot \frac{1+z}{\Delta t}, 
\end{equation} 

where $S_\mathrm{max}$ and $S_\mathrm{min}$ are the maximum and minimum measured flux densities, respectively, and $\Delta t$ is the elapsed time between those epochs. Since this estimator is highly sensitive to flux-density uncertainties, we propagated the $1\sigma$ measurement errors to derive conservative lower and upper limits on $V$. Only sources with known redshifts were considered. 

Using the lower limits on the variability, only J1632+3547 and J2347-1856 remain formally consistent with the $\leq20\%$~yr$^{-1}$ threshold proposed by \citet{Kiehlmann2024} for CSOs. However, given the calibration effects described above, and because that criterion was established using integrated source flux densities rather than individual VLBI components, the derived values of $V$ cannot be regarded as robust estimates of the intrinsic variability and do not provide a reliable discriminator between CSOs and milli-lenses in our dataset. Nevertheless, the absence of statistically significant variability in any of the sources remains fully compatible with the CSO interpretation. A definitive assessment of the intrinsic variability of these candidates will require more densely sampled monitoring observations with improved absolute flux calibration.

\begin{figure}[h]
     \centering
\includegraphics[width=0.45\textwidth]{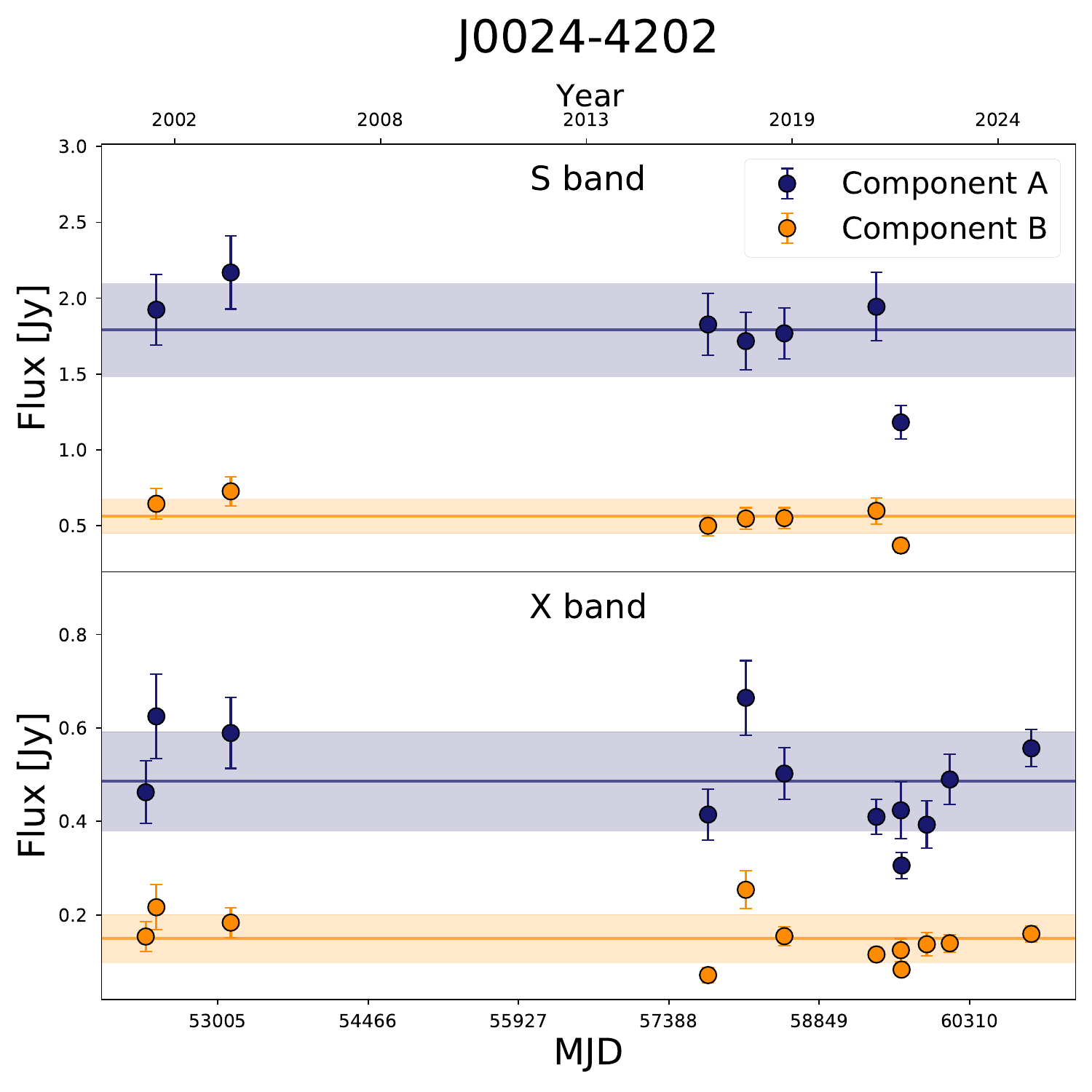}
\caption{VLBI light curves of the individual components for J0024-4202. The shaded areas correspond to the 1-$\sigma$ deviation from the mean flux of each component at each band.}
\label{fig:timevarexample}
\end{figure}


\subsubsection{Magnetic fields}\label{sec:magfields}

We estimated the magnetic field of each component with the two methods described in Sect.~\ref{sec:magmethods}. The equipartition field, $B_\mathrm{eq}$, was computed for all sources with known redshifts. The SSA field, $B_\mathrm{SSA}$, was computed only for the three sources with redshifts and whose VLBI spectra constrain the turnover (J0024-4202, J1143+1834, and J2347-1856). The resulting magnetic field ranges are listed in Table~\ref{tab:Bfield}, together with the number of datasets retained for each component.

With the exception of component B of J1653+3503, the equipartition fields lie between $\sim$20 and $\sim$320\,mG, with typical medians of 50--150\,mG, comparable to values reported for CSO hot spots (e.g., \citealt{Kameno2000,Orienti2008,Orienti2012,Surti2026}). The two components of each source have similar equipartition fields, as expected for the symmetric mini-lobes of a CSO. The much higher field of component B of J1653+3503 (414--1667\,mG) should be regarded as unreliable. The very steep spectral index ($\alpha = -3.03 \pm 0.35$) and the lack of data at low frequencies lead to an overestimate of the luminosity (Eq.~\ref{eq:luminosity}) and, consequently, of the magnetic field.

The comparison between the two methods differs from source to source. In J1143+1834, the SSA and equipartition ranges overlap for both components, and the same holds for component A of J0024-4202. For these components, the two independent estimates are therefore consistent for $\delta \approx 1$, as expected for the mildly relativistic, nearly symmetric lobes of a CSO. In contrast, $B_\mathrm{SSA}\,\delta^{-1}$ exceeds $B_\mathrm{eq}$ for component B of J0024-4202 by a factor of $\sim$8 between the medians, and for both components of J2347-1856, by factors of $\sim$23 and $\sim$18 for components A and B, respectively.

One possible explanation for these discrepancies is relativistic beaming. Matching the medians of the two estimates would require $\delta \approx 0.12$ for component B of J0024-4202, while component A is consistent with $\delta \approx 1$. Such differential beaming would require the approaching lobe to exhibit a blueshifted turnover frequency relative to the receding lobe \citep{Kameno2000}, which we do not observe. For J2347-1856, matching the SSA and equipartition estimates would require $\delta \approx 0.044$ and 0.055 for components A and B, respectively. The nearly identical flux densities of the two components suggest a viewing angle close to $90^\circ$, for which these Doppler factors correspond to $\beta \simeq 0.999$. Since the two lobes would then move in opposite directions, the expected separation speed is $\sim$2$c$, about five times our $3\sigma$ upper limit of 0.41$c$ (Table~\ref{tab:distances}). Relativistic beaming therefore cannot account for the discrepancies in either source.

A more plausible explanation for J2347-1856 is free-free absorption by a surrounding ionized medium, for which its integrated spectrum shows clear evidence (Sect.~\ref{sec:intfluxspectra}). If the observed spectral peak is shaped by FFA rather than by synchrotron opacity, the turnover parameters do not describe an SSA turnover and Eq.~\ref{eq:pushkarevB} does not yield a valid magnetic field. Unfortunately, the sparse frequency coverage of our VLBI data in the optically thick regime prevents us from confirming FFA on parsec scales. Different amounts of free-free absorption along the lines of sight to the two lobes could also explain the different behavior of the two components of J0024-4202 \citep{Kameno2000}.

In addition, $B_\mathrm{SSA}$ is extremely sensitive to the measured parameters. Because of its $\theta^4$ dependence, approximating the size of the emitting region as 1.8 times the FWHM of the fitted Gaussian component can introduce uncertainties of up to an order of magnitude for non-spherical or partially resolved emitting regions, and an overestimated size leads to an overestimated field. Likewise, the $\nu_m^5$ dependence turns modest uncertainties in the turnover frequency into large uncertainties in $B_\mathrm{SSA}$. These effects are reflected in the much wider SSA ranges in Table~\ref{tab:Bfield}, which span up to more than an order of magnitude for a single component. We therefore regard the equipartition estimates as the more reliable measure of the magnetic field in these sources.

\begin{table}
\caption{Magnetic field ranges of the source components.}
\label{tab:Bfield}
\centering
\begin{tabular}{llccc}
\hline\hline
Source & Method & Comp. & $B$ (mG) & $N$ \\
\hline
J0024$-$4202 & SSA & A & 20--138$\times\delta$    & 6/7 \\
             &  & B & 167--4050$\times\delta$  & 7/7 \\
             & Equip.  & A & 30--119    & 21/24 \\
             &   & B & 25--130    & 23/24 \\
\hline
J0210$-$2213 & Equip.  & A & 68--158    & 9/10 \\
             &   & B & 74--230    & 11/11 \\
\hline
J1143+1834   & SSA & A & 48--409$\times\delta$    & 5/5 \\
             &  & B & 157--1392$\times\delta$  & 5/5 \\
             & Equip.  & A & 32--269    & 31/31 \\
             &   & B & 44--284    & 29/31 \\
\hline
J1632+3547   & Equip.  & A & 60--315    & 10/10 \\
             &   & B & 50--202    & 8/8 \\
\hline
J1653+3503   & Equip.  & A & 33--263    & 5/5 \\
             &   & B & 414--1667  & 4/4 \\
\hline
J2347$-$1856 & SSA & A & 765--1881$\times\delta$  & 5/5 \\
             &  & B & 212--2532$\times\delta$  & 5/5 \\
             & Equip. & A & 26--161    & 23/23 \\
             &   & B & 21--122    & 22/23 \\
\hline
\end{tabular}
\tablefoot{$N$ denotes the number of datasets retained for the analysis relative to the total number available. Datasets for which the median $B$ deviated by more than $5\sigma$ from the median across datasets for the corresponding component were excluded.}
\end{table}


\subsection{Individual sources}\label{sec:ressources}

In this subsection we comment on the physical nature of each of the nine milli-lens candidates by integrating the results of our morphological, spectral, and kinematic studies. As anticipated from the preliminary classifications presented in \citet{Potzl2025}, and given the known contaminant populations in the search for milli-lenses, our analysis confirms that the vast majority of these sources are CSOs. Below, we summarize the principal observational evidence supporting the classification of each candidate, highlighting the diagnostics that distinguish intrinsic compact radio sources from genuine milli-lenses.

\subsubsection{J0024-4202}

J0024-4202 is best interpreted as a CSO despite satisfying all initial milli-lens selection criteria. The source exhibits two steep-spectrum lobes with equipartition magnetic fields consistent with typical CSO hot spots. The gravitational lens selection criteria are formally satisfied across all available epochs and frequencies. The very high-sensitivity observations (rms = 0.13 mJy\,beam$^{-1}$, dynamic range $>450$) reveal no central core-like component. However, the main evidence against the milli-lens hypothesis is the substantial difference in the morphologies of the two components. Notably, component B exhibits significant changes between epochs, occasionally appearing highly elongated along the North--South direction. This apparent distortion is likely exacerbated by the source's low declination and the resulting highly elongated beam. For component A, the SSA and equipartition magnetic fields agree for $\delta \approx 1$, supporting the interpretation of the components as intrinsic radio hot spots. For component B, the SSA estimate exceeds the equipartition value by a factor of $\sim$8, which is more plausibly explained by the strong sensitivity of $B_\mathrm{SSA}$ to the component size and turnover frequency than by relativistic beaming; free-free absorption is not required by the integrated spectrum, which is well described by a single SSA component. We therefore classify J0024-4202 as a CSO.

\subsubsection{J0210-2213}

J0210-2213 is a clear example in our sample of a CSO affected by free-free absorption. The morphology observed in the most recent, high-sensitivity radio maps clearly identifies this source as a CSO, definitively ruling out a gravitational lensing interpretation. The images reveal a central component with a flatter spectral index flanked by extended emission, an overall structure perfectly consistent with an edge-brightened CSO 2 \citep{Kiehlmann2024}. Furthermore, the integrated radio spectrum shows a distinct flattening in the optically thick regime, deviating from the canonical $+5/2$ slope expected for homogeneous SSA emission. The fit of alternative FFA models supports the presence of an ionized absorbing medium, a feature previously documented in a select number of CSOs \citep{Kameno2000, Marr2001}. An inhomogeneous SSA model consisting of two synchrotron components is strongly disfavored ($\Delta\mathrm{BIC} = 25$), reinforcing the FFA interpretation.


\subsubsection{J0923-3435}

J0923-3435 remains the only source for which the available observations do not allow a definitive classification. The source was previously unclassified in the literature, and, based on the limited dataset currently available, all empirical lens selection criteria remain satisfied. The only morphological evidence challenging the milli-lens scenario is a tentative third component located east of component B, faintly visible in an archival Astrogeo image. However, the limited sensitivity of this specific map makes the detection highly uncertain. New high-sensitivity observations have been awarded to directly test this scenario: HSA observations at C- and X-band in semester 2026B (BA201; PI: \'Alvarez-Ortega), with expected image rms noise levels of 4.4 and 4.9\,$\mu$Jy\,beam$^{-1}$, respectively, and Long Baseline Array (LBA) observations at X-band in the second semester of 2026 (V701a; PI: \'Alvarez-Ortega), with an expected rms noise of 9.8\,$\mu$Jy\,beam$^{-1}$. The HSA data at two frequencies will provide spectral indices for all components, while the LBA data improve the $(u,v)$ coverage for this southern source. Confirmation of the third component as a flat-spectrum radio core would unambiguously point to a CSO classification. Its integrated spectrum is poorly described by SSA and favors free-free absorption, although internal and inhomogeneous external absorbers cannot be distinguished ($\Delta\mathrm{BIC} = 3.4$).

\subsubsection{J1143+1834}

The newly detected bridging emission provides decisive evidence that J1143+1834 is a CSO rather than a milli-lens. The source had previously been proposed as either a CSO or a SMBHB, while also satisfying all the selection criteria for a milli-lens candidate. Recent high-frequency (U- and K-band) observations reveal clear, extended emission connecting component A to component B. Together with the individual VLBI spectra, this confirms its classification as a CSO. The magnetic fields support this interpretation: for both components, the SSA and equipartition estimates agree for $\delta \approx 1$, as expected for CSO hot spots (Sect.~\ref{sec:magfields}). The only remaining caveat is the apparent K-band variability inferred from observations obtained with different VLBI arrays. Although this would formally be inconsistent with the variability criterion proposed by \citet{Kiehlmann2024}, the comparison is not straightforward because that criterion was defined using integrated source flux densities, whereas our analysis is based on individual VLBI components. Moreover, as discussed in Sect.~\ref{results:variability}, residual systematic effects cannot be excluded with the available data.

\subsubsection{J1218-2159}

\begin{figure}[h]
     \centering
\includegraphics[width=0.49\textwidth]{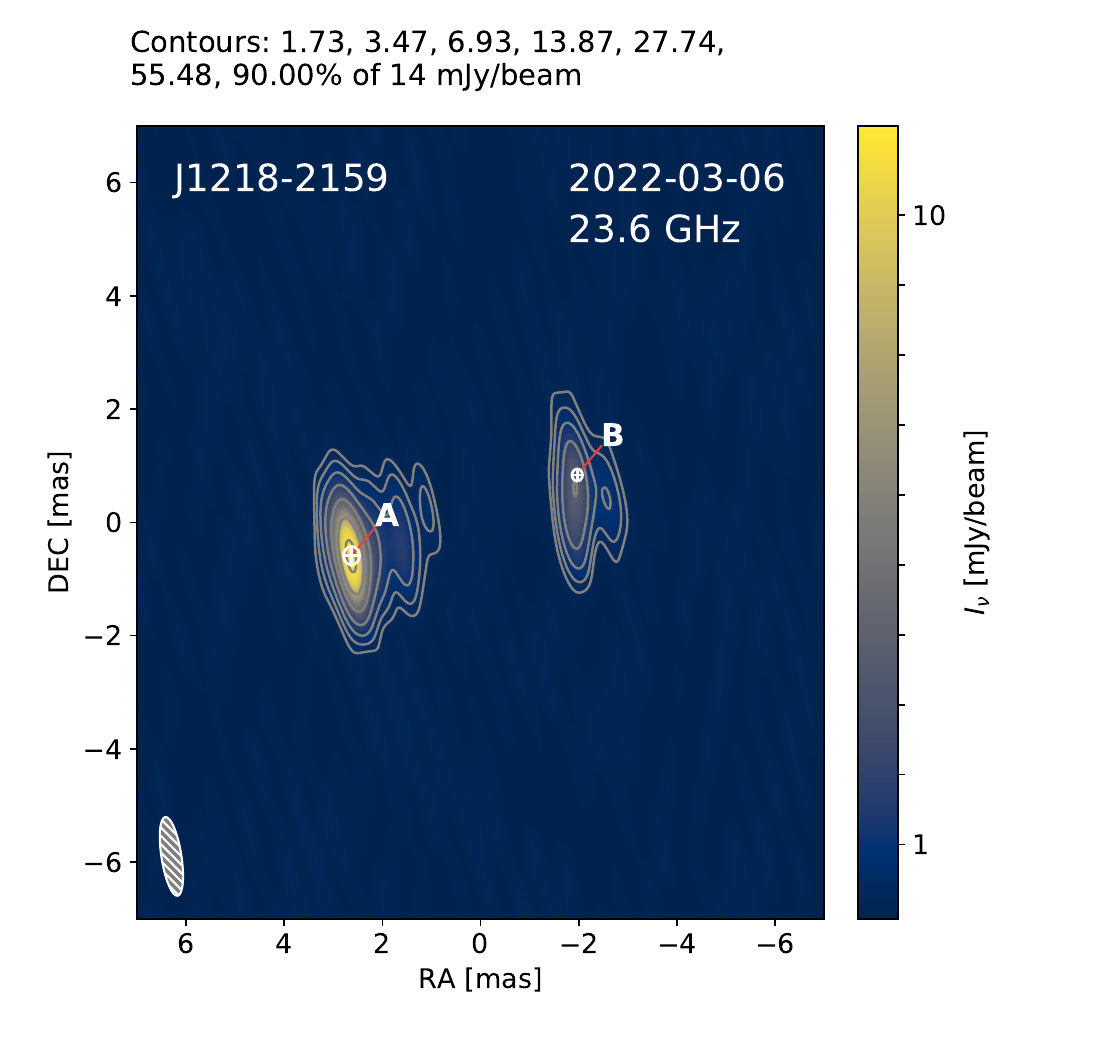}
\caption{VLBI intensity map of J1218-2159 at K-band. Lowest contours are drawn at four times the rms of the residuals from the CLEAN model. The map shows a continuous emission feature extending from component A to component B.}
\label{fig:J1218KBand}
\end{figure}

\begin{figure}[h]
     \centering
\includegraphics[width=0.45\textwidth]{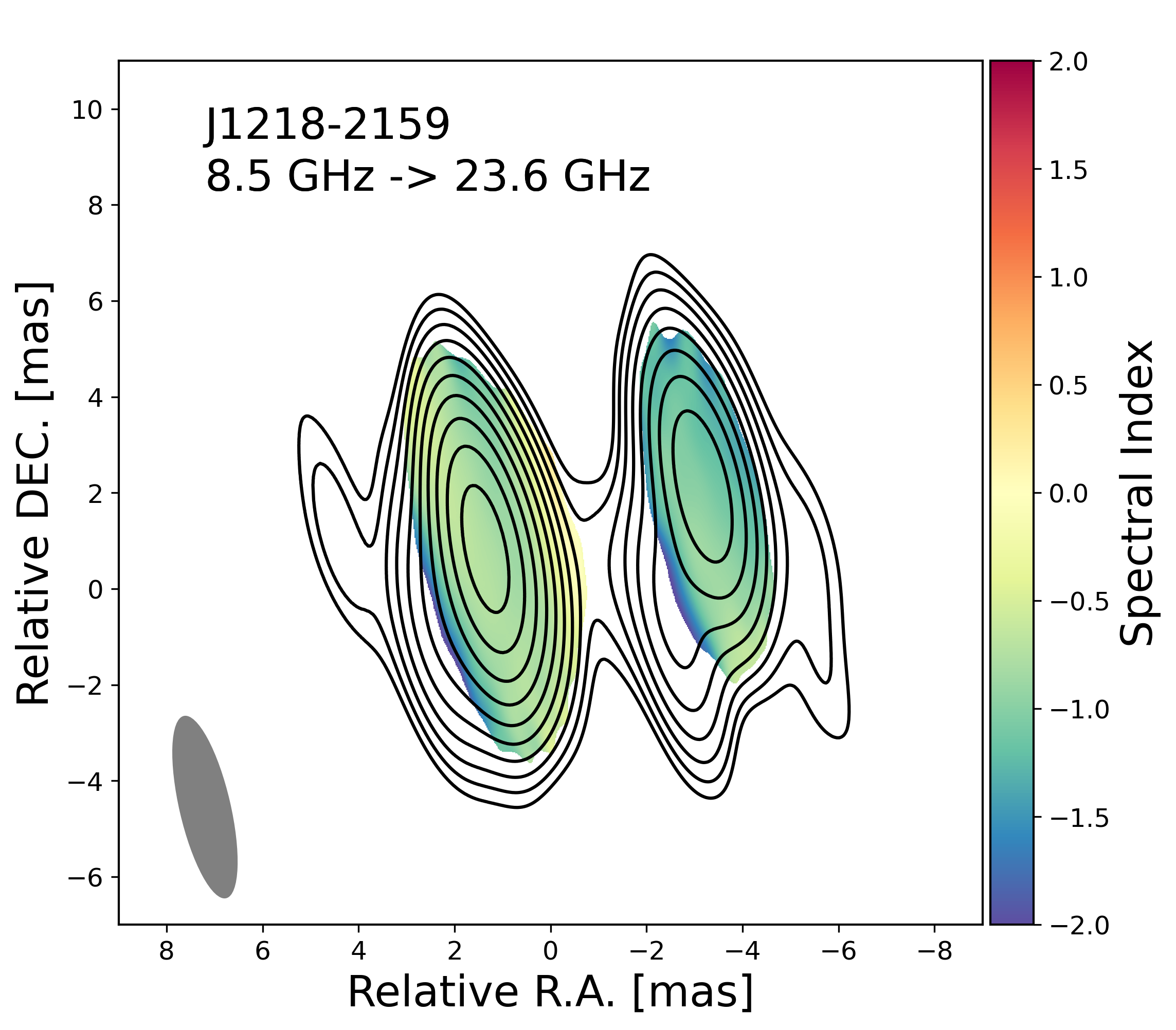}
\caption{Spectral index map of J1218-2159 between X- and K-band. The central emission shows a flatter spectral index, suggesting the presence of a radio core.}
\label{fig:J1218KBandspix}
\end{figure}

The detection of a central flat-spectrum component and bridging emission provides compelling evidence that J1218-2159 hosts an active radio core. While previously unclassified, the combination of its VLBI morphology and spatially resolved spectral properties now identifies it as a CSO. Although the source formally satisfies the lensing criteria across all epochs, its component spectra differ slightly, with component A exhibiting a flatter slope. Crucially, in two epochs (2019-05-28 at X-band; 2022-03-06 at C-band), a weak central component of a few mJy is detected between the main lobes. The sole available K-band observation corroborates this, showing emission that extends from component A toward component B (Fig.~\ref{fig:J1218KBand}). The spectral index map generated between the X- and K-bands isolates this central bridging emission, revealing it has a significantly flatter spectrum than either of the outer lobes (Fig.~\ref{fig:J1218KBandspix}). This morphology identifies a radio core situated between the two lobes of a CSO. Similar to other sources in the sample, the optically thick portion of the integrated spectrum is flatter than predicted by homogeneous SSA, hinting at free-free absorption, although the sparse frequency coverage currently limits the reliability of the spectral fitting.


\subsubsection{J1632+3547}

The resolved spectral-index map clearly identifies J1632+3547 as a CSO rather than a supermassive black hole binary or a gravitational lens. The source had previously been suggested to host a SMBHB. Although it preserves flux and surface brightness ratios within measurement errors, the spectral indices of components A and B differ significantly enough to rule out a gravitational lensing scenario. Instead, the resolved spectral-index structure (Fig.~\ref{fig:spixmapsexample}; see also the VLBI component maps in Fig.~\ref{fig:vlbimaps}) reveals a flat-spectrum core (sub-component A0) with steep-spectrum jet or lobe emission on both sides (A1, A2, and B), the characteristic structure of a CSO. The integrated spectrum of J1632+3547 is also poorly described by a single SSA component ($\chi^2_\mathrm{red} = 46.1$), whereas both the internal and the inhomogeneous external FFA models provide statistically equivalent and much better fits ($\chi^2_\mathrm{red} = 4.7$ and 5.3; $\Delta\mathrm{BIC} = 0.5$ between them and $\Delta\mathrm{BIC} = 208$ for SSA; Table~\ref{tab:FFAtable}), suggesting that free-free absorption also shapes its low-frequency spectrum. 

\subsubsection{J1653+3503}
J1653+3503 provides a clear example of how strong spectral asymmetries can distinguish intrinsic CSOs from milli-lenses. Previously unclassified in the literature, the source explicitly violates the spectral-index conservation criterion required for gravitational lensing. The two components exhibit significantly different spectral slopes ($\alpha_A=-1.07$ versus $\alpha_B=-3.03$), confirming they are distinct physical regions rather than lensed images. A plausible explanation is that the two advancing lobes are expanding into a highly asymmetric ambient medium. Variations in the surrounding density naturally affect particle acceleration and radiative losses, producing the observed spectral differences. Similar severe spectral asymmetries between lobes have been documented in other CSOs (e.g., J1700+3830; \citealt{Tremblay2016}). Additionally, the integrated spectrum shows a significant flux excess compared to the sum of the VLBI components, particularly at higher frequencies, suggesting the presence of extended, low-surface-brightness emission resolved out by the VLBI baselines. The equipartition magnetic field of component A (33--263\,mG) is typical of CSO hot spots, whereas the estimate for component B is unreliable because of its extremely steep spectrum (Sect.~\ref{sec:magfields}).

\subsubsection{J1805-0438}

The combined morphological and spectral analyses (Figs.~\ref{fig:vlbimaps} and~\ref{fig:vlbispectra}) demonstrate that the three main components of J1805-0438 are intrinsic to the source rather than multiple lensed images, and identify it as a CSO. The eastern component is well described by an SSA spectrum, while the central and western components follow steep power laws. The central component has the flattest spectrum of the three and most likely marks the region of the core, flanked on both sides by steep-spectrum emission, the characteristic structure of a CSO. Although the source was previously listed as a BL Lac candidate \citep{DAbrusco2019}, its two-sided morphology, steep component spectra, and the absence of significant variability argue against strong relativistic beaming.

\subsubsection{J2347-1856}

J2347-1856 provides one of the strongest confirmations of the CSO interpretation through the combination of bridging emission, remarkable flux symmetry, and long-term flux stability. Previously identified as a CSO, our new observations provide strong independent confirmation of this classification. Similar to J1143+1834, while the preliminary, lower-sensitivity VLBI data satisfied all milli-lens selection criteria, our recent high-sensitivity observations unambiguously reveal extended, contiguous emission bridging the two main components. This morphology strongly indicates that both components are lobes belonging to a single radio source. The near-perfect flux symmetry between the two lobes suggests a viewing angle very close to $90^\circ$. Furthermore, over the 27-year observing baseline, neither component exhibits statistically significant flux density variability, with all measured variations remaining below the $2\sigma$ level (Fig.~\ref{fig:timevar}). Finally, the SSA magnetic fields of both components exceed the equipartition values by factors of $\sim$20, a discrepancy that relativistic beaming cannot explain given the upper limit on the separation speed of the components (Sect.~\ref{sec:magfields}). Combined with the pronounced low-frequency flattening in the integrated spectrum, this indicates that the spectral turnover is substantially modified by free-free absorption, rendering the magnetic fields inferred under the assumption of homogeneous SSA unreliable. This interpretation is further supported by the model comparison: neither a single nor a double SSA component can reproduce the flat low-frequency spectrum ($\Delta\mathrm{BIC} = 290$ and 16 with respect to the inhomogeneous external FFA model).


\section{Discussion}\label{sec:discussion}

In this section, we place our results in a broader context. We first discuss the implications of the pilot study for the search for milli-lenses and for the main SMILE survey (Sect.~\ref{sec:disc_smile}), and then what this study has taught us about CSOs (Sect.~\ref{sec:disc_cso}).


\subsection{Milli-lenses and the SMILE survey}\label{sec:disc_smile}

The absence of confirmed milli-lenses in the pilot sample is consistent with theoretical expectations. Milli-lens systems are intrinsically rare: \citet{Wilkinson2001} found none among 300 compact radio sources, and the predictions of \citet{Loudas2022} for the $\sim$5000 sources of SMILE range from zero to a few lensed systems when considering dark matter halos as lenses. However, we cannot directly extrapolate this result to the expected outcome in the SMILE survey, given the differences between the samples used in the pilot study and the main survey. The pilot sample was drawn from the Astrogeo VLBI image database, which combines observations from many different programmes, with heterogeneous sensitivities, frequencies, and $(u,v)$ coverages, and has no uniform selection criteria. Candidates were identified by visual inspection of these images, and many sources were observed in only a few epochs or with limited sensitivity. In contrast, the main SMILE survey is drawn from a complete, flux-density-limited sample of flat-spectrum sources, with VLBI data selected to guarantee a minimum sensitivity \citep{Casadio2021}, which will allow the detection rate, or an upper limit on it, to be translated directly into constraints on the abundance of low-mass halos \citep{Loudas2022}.

Building on the classifications of \citet{Potzl2025}, we have established the nature of four previously unclassified or differently classified sources in the pilot study (J0024-4202, J1218-2159, J1653+3503, and J1805-0438) and ruled out the SMBHB interpretation proposed for J1143+1834 and J1632+3547. While the flux-ratio, surface-brightness, and proper-motion criteria were effective in reducing the initial 40 candidates to nine \citep{Potzl2025}, most of the remaining candidates satisfy them at all epochs, so they cannot distinguish these sources from milli-lenses. The decisive diagnostics were instead i) high-sensitivity imaging, which revealed bridging emission or central cores in J0210-2213, J1143+1834, J1218-2159, and J2347-1856, and ii) spatially resolved or component-wise spectral information, which showed incompatible spectral indices in J1653+3503, J1632+3547, and J1805-0438. For the main SMILE survey, the follow-up of candidates should therefore prioritize deep, multi-frequency VLBI observations over the accumulation of additional epochs.


\subsection{The sample in the context of CSO studies}\label{sec:disc_cso}

The CSOs confirmed in this work add to the still limited population of well-characterized CSOs. The most recent compilations comprise 144 bona fide CSOs \citep{Kiehlmann2024, Sheldahl2025}, and the four sources that we classify as CSOs for the first time (J0024-4202, J1218-2159, J1653+3503, and J1805-0438) increase this number by $\sim$3\%. More importantly, rather than merely classifying these sources, we characterize their physical properties by combining multi-epoch, multi-frequency VLBI imaging with spatially resolved spectral indices, broadband spectral modeling of the absorption mechanisms, limits on expansion speeds and kinematic ages, and magnetic field estimates.

These properties can be compared with the predictions of the two classic scenarios introduced in Sect.~\ref{sec:introduction}. The equipartition magnetic fields of the lobes (20--320\,mG) are comparable to those reported for CSO hot spots \citep{Kameno2000, Orienti2008, Orienti2012}, and in J1143+1834 and component A of J0024-4202 the independent SSA estimates agree with them for $\delta \approx 1$. This indicates hot spots close to equipartition, as expected for young, freely expanding sources \citep{Orienti2008}. On the other hand, the integrated spectra of J0210-2213 and J2347-1856 are best described by an inhomogeneous free-free absorbing screen, the model predicted for radio lobes expanding into a dense, clumpy medium \citep{Bicknell1997} and previously found for other compact sources \citep{Kameno2000, Marr2001, Tingay2015, Callingham2015}. The strong spectral asymmetry between the lobes of J1653+3503 also points to an inhomogeneous environment, as observed in other CSOs \citep{Tremblay2016}. The two scenarios are, however, not mutually exclusive: a dense ionized medium can shape the low-frequency spectrum without necessarily confining the source. The kinematic constraints cannot yet discriminate between them, since our upper limits on the separation speeds (0.41--14.7\,$c$) remain well above the typical hot spot separation speeds of a few tenths of the speed of light \citep{Polatidis2003, An2012}. Overall, our results are consistent with CSOs being compact sources whose lobes are close to equipartition but embedded, at least in some cases, in a dense, inhomogeneous medium. They are also consistent with the absence of strong variability and superluminal motion that characterizes the bona fide CSO population \citep{Kiehlmann2024}. Tighter constraints on the expansion speeds, from longer time baselines and more uniform VLBI monitoring, will be required to test whether these sources are young or confined.

Moreover, the nine sources analyzed here are the milli-lens candidates that remained after the analysis of the 40 initial candidates in the SMILE pilot study presented in \citet{Potzl2025}. Of the other 31 sources rejected as milli-lenses, 27 exhibited CSO-like properties and could be further investigated using the same multi-epoch, multi-frequency approach presented here. This would help confirm their nature and expand the sample of well-characterized CSOs.

\section{Conclusions}\label{sec:conclusions}

We have presented a comprehensive follow-up of the nine final milli-lens candidates of the SMILE pilot project, combining all available multi-epoch, multi-frequency VLBI observations with single-dish and connected-element interferometer data. Our main conclusions are as follows:

\begin{enumerate}
    \item Eight of the nine candidates are not milli-lenses but CSOs. Four of them (J0024-4202, J1218-2159, J1653+3503, and J1805-0438) are classified as CSOs for the first time, and the SMBHB interpretation previously proposed for J1143+1834 and J1632+3547 is ruled out. All eight share the characteristic properties of CSOs: compact, two-sided morphologies with sub-kpc projected sizes, steep optically thin spectra, and no significant proper motion or flux-density variability over baselines of up to three decades.
     \item High-sensitivity imaging and component-wise spectral information provide diagnostics complementary to the lensing criteria of \citet{Potzl2025}: they revealed bridging emission or central cores, and incompatible spectral indices, in candidates that satisfy the flux-ratio, surface-brightness, and proper-motion criteria.
    \item J0923-3435 remains the only viable milli-lens candidate in the pilot sample. Deep HSA and LBA observations scheduled for 2026 will test the reality of the tentative third component that would argue against lensing.
    \item The equipartition magnetic fields of the lobes are 20--320\,mG, typical of CSO hot spots. The SSA estimates are consistent with these values for J1143+1834 and for component A of J0024-4202, but exceed them for component B of J0024-4202 and for J2347-1856.
    \item The integrated spectra of J0210-2213 and J2347-1856 cannot be explained by a single homogeneous SSA component and are best described by an inhomogeneous free-free absorbing screen, pointing to a dense, clumpy ionized medium around the lobes. The upper limits on the separation speeds (0.41--14.7\,$c$) imply kinematic ages of at least 54--951\,yr, but are insufficient to determine whether these sources are young or confined.
\end{enumerate}

The absence of milli-lenses in the pilot sample is consistent with theoretical expectations, and the main SMILE survey, drawn from a complete sample, will apply the framework established here both to search for milli-lenses and to characterize the large population of CSOs it is expected to contain.


\begin{acknowledgements}
D.A., F.P. \& C.C. acknowledge support from the European Research Council (ERC) under the Horizon ERC Grants 2021 programme under grant agreement No. 101040021.
\end{acknowledgements}

\bibliographystyle{aa}
\bibliography{mybib.bib}

\begin{appendix}

\section{Used data}
\subsection{VLBI observations}
\begin{table*}
\centering
\caption{VLBI datasets used in the study. Superscript $A$ indicate data obtained from the Astrogeo database, superscript $E$ are data observed with the EVN.}
\label{table:usedvlbidata}
\renewcommand{\arraystretch}{1.2}
\begin{adjustbox}{max width=1\textwidth}
\begin{tabular}{ccccccccccccc}
\multicolumn{6}{c}{J0024-4202} & \multicolumn{1}{c}{} & \multicolumn{6}{c}{J1143+1834} \\ \cline{1-6} \cline{8-13}
\multicolumn{1}{c}{\raisebox{-0.6\normalbaselineskip}[0pt][0pt]{Project}} & \multicolumn{1}{c}{Freq.} & \multicolumn{1}{c}{\raisebox{-0.6\normalbaselineskip}[0pt][0pt]{Date}} & \multicolumn{1}{c}{Exp.} & \multicolumn{1}{c}{rms} & \multicolumn{1}{c}{Beam} & \multicolumn{1}{c}{} & \multicolumn{1}{c}{\raisebox{-0.6\normalbaselineskip}[0pt][0pt]{Project}} & \multicolumn{1}{c}{Freq.} & \multicolumn{1}{c}{\raisebox{-0.6\normalbaselineskip}[0pt][0pt]{Date}} & \multicolumn{1}{c}{Exp.} & \multicolumn{1}{c}{rms} & \multicolumn{1}{c}{Beam} \\ 
\multicolumn{1}{c}{} & \multicolumn{1}{c}{[GHz]} & \multicolumn{1}{c}{} & \multicolumn{1}{c}{[s]} & \multicolumn{1}{c}{[mJy\,beam$^{-1}$]} & \multicolumn{1}{c}{[mas$\times$mas]} & \multicolumn{1}{c}{} & \multicolumn{1}{c}{} & \multicolumn{1}{c}{[GHz]} & \multicolumn{1}{c}{} & \multicolumn{1}{c}{[s]} & \multicolumn{1}{c}{[mJy\,beam$^{-1}$]} & \multicolumn{1}{c}{[mas$\times$mas]} \\ \cline{1-6} \cline{8-13}
\multicolumn{1}{c}{BF071$^{A}$} & \multicolumn{1}{c}{8.65} & \multicolumn{1}{c}{2002-01-31} & \multicolumn{1}{c}{180} & \multicolumn{1}{c}{1.59} & \multicolumn{1}{c}{2.21$\times$0.79} & \multicolumn{1}{c}{} & \multicolumn{1}{c}{BB023$^{A}$} & \multicolumn{1}{c}{2.27} & \multicolumn{1}{c}{1996-03-13} & \multicolumn{1}{c}{284} & \multicolumn{1}{c}{0.98} & \multicolumn{1}{c}{8.35$\times$3.85} \\ 
\multicolumn{1}{c}{BF071$^{A}$} & \multicolumn{1}{c}{2.30} & \multicolumn{1}{c}{2002-05-14} & \multicolumn{1}{c}{189} & \multicolumn{1}{c}{2.91} & \multicolumn{1}{c}{7.86$\times$2.80} & \multicolumn{1}{c}{} & \multicolumn{1}{c}{BB023$^{A}$} & \multicolumn{1}{c}{8.34} & \multicolumn{1}{c}{1996-03-13} & \multicolumn{1}{c}{284} & \multicolumn{1}{c}{1.44} & \multicolumn{1}{c}{2.19$\times$1.00} \\ 
\multicolumn{1}{c}{BF071$^{A}$} & \multicolumn{1}{c}{8.65} & \multicolumn{1}{c}{2002-05-14} & \multicolumn{1}{c}{189} & \multicolumn{1}{c}{1.72} & \multicolumn{1}{c}{2.26$\times$0.80} & \multicolumn{1}{c}{} & \multicolumn{1}{c}{BP041} & \multicolumn{1}{c}{4.99} & \multicolumn{1}{c}{1998-03-17} & \multicolumn{1}{c}{1800} & \multicolumn{1}{c}{0.29} & \multicolumn{1}{c}{3.36$\times$1.80} \\ 
\multicolumn{1}{c}{RDV44} & \multicolumn{1}{c}{2.31} & \multicolumn{1}{c}{2004-05-06} & \multicolumn{1}{c}{1560} & \multicolumn{1}{c}{1.04} & \multicolumn{1}{c}{8.20$\times$2.86} & \multicolumn{1}{c}{} & \multicolumn{1}{c}{BP041} & \multicolumn{1}{c}{8.42} & \multicolumn{1}{c}{1998-03-17} & \multicolumn{1}{c}{1800} & \multicolumn{1}{c}{0.26} & \multicolumn{1}{c}{2.08$\times$1.17} \\ 
\multicolumn{1}{c}{RDV44} & \multicolumn{1}{c}{8.65} & \multicolumn{1}{c}{2004-05-06} & \multicolumn{1}{c}{1560} & \multicolumn{1}{c}{0.72} & \multicolumn{1}{c}{9.25$\times$1.06} & \multicolumn{1}{c}{} & \multicolumn{1}{c}{BP065} & \multicolumn{1}{c}{8.42} & \multicolumn{1}{c}{2000-03-25} & \multicolumn{1}{c}{3807} & \multicolumn{1}{c}{0.18} & \multicolumn{1}{c}{2.20$\times$1.07} \\ 
\multicolumn{1}{c}{BC216} & \multicolumn{1}{c}{1.55} & \multicolumn{1}{c}{2014-02-13} & \multicolumn{1}{c}{3134} & \multicolumn{1}{c}{0.97} & \multicolumn{1}{c}{38.73$\times$5.79} & \multicolumn{1}{c}{} & \multicolumn{1}{c}{BT055} & \multicolumn{1}{c}{15.36} & \multicolumn{1}{c}{2001-01-05} & \multicolumn{1}{c}{1419} & \multicolumn{1}{c}{0.20} & \multicolumn{1}{c}{1.11$\times$0.57} \\ 
\multicolumn{1}{c}{BC216} & \multicolumn{1}{c}{4.87} & \multicolumn{1}{c}{2014-02-17} & \multicolumn{1}{c}{3134} & \multicolumn{1}{c}{1.00} & \multicolumn{1}{c}{12.33$\times$1.84} & \multicolumn{1}{c}{} & \multicolumn{1}{c}{BP104} & \multicolumn{1}{c}{8.42} & \multicolumn{1}{c}{2002-12-02} & \multicolumn{1}{c}{3807} & \multicolumn{1}{c}{0.15} & \multicolumn{1}{c}{3.44$\times$1.19} \\ 
\multicolumn{1}{c}{UF001$^{A}$} & \multicolumn{1}{c}{2.25} & \multicolumn{1}{c}{2017-01-16} & \multicolumn{1}{c}{148} & \multicolumn{1}{c}{1.04} & \multicolumn{1}{c}{7.45$\times$2.52} & \multicolumn{1}{c}{} & \multicolumn{1}{c}{BO015} & \multicolumn{1}{c}{1.67} & \multicolumn{1}{c}{2003-02-05} & \multicolumn{1}{c}{174} & \multicolumn{1}{c}{0.38} & \multicolumn{1}{c}{11.90$\times$5.11} \\ 
\multicolumn{1}{c}{UF001$^{A}$} & \multicolumn{1}{c}{8.65} & \multicolumn{1}{c}{2017-01-16} & \multicolumn{1}{c}{148} & \multicolumn{1}{c}{0.64} & \multicolumn{1}{c}{2.27$\times$0.77} & \multicolumn{1}{c}{} & \multicolumn{1}{c}{BT085$^{A}$} & \multicolumn{1}{c}{4.84} & \multicolumn{1}{c}{2006-02-09} & \multicolumn{1}{c}{543} & \multicolumn{1}{c}{0.29} & \multicolumn{1}{c}{3.00$\times$1.80} \\ 
\multicolumn{1}{c}{UG002} & \multicolumn{1}{c}{2.29} & \multicolumn{1}{c}{2018-01-18} & \multicolumn{1}{c}{266} & \multicolumn{1}{c}{0.72} & \multicolumn{1}{c}{18.05$\times$3.69} & \multicolumn{1}{c}{} & \multicolumn{1}{c}{BT094$^{A}$} & \multicolumn{1}{c}{15.14} & \multicolumn{1}{c}{2008-05-15} & \multicolumn{1}{c}{1571} & \multicolumn{1}{c}{0.58} & \multicolumn{1}{c}{1.33$\times$0.68} \\ 
\multicolumn{1}{c}{UG002} & \multicolumn{1}{c}{8.67} & \multicolumn{1}{c}{2018-01-18} & \multicolumn{1}{c}{266} & \multicolumn{1}{c}{0.76} & \multicolumn{1}{c}{9.41$\times$2.66} & \multicolumn{1}{c}{} & \multicolumn{1}{c}{BC201$^{A}$} & \multicolumn{1}{c}{8.35} & \multicolumn{1}{c}{2012-02-20} & \multicolumn{1}{c}{117} & \multicolumn{1}{c}{0.21} & \multicolumn{1}{c}{2.32$\times$1.00} \\ 
\multicolumn{1}{c}{UG003} & \multicolumn{1}{c}{2.29} & \multicolumn{1}{c}{2019-01-27} & \multicolumn{1}{c}{118} & \multicolumn{1}{c}{2.78} & \multicolumn{1}{c}{6.96$\times$2.87} & \multicolumn{1}{c}{} & \multicolumn{1}{c}{BL196} & \multicolumn{1}{c}{2.32} & \multicolumn{1}{c}{2014-09-06} & \multicolumn{1}{c}{952} & \multicolumn{1}{c}{0.22} & \multicolumn{1}{c}{8.77$\times$3.53} \\ 
\multicolumn{1}{c}{UG003} & \multicolumn{1}{c}{8.67} & \multicolumn{1}{c}{2019-01-27} & \multicolumn{1}{c}{118} & \multicolumn{1}{c}{0.32} & \multicolumn{1}{c}{3.46$\times$1.10} & \multicolumn{1}{c}{} & \multicolumn{1}{c}{BL196} & \multicolumn{1}{c}{4.26} & \multicolumn{1}{c}{2014-09-06} & \multicolumn{1}{c}{952} & \multicolumn{1}{c}{0.06} & \multicolumn{1}{c}{4.33$\times$1.66} \\ 
\multicolumn{1}{c}{UH007} & \multicolumn{1}{c}{2.30} & \multicolumn{1}{c}{2021-07-10} & \multicolumn{1}{c}{424} & \multicolumn{1}{c}{0.92} & \multicolumn{1}{c}{9.04$\times$2.82} & \multicolumn{1}{c}{} & \multicolumn{1}{c}{BL196} & \multicolumn{1}{c}{6.56} & \multicolumn{1}{c}{2014-09-06} & \multicolumn{1}{c}{952} & \multicolumn{1}{c}{0.10} & \multicolumn{1}{c}{2.86$\times$1.12} \\ 
\multicolumn{1}{c}{UH007} & \multicolumn{1}{c}{8.67} & \multicolumn{1}{c}{2021-07-10} & \multicolumn{1}{c}{424} & \multicolumn{1}{c}{0.31} & \multicolumn{1}{c}{3.91$\times$0.89} & \multicolumn{1}{c}{} & \multicolumn{1}{c}{BL196} & \multicolumn{1}{c}{8.82} & \multicolumn{1}{c}{2014-09-06} & \multicolumn{1}{c}{952} & \multicolumn{1}{c}{0.09} & \multicolumn{1}{c}{2.14$\times$0.82} \\ 
\multicolumn{1}{c}{UH007} & \multicolumn{1}{c}{2.30} & \multicolumn{1}{c}{2022-03-04} & \multicolumn{1}{c}{466} & \multicolumn{1}{c}{0.66} & \multicolumn{1}{c}{8.56$\times$3.08} & \multicolumn{1}{c}{} & \multicolumn{1}{c}{BG219$^{A}$} & \multicolumn{1}{c}{2.29} & \multicolumn{1}{c}{2015-01-23} & \multicolumn{1}{c}{164} & \multicolumn{1}{c}{0.23} & \multicolumn{1}{c}{6.99$\times$3.10} \\ 
\multicolumn{1}{c}{UH007} & \multicolumn{1}{c}{8.67} & \multicolumn{1}{c}{2022-03-04} & \multicolumn{1}{c}{466} & \multicolumn{1}{c}{0.28} & \multicolumn{1}{c}{11.56$\times$0.89} & \multicolumn{1}{c}{} & \multicolumn{1}{c}{BG219$^{A}$} & \multicolumn{1}{c}{8.67} & \multicolumn{1}{c}{2015-01-23} & \multicolumn{1}{c}{164} & \multicolumn{1}{c}{0.14} & \multicolumn{1}{c}{1.99$\times$0.90} \\ 
\multicolumn{1}{c}{BL292} & \multicolumn{1}{c}{4.87} & \multicolumn{1}{c}{2022-03-11} & \multicolumn{1}{c}{708} & \multicolumn{1}{c}{0.30} & \multicolumn{1}{c}{3.89$\times$1.43} & \multicolumn{1}{c}{} & \multicolumn{1}{c}{UF001$^{A}$} & \multicolumn{1}{c}{8.67} & \multicolumn{1}{c}{2017-02-24} & \multicolumn{1}{c}{290} & \multicolumn{1}{c}{0.17} & \multicolumn{1}{c}{1.86$\times$0.97} \\ 
\multicolumn{1}{c}{BL292} & \multicolumn{1}{c}{8.50} & \multicolumn{1}{c}{2022-03-11} & \multicolumn{1}{c}{708} & \multicolumn{1}{c}{0.22} & \multicolumn{1}{c}{2.58$\times$0.79} & \multicolumn{1}{c}{} & \multicolumn{1}{c}{UG002$^{A}$} & \multicolumn{1}{c}{2.25} & \multicolumn{1}{c}{2018-05-07} & \multicolumn{1}{c}{292} & \multicolumn{1}{c}{0.20} & \multicolumn{1}{c}{7.13$\times$3.36} \\ 
\multicolumn{1}{c}{BL292} & \multicolumn{1}{c}{23.57} & \multicolumn{1}{c}{2022-03-11} & \multicolumn{1}{c}{576} & \multicolumn{1}{c}{0.12} & \multicolumn{1}{c}{10.25$\times$2.19} & \multicolumn{1}{c}{} & \multicolumn{1}{c}{UG002$^{A}$} & \multicolumn{1}{c}{8.65} & \multicolumn{1}{c}{2018-05-07} & \multicolumn{1}{c}{292} & \multicolumn{1}{c}{0.11} & \multicolumn{1}{c}{1.84$\times$0.91} \\ 
\multicolumn{1}{c}{UH007} & \multicolumn{1}{c}{8.67} & \multicolumn{1}{c}{2022-11-11} & \multicolumn{1}{c}{468} & \multicolumn{1}{c}{0.28} & \multicolumn{1}{c}{8.02$\times$0.81} & \multicolumn{1}{c}{} & \multicolumn{1}{c}{EC071$^{E}$} & \multicolumn{1}{c}{4.93} & \multicolumn{1}{c}{2020-02-26} & \multicolumn{1}{c}{1050} & \multicolumn{1}{c}{0.40} & \multicolumn{1}{c}{2.00$\times$1.76} \\ 
\multicolumn{1}{c}{UG004} & \multicolumn{1}{c}{8.67} & \multicolumn{1}{c}{2023-06-23} & \multicolumn{1}{c}{382} & \multicolumn{1}{c}{0.26} & \multicolumn{1}{c}{2.38$\times$0.72} & \multicolumn{1}{c}{} & \multicolumn{1}{c}{EC071$^{E}$} & \multicolumn{1}{c}{22.24} & \multicolumn{1}{c}{2020-02-26} & \multicolumn{1}{c}{3600} & \multicolumn{1}{c}{0.09} & \multicolumn{1}{c}{1.17$\times$0.98} \\ 
\multicolumn{1}{c}{BA189} & \multicolumn{1}{c}{8.67} & \multicolumn{1}{c}{2025-08-23} & \multicolumn{1}{c}{4240} & \multicolumn{1}{c}{0.13} & \multicolumn{1}{c}{2.86$\times$0.89} & \multicolumn{1}{c}{} & \multicolumn{1}{c}{BT152} & \multicolumn{1}{c}{4.87} & \multicolumn{1}{c}{2022-01-23} & \multicolumn{1}{c}{224} & \multicolumn{1}{c}{0.19} & \multicolumn{1}{c}{4.17$\times$1.80} \\ 
\multicolumn{1}{c}{BA189} & \multicolumn{1}{c}{4.87} & \multicolumn{1}{c}{2025-08-24} & \multicolumn{1}{c}{4240} & \multicolumn{1}{c}{0.31} & \multicolumn{1}{c}{3.53$\times$1.35} & \multicolumn{1}{c}{} & \multicolumn{1}{c}{BT152} & \multicolumn{1}{c}{8.37} & \multicolumn{1}{c}{2022-01-23} & \multicolumn{1}{c}{576} & \multicolumn{1}{c}{0.12} & \multicolumn{1}{c}{2.02$\times$1.07} \\ \cline{1-6}
 &  &  &  &  &  & \multicolumn{1}{c}{} & \multicolumn{1}{c}{BT152} & \multicolumn{1}{c}{15.17} & \multicolumn{1}{c}{2022-01-23} & \multicolumn{1}{c}{1600} & \multicolumn{1}{c}{0.12} & \multicolumn{1}{c}{0.99$\times$0.55} \\ 
 &  &  &  &  &  & \multicolumn{1}{c}{} & \multicolumn{1}{c}{BL292} & \multicolumn{1}{c}{4.87} & \multicolumn{1}{c}{2022-03-06} & \multicolumn{1}{c}{708} & \multicolumn{1}{c}{0.10} & \multicolumn{1}{c}{3.94$\times$1.55} \\ 
 &  &  &  &  &  & \multicolumn{1}{c}{} & \multicolumn{1}{c}{BL292} & \multicolumn{1}{c}{23.57} & \multicolumn{1}{c}{2022-03-06} & \multicolumn{1}{c}{578} & \multicolumn{1}{c}{0.07} & \multicolumn{1}{c}{0.83$\times$0.35} \\ 
\multicolumn{6}{c}{J0210-2213} & \multicolumn{1}{c}{} & \multicolumn{1}{c}{UC003} & \multicolumn{1}{c}{23.46} & \multicolumn{1}{c}{2023-01-30} & \multicolumn{1}{c}{114} & \multicolumn{1}{c}{0.13} & \multicolumn{1}{c}{0.84$\times$0.32} \\ \cline{1-6}
\multicolumn{1}{c}{\raisebox{-0.6\normalbaselineskip}[0pt][0pt]{Project}} & \multicolumn{1}{c}{Freq.} & \multicolumn{1}{c}{\raisebox{-0.6\normalbaselineskip}[0pt][0pt]{Date}} & \multicolumn{1}{c}{Exp.} & \multicolumn{1}{c}{rms} & \multicolumn{1}{c}{Beam} & \multicolumn{1}{c}{} & \multicolumn{1}{c}{UC003} & \multicolumn{1}{c}{2.30} & \multicolumn{1}{c}{2024-02-04} & \multicolumn{1}{c}{113} & \multicolumn{1}{c}{0.26} & \multicolumn{1}{c}{9.29$\times$2.63} \\ 
\multicolumn{1}{c}{} & \multicolumn{1}{c}{[GHz]} & \multicolumn{1}{c}{} & \multicolumn{1}{c}{[s]} & \multicolumn{1}{c}{[mJy\,beam$^{-1}$]} & \multicolumn{1}{c}{[mas$\times$mas]} & \multicolumn{1}{c}{} & \multicolumn{1}{c}{UC003} & \multicolumn{1}{c}{8.67} & \multicolumn{1}{c}{2024-02-04} & \multicolumn{1}{c}{113} & \multicolumn{1}{c}{0.16} & \multicolumn{1}{c}{2.31$\times$0.70} \\ \cline{1-6}
\multicolumn{1}{c}{BE043} & \multicolumn{1}{c}{2.27} & \multicolumn{1}{c}{2005-12-11} & \multicolumn{1}{c}{900} & \multicolumn{1}{c}{1.19} & \multicolumn{1}{c}{8.83$\times$3.08} & \multicolumn{1}{c}{} & \multicolumn{1}{c}{EA075$^{E}$} & \multicolumn{1}{c}{22.25} & \multicolumn{1}{c}{2024-03-13} & \multicolumn{1}{c}{16426} & \multicolumn{1}{c}{0.05} & \multicolumn{1}{c}{0.51$\times$0.28} \\ \cline{8-13}
\multicolumn{1}{c}{BP192$^{A}$} & \multicolumn{1}{c}{4.34} & \multicolumn{1}{c}{2015-08-31} & \multicolumn{1}{c}{62} & \multicolumn{1}{c}{0.16} & \multicolumn{1}{c}{4.38$\times$1.73} &  &  &  &  &  &  &  \\
\multicolumn{1}{c}{BP192$^{A}$} & \multicolumn{1}{c}{7.62} & \multicolumn{1}{c}{2015-08-31} & \multicolumn{1}{c}{62} & \multicolumn{1}{c}{0.18} & \multicolumn{1}{c}{2.53$\times$1.01} & \multicolumn{1}{c}{} & \multicolumn{6}{c}{J1218-2159} \\ \cline{8-13}
\multicolumn{1}{c}{EC071$^{E}$} & \multicolumn{1}{c}{4.93} & \multicolumn{1}{c}{2020-10-29} & \multicolumn{1}{c}{1320} & \multicolumn{1}{c}{0.11} & \multicolumn{1}{c}{4.95$\times$1.41} & \multicolumn{1}{c}{} & \multicolumn{1}{c}{\raisebox{-0.6\normalbaselineskip}[0pt][0pt]{Project}} & \multicolumn{1}{c}{Freq.} & \multicolumn{1}{c}{\raisebox{-0.6\normalbaselineskip}[0pt][0pt]{Date}} & \multicolumn{1}{c}{Exp.} & \multicolumn{1}{c}{rms} & \multicolumn{1}{c}{Beam} \\ 
\multicolumn{1}{c}{BT152} & \multicolumn{1}{c}{4.87} & \multicolumn{1}{c}{2022-01-23} & \multicolumn{1}{c}{224} & \multicolumn{1}{c}{0.51} & \multicolumn{1}{c}{8.59$\times$1.93} & \multicolumn{1}{c}{} & \multicolumn{1}{c}{} & \multicolumn{1}{c}{[GHz]} & \multicolumn{1}{c}{} & \multicolumn{1}{c}{[s]} & \multicolumn{1}{c}{[mJy\,beam$^{-1}$]} & \multicolumn{1}{c}{[mas$\times$mas]} \\ \cline{8-13}
\multicolumn{1}{c}{BT152} & \multicolumn{1}{c}{8.37} & \multicolumn{1}{c}{2022-01-23} & \multicolumn{1}{c}{576} & \multicolumn{1}{c}{0.58} & \multicolumn{1}{c}{2.99$\times$1.06} & \multicolumn{1}{c}{} & \multicolumn{1}{c}{BG219$^{A}$} & \multicolumn{1}{c}{2.29} & \multicolumn{1}{c}{2014-12-20} & \multicolumn{1}{c}{248} & \multicolumn{1}{c}{0.19} & \multicolumn{1}{c}{7.73$\times$3.01} \\ 
\multicolumn{1}{c}{BT152} & \multicolumn{1}{c}{15.17} & \multicolumn{1}{c}{2022-01-23} & \multicolumn{1}{c}{1600} & \multicolumn{1}{c}{0.12} & \multicolumn{1}{c}{2.68$\times$1.03} & \multicolumn{1}{c}{} & \multicolumn{1}{c}{BG219$^{A}$} & \multicolumn{1}{c}{8.67} & \multicolumn{1}{c}{2014-12-20} & \multicolumn{1}{c}{248} & \multicolumn{1}{c}{0.11} & \multicolumn{1}{c}{2.09$\times$0.83} \\ 
\multicolumn{1}{c}{BL292} & \multicolumn{1}{c}{4.87} & \multicolumn{1}{c}{2022-03-25} & \multicolumn{1}{c}{710} & \multicolumn{1}{c}{0.20} & \multicolumn{1}{c}{3.81$\times$1.42} & \multicolumn{1}{c}{} & \multicolumn{1}{c}{UF001$^{A}$} & \multicolumn{1}{c}{2.25} & \multicolumn{1}{c}{2017-03-27} & \multicolumn{1}{c}{194} & \multicolumn{1}{c}{0.24} & \multicolumn{1}{c}{7.30$\times$3.31} \\ 
\multicolumn{1}{c}{BL292} & \multicolumn{1}{c}{8.50} & \multicolumn{1}{c}{2022-03-25} & \multicolumn{1}{c}{710} & \multicolumn{1}{c}{0.09} & \multicolumn{1}{c}{2.09$\times$0.80} & \multicolumn{1}{c}{} & \multicolumn{1}{c}{UF001$^{A}$} & \multicolumn{1}{c}{8.65} & \multicolumn{1}{c}{2017-03-27} & \multicolumn{1}{c}{194} & \multicolumn{1}{c}{0.13} & \multicolumn{1}{c}{1.94$\times$0.88} \\ 
\multicolumn{1}{c}{BL292} & \multicolumn{1}{c}{23.57} & \multicolumn{1}{c}{2022-03-25} & \multicolumn{1}{c}{578} & \multicolumn{1}{c}{0.08} & \multicolumn{1}{c}{1.88$\times$0.43} & \multicolumn{1}{c}{} & \multicolumn{1}{c}{UG002$^{A}$} & \multicolumn{1}{c}{2.25} & \multicolumn{1}{c}{2018-12-04} & \multicolumn{1}{c}{196} & \multicolumn{1}{c}{0.24} & \multicolumn{1}{c}{8.11$\times$3.02} \\ 
\multicolumn{1}{c}{BL301} & \multicolumn{1}{c}{4.87} & \multicolumn{1}{c}{2022-08-16} & \multicolumn{1}{c}{56} & \multicolumn{1}{c}{0.25} & \multicolumn{1}{c}{3.60$\times$1.39} & \multicolumn{1}{c}{} & \multicolumn{1}{c}{UG002$^{A}$} & \multicolumn{1}{c}{8.65} & \multicolumn{1}{c}{2018-12-04} & \multicolumn{1}{c}{196} & \multicolumn{1}{c}{0.13} & \multicolumn{1}{c}{2.04$\times$0.82} \\ \cline{1-6}
 &  &  &  &  &  & \multicolumn{1}{c}{} & \multicolumn{1}{c}{UG003} & \multicolumn{1}{c}{8.67} & \multicolumn{1}{c}{2019-05-28} & \multicolumn{1}{c}{236} & \multicolumn{1}{c}{0.17} & \multicolumn{1}{c}{4.31$\times$1.14} \\ 
 &  &  &  &  &  & \multicolumn{1}{c}{} & \multicolumn{1}{c}{EC071$^{E}$} & \multicolumn{1}{c}{4.93} & \multicolumn{1}{c}{2021-06-17} & \multicolumn{1}{c}{1050} & \multicolumn{1}{c}{0.11} & \multicolumn{1}{c}{3.99$\times$1.71} \\ 
 &  &  &  &  &  & \multicolumn{1}{c}{} & \multicolumn{1}{c}{BL292} & \multicolumn{1}{c}{4.87} & \multicolumn{1}{c}{2022-03-06} & \multicolumn{1}{c}{704} & \multicolumn{1}{c}{0.09} & \multicolumn{1}{c}{7.48$\times$1.85} \\ 
\multicolumn{6}{c}{J0923-3435} & \multicolumn{1}{c}{} & \multicolumn{1}{c}{BL292} & \multicolumn{1}{c}{8.50} & \multicolumn{1}{c}{2022-03-06} & \multicolumn{1}{c}{708} & \multicolumn{1}{c}{0.07} & \multicolumn{1}{c}{3.90$\times$1.07} \\ \cline{1-6}
\multicolumn{1}{c}{\raisebox{-0.6\normalbaselineskip}[0pt][0pt]{Project}} & \multicolumn{1}{c}{Freq.} & \multicolumn{1}{c}{\raisebox{-0.6\normalbaselineskip}[0pt][0pt]{Date}} & \multicolumn{1}{c}{Exp.} & \multicolumn{1}{c}{rms} & \multicolumn{1}{c}{Beam} & \multicolumn{1}{c}{} & \multicolumn{1}{c}{BL292} & \multicolumn{1}{c}{23.57} & \multicolumn{1}{c}{2022-03-06} & \multicolumn{1}{c}{576} & \multicolumn{1}{c}{0.08} & \multicolumn{1}{c}{1.41$\times$0.40} \\ 
\multicolumn{1}{c}{} & \multicolumn{1}{c}{[GHz]} & \multicolumn{1}{c}{} & \multicolumn{1}{c}{[s]} & \multicolumn{1}{c}{[mJy\,beam$^{-1}$]} & \multicolumn{1}{c}{[mas$\times$mas]} & \multicolumn{1}{c}{} & \multicolumn{1}{c}{UC003} & \multicolumn{1}{c}{2.30} & \multicolumn{1}{c}{2023-04-10} & \multicolumn{1}{c}{111} & \multicolumn{1}{c}{0.45} & \multicolumn{1}{c}{16.22$\times$4.44} \\ \cline{1-6}
\multicolumn{1}{c}{BP171$^{A}$} & \multicolumn{1}{c}{4.34} & \multicolumn{1}{c}{2013-02-22} & \multicolumn{1}{c}{74} & \multicolumn{1}{c}{0.52} & \multicolumn{1}{c}{8.88$\times$2.24} & \multicolumn{1}{c}{} & \multicolumn{1}{c}{UC003} & \multicolumn{1}{c}{8.67} & \multicolumn{1}{c}{2023-04-10} & \multicolumn{1}{c}{111} & \multicolumn{1}{c}{0.22} & \multicolumn{1}{c}{5.18$\times$1.62} \\ 
\multicolumn{1}{c}{BP171$^{A}$} & \multicolumn{1}{c}{7.62} & \multicolumn{1}{c}{2013-02-22} & \multicolumn{1}{c}{74} & \multicolumn{1}{c}{0.20} & \multicolumn{1}{c}{5.93$\times$1.05} & \multicolumn{1}{c}{} & \multicolumn{1}{c}{UC003} & \multicolumn{1}{c}{8.67} & \multicolumn{1}{c}{2023-12-20} & \multicolumn{1}{c}{111} & \multicolumn{1}{c}{0.11} & \multicolumn{1}{c}{12.32$\times$2.41} \\ 
\multicolumn{1}{c}{BL292} & \multicolumn{1}{c}{4.87} & \multicolumn{1}{c}{2022-03-20} & \multicolumn{1}{c}{708} & \multicolumn{1}{c}{0.07} & \multicolumn{1}{c}{4.00$\times$1.44} & \multicolumn{1}{c}{} & \multicolumn{1}{c}{UC003} & \multicolumn{1}{c}{8.67} & \multicolumn{1}{c}{2024-06-23} & \multicolumn{1}{c}{111} & \multicolumn{1}{c}{0.13} & \multicolumn{1}{c}{4.88$\times$1.01} \\ \cline{1-6} \cline{8-13}
\end{tabular}
\end{adjustbox}
\end{table*}

\begin{table*}[t!]
\ContinuedFloat
\caption{continued}
\label{table:usedvlbidata2}
\renewcommand{\arraystretch}{1.2}
\centering
\begin{adjustbox}{max width=1\textwidth}
\begin{tabular}{ccccccccccccc}
\multicolumn{6}{c}{J1632+3547} &  &  &  &  &  &  &  \\ \cline{1-6}
\multicolumn{1}{c}{\raisebox{-0.6\normalbaselineskip}[0pt][0pt]{Project}} & \multicolumn{1}{c}{Freq.} & \multicolumn{1}{c}{\raisebox{-0.6\normalbaselineskip}[0pt][0pt]{Date}} & \multicolumn{1}{c}{Exp.} & \multicolumn{1}{c}{rms} & \multicolumn{1}{c}{Beam} &  &  &  &  &  &  &  \\ 
\multicolumn{1}{c}{} & \multicolumn{1}{c}{[GHz]} & \multicolumn{1}{c}{} & \multicolumn{1}{c}{[s]} & \multicolumn{1}{c}{[mJy\,beam$^{-1}$]} & \multicolumn{1}{c}{[mas$\times$mas]} &  &  &  &  &  &  &  \\ \cline{1-6}
\multicolumn{1}{c}{BT085} & \multicolumn{1}{c}{4.84} & \multicolumn{1}{c}{2006-01-03} & \multicolumn{1}{c}{558} & \multicolumn{1}{c}{0.26} & \multicolumn{1}{c}{2.79$\times$1.76} &  &  &  &  &  &  &  \\ 
\multicolumn{1}{c}{BT088} & \multicolumn{1}{c}{4.84} & \multicolumn{1}{c}{2006-11-22} & \multicolumn{1}{c}{10995} & \multicolumn{1}{c}{0.21} & \multicolumn{1}{c}{2.91$\times$1.68} &  &  &  &  &  &  &  \\ 
\multicolumn{1}{c}{BT088} & \multicolumn{1}{c}{15.14} & \multicolumn{1}{c}{2006-11-22} & \multicolumn{1}{c}{10995} & \multicolumn{1}{c}{0.15} & \multicolumn{1}{c}{1.28$\times$0.62} &  &  &  &  &  &  &  \\ 
\multicolumn{1}{c}{BT088} & \multicolumn{1}{c}{4.84} & \multicolumn{1}{c}{2007-02-19} & \multicolumn{1}{c}{3689} & \multicolumn{1}{c}{0.10} & \multicolumn{1}{c}{2.91$\times$1.83} &  &  &  &  &  &  &  \\ 
\multicolumn{1}{c}{BT152} & \multicolumn{1}{c}{4.87} & \multicolumn{1}{c}{2022-01-23} & \multicolumn{1}{c}{196} & \multicolumn{1}{c}{0.16} & \multicolumn{1}{c}{4.00$\times$2.09} &  &  &  &  &  &  &  \\ 
\multicolumn{1}{c}{BT152} & \multicolumn{1}{c}{8.37} & \multicolumn{1}{c}{2022-01-23} & \multicolumn{1}{c}{504} & \multicolumn{1}{c}{0.12} & \multicolumn{1}{c}{1.92$\times$1.08} & \multicolumn{1}{c}{} & \multicolumn{6}{c}{J2347-1856} \\ \cline{8-13}
\multicolumn{1}{c}{BT152} & \multicolumn{1}{c}{15.17} & \multicolumn{1}{c}{2022-01-23} & \multicolumn{1}{c}{1400} & \multicolumn{1}{c}{0.12} & \multicolumn{1}{c}{0.92$\times$0.55} & \multicolumn{1}{c}{} & \multicolumn{1}{c}{\raisebox{-0.6\normalbaselineskip}[0pt][0pt]{Project}} & \multicolumn{1}{c}{Freq.} & \multicolumn{1}{c}{\raisebox{-0.6\normalbaselineskip}[0pt][0pt]{Date}} & \multicolumn{1}{c}{Exp.} & \multicolumn{1}{c}{rms} & \multicolumn{1}{c}{Beam} \\ 
\multicolumn{1}{c}{BL292} & \multicolumn{1}{c}{4.87} & \multicolumn{1}{c}{2022-06-18} & \multicolumn{1}{c}{708} & \multicolumn{1}{c}{0.07} & \multicolumn{1}{c}{3.94$\times$1.61} & \multicolumn{1}{c}{} & \multicolumn{1}{c}{} & \multicolumn{1}{c}{[GHz]} & \multicolumn{1}{c}{} & \multicolumn{1}{c}{[s]} & \multicolumn{1}{c}{[mJy\,beam$^{-1}$]} & \multicolumn{1}{c}{[mas$\times$mas]} \\ \cline{8-13}
\multicolumn{1}{c}{BL292} & \multicolumn{1}{c}{8.50} & \multicolumn{1}{c}{2022-06-18} & \multicolumn{1}{c}{710} & \multicolumn{1}{c}{0.13} & \multicolumn{1}{c}{2.24$\times$0.81} & \multicolumn{1}{c}{} & \multicolumn{1}{c}{BB023$^{A}$} & \multicolumn{1}{c}{2.27} & \multicolumn{1}{c}{1997-07-02} & \multicolumn{1}{c}{489} & \multicolumn{1}{c}{0.94} & \multicolumn{1}{c}{9.30$\times$3.57} \\ 
\multicolumn{1}{c}{BL292} & \multicolumn{1}{c}{23.57} & \multicolumn{1}{c}{2022-06-18} & \multicolumn{1}{c}{580} & \multicolumn{1}{c}{0.12} & \multicolumn{1}{c}{0.85$\times$0.32} & \multicolumn{1}{c}{} & \multicolumn{1}{c}{BB023$^{A}$} & \multicolumn{1}{c}{8.34} & \multicolumn{1}{c}{1997-07-02} & \multicolumn{1}{c}{489} & \multicolumn{1}{c}{0.54} & \multicolumn{1}{c}{2.60$\times$1.04} \\ \cline{1-6}
 &  &  &  &  &  & \multicolumn{1}{c}{} & \multicolumn{1}{c}{BB023$^{A}$} & \multicolumn{1}{c}{2.27} & \multicolumn{1}{c}{1997-08-27} & \multicolumn{1}{c}{325} & \multicolumn{1}{c}{1.51} & \multicolumn{1}{c}{21.75$\times$5.82} \\
\multicolumn{6}{c}{J1653+3503} & \multicolumn{1}{c}{} & \multicolumn{1}{c}{BB023$^{A}$} & \multicolumn{1}{c}{8.34} & \multicolumn{1}{c}{1997-08-28} & \multicolumn{1}{c}{325} & \multicolumn{1}{c}{0.67} & \multicolumn{1}{c}{3.57$\times$1.16} \\ \cline{1-6}
\multicolumn{1}{c}{\raisebox{-0.6\normalbaselineskip}[0pt][0pt]{Project}} & \multicolumn{1}{c}{Freq.} & \multicolumn{1}{c}{\raisebox{-0.6\normalbaselineskip}[0pt][0pt]{Date}} & \multicolumn{1}{c}{Exp.} & \multicolumn{1}{c}{rms} & \multicolumn{1}{c}{Beam} & \multicolumn{1}{c}{} & \multicolumn{1}{c}{BP091} & \multicolumn{1}{c}{4.99} & \multicolumn{1}{c}{2002-04-26} & \multicolumn{1}{c}{1620} & \multicolumn{1}{c}{0.38} & \multicolumn{1}{c}{4.45$\times$1.73} \\ 
\multicolumn{1}{c}{} & \multicolumn{1}{c}{[GHz]} & \multicolumn{1}{c}{} & \multicolumn{1}{c}{[s]} & \multicolumn{1}{c}{[mJy\,beam$^{-1}$]} & \multicolumn{1}{c}{[mas$\times$mas]} & \multicolumn{1}{c}{} & \multicolumn{1}{c}{BP091} & \multicolumn{1}{c}{15.17} & \multicolumn{1}{c}{2002-04-26} & \multicolumn{1}{c}{3600} & \multicolumn{1}{c}{0.31} & \multicolumn{1}{c}{2.13$\times$0.71} \\ \cline{1-6}
\multicolumn{1}{c}{BT085$^{A}$} & \multicolumn{1}{c}{4.84} & \multicolumn{1}{c}{2006-08-07} & \multicolumn{1}{c}{600} & \multicolumn{1}{c}{0.29} & \multicolumn{1}{c}{3.00$\times$1.80} & \multicolumn{1}{c}{} & \multicolumn{1}{c}{BG219$^{A}$} & \multicolumn{1}{c}{2.29} & \multicolumn{1}{c}{2014-06-09} & \multicolumn{1}{c}{186} & \multicolumn{1}{c}{0.25} & \multicolumn{1}{c}{13.26$\times$4.02} \\ 
\multicolumn{1}{c}{BT094} & \multicolumn{1}{c}{4.84} & \multicolumn{1}{c}{2007-12-31} & \multicolumn{1}{c}{2828} & \multicolumn{1}{c}{0.37} & \multicolumn{1}{c}{5.23$\times$2.44} & \multicolumn{1}{c}{} & \multicolumn{1}{c}{BG219$^{A}$} & \multicolumn{1}{c}{8.67} & \multicolumn{1}{c}{2014-06-09} & \multicolumn{1}{c}{186} & \multicolumn{1}{c}{0.12} & \multicolumn{1}{c}{3.45$\times$1.01} \\ 
\multicolumn{1}{c}{BL292} & \multicolumn{1}{c}{4.87} & \multicolumn{1}{c}{2022-06-18} & \multicolumn{1}{c}{708} & \multicolumn{1}{c}{0.10} & \multicolumn{1}{c}{3.96$\times$1.57} & \multicolumn{1}{c}{} & \multicolumn{1}{c}{UF001$^{A}$} & \multicolumn{1}{c}{2.26} & \multicolumn{1}{c}{2017-05-01} & \multicolumn{1}{c}{232} & \multicolumn{1}{c}{0.28} & \multicolumn{1}{c}{7.72$\times$3.22} \\ 
\multicolumn{1}{c}{BL292} & \multicolumn{1}{c}{8.50} & \multicolumn{1}{c}{2022-06-18} & \multicolumn{1}{c}{708} & \multicolumn{1}{c}{0.09} & \multicolumn{1}{c}{2.27$\times$0.77} & \multicolumn{1}{c}{} & \multicolumn{1}{c}{UF001$^{A}$} & \multicolumn{1}{c}{8.65} & \multicolumn{1}{c}{2017-05-01} & \multicolumn{1}{c}{232} & \multicolumn{1}{c}{0.13} & \multicolumn{1}{c}{2.06$\times$0.91} \\ 
\multicolumn{1}{c}{BL292} & \multicolumn{1}{c}{23.57} & \multicolumn{1}{c}{2022-06-18} & \multicolumn{1}{c}{578} & \multicolumn{1}{c}{0.11} & \multicolumn{1}{c}{0.83$\times$0.30} & \multicolumn{1}{c}{} & \multicolumn{1}{c}{UG002$^{A}$} & \multicolumn{1}{c}{2.26} & \multicolumn{1}{c}{2018-04-08} & \multicolumn{1}{c}{236} & \multicolumn{1}{c}{0.24} & \multicolumn{1}{c}{7.87$\times$3.17} \\ \cline{1-6}
 &  &  &  &  &  & \multicolumn{1}{c}{} & \multicolumn{1}{c}{UG002$^{A}$} & \multicolumn{1}{c}{8.65} & \multicolumn{1}{c}{2018-04-08} & \multicolumn{1}{c}{236} & \multicolumn{1}{c}{0.12} & \multicolumn{1}{c}{2.01$\times$0.85} \\ 
\multicolumn{6}{c}{J1805-0438} & \multicolumn{1}{c}{} & \multicolumn{1}{c}{UG003} & \multicolumn{1}{c}{2.30} & \multicolumn{1}{c}{2019-12-12} & \multicolumn{1}{c}{296} & \multicolumn{1}{c}{0.41} & \multicolumn{1}{c}{8.38$\times$2.99} \\ \cline{1-6}
\multicolumn{1}{c}{\raisebox{-0.6\normalbaselineskip}[0pt][0pt]{Project}} & \multicolumn{1}{c}{Freq.} & \multicolumn{1}{c}{\raisebox{-0.6\normalbaselineskip}[0pt][0pt]{Date}} & \multicolumn{1}{c}{Exp.} & \multicolumn{1}{c}{rms} & \multicolumn{1}{c}{Beam} & \multicolumn{1}{c}{} & \multicolumn{1}{c}{UG003} & \multicolumn{1}{c}{8.67} & \multicolumn{1}{c}{2019-12-12} & \multicolumn{1}{c}{296} & \multicolumn{1}{c}{0.24} & \multicolumn{1}{c}{3.39$\times$0.95} \\ 
\multicolumn{1}{c}{} & \multicolumn{1}{c}{[GHz]} & \multicolumn{1}{c}{} & \multicolumn{1}{c}{[s]} & \multicolumn{1}{c}{[mJy\,beam$^{-1}$]} & \multicolumn{1}{c}{[mas$\times$mas]} & \multicolumn{1}{c}{} & \multicolumn{1}{c}{EC071$^{E}$} & \multicolumn{1}{c}{4.93} & \multicolumn{1}{c}{2021-03-04} & \multicolumn{1}{c}{1050} & \multicolumn{1}{c}{0.30} & \multicolumn{1}{c}{4.47$\times$1.29} \\ \cline{1-6}
\multicolumn{1}{c}{RDV91$^{A}$} & \multicolumn{1}{c}{2.31} & \multicolumn{1}{c}{2012-02-08} & \multicolumn{1}{c}{532} & \multicolumn{1}{c}{0.24} & \multicolumn{1}{c}{8.32$\times$3.57} & \multicolumn{1}{c}{} & \multicolumn{1}{c}{BL292} & \multicolumn{1}{c}{4.87} & \multicolumn{1}{c}{2022-03-22} & \multicolumn{1}{c}{708} & \multicolumn{1}{c}{0.17} & \multicolumn{1}{c}{3.92$\times$1.55} \\ 
\multicolumn{1}{c}{RDV91$^{A}$} & \multicolumn{1}{c}{8.64} & \multicolumn{1}{c}{2012-02-08} & \multicolumn{1}{c}{532} & \multicolumn{1}{c}{0.14} & \multicolumn{1}{c}{3.79$\times$1.34} & \multicolumn{1}{c}{} & \multicolumn{1}{c}{BL292} & \multicolumn{1}{c}{8.50} & \multicolumn{1}{c}{2022-03-22} & \multicolumn{1}{c}{710} & \multicolumn{1}{c}{0.18} & \multicolumn{1}{c}{2.32$\times$0.86} \\ 
\multicolumn{1}{c}{UF001$^{A}$} & \multicolumn{1}{c}{2.25} & \multicolumn{1}{c}{2017-07-09} & \multicolumn{1}{c}{438} & \multicolumn{1}{c}{0.16} & \multicolumn{1}{c}{8.18$\times$3.27} & \multicolumn{1}{c}{} & \multicolumn{1}{c}{BL292} & \multicolumn{1}{c}{23.57} & \multicolumn{1}{c}{2022-03-22} & \multicolumn{1}{c}{574} & \multicolumn{1}{c}{0.06} & \multicolumn{1}{c}{2.57$\times$1.08} \\ 
\multicolumn{1}{c}{UF001$^{A}$} & \multicolumn{1}{c}{8.65} & \multicolumn{1}{c}{2017-07-09} & \multicolumn{1}{c}{438} & \multicolumn{1}{c}{0.11} & \multicolumn{1}{c}{2.10$\times$0.87} & \multicolumn{1}{c}{} & \multicolumn{1}{c}{UC003} & \multicolumn{1}{c}{8.67} & \multicolumn{1}{c}{2023-03-02} & \multicolumn{1}{c}{108} & \multicolumn{1}{c}{0.19} & \multicolumn{1}{c}{5.18$\times$1.20} \\ 
\multicolumn{1}{c}{UG002$^{A}$} & \multicolumn{1}{c}{2.28} & \multicolumn{1}{c}{2018-09-02} & \multicolumn{1}{c}{352} & \multicolumn{1}{c}{0.16} & \multicolumn{1}{c}{8.88$\times$4.02} & \multicolumn{1}{c}{} & \multicolumn{1}{c}{UC003} & \multicolumn{1}{c}{2.30} & \multicolumn{1}{c}{2023-05-12} & \multicolumn{1}{c}{110} & \multicolumn{1}{c}{0.52} & \multicolumn{1}{c}{17.99$\times$4.36} \\ 
\multicolumn{1}{c}{UG002$^{A}$} & \multicolumn{1}{c}{8.65} & \multicolumn{1}{c}{2018-09-02} & \multicolumn{1}{c}{352} & \multicolumn{1}{c}{0.09} & \multicolumn{1}{c}{4.24$\times$1.71} & \multicolumn{1}{c}{} & \multicolumn{1}{c}{UC003} & \multicolumn{1}{c}{8.67} & \multicolumn{1}{c}{2023-05-12} & \multicolumn{1}{c}{110} & \multicolumn{1}{c}{0.22} & \multicolumn{1}{c}{4.60$\times$1.26} \\ 
\multicolumn{1}{c}{UG003} & \multicolumn{1}{c}{2.29} & \multicolumn{1}{c}{2019-01-29} & \multicolumn{1}{c}{534} & \multicolumn{1}{c}{0.39} & \multicolumn{1}{c}{7.61$\times$2.69} & \multicolumn{1}{c}{} & \multicolumn{1}{c}{BA189} & \multicolumn{1}{c}{8.67} & \multicolumn{1}{c}{2025-08-23} & \multicolumn{1}{c}{3600} & \multicolumn{1}{c}{0.05} & \multicolumn{1}{c}{2.76$\times$1.02} \\ 
\multicolumn{1}{c}{UG003} & \multicolumn{1}{c}{8.67} & \multicolumn{1}{c}{2019-01-29} & \multicolumn{1}{c}{534} & \multicolumn{1}{c}{0.19} & \multicolumn{1}{c}{5.64$\times$3.19} & \multicolumn{1}{c}{} & \multicolumn{1}{c}{BA189} & \multicolumn{1}{c}{4.87} & \multicolumn{1}{c}{2025-08-24} & \multicolumn{1}{c}{3600} & \multicolumn{1}{c}{0.03} & \multicolumn{1}{c}{5.06$\times$1.88} \\ \cline{8-13}
\multicolumn{1}{c}{EC071$^{E}$} & \multicolumn{1}{c}{4.93} & \multicolumn{1}{c}{2021-06-17} & \multicolumn{1}{c}{1050} & \multicolumn{1}{c}{0.07} & \multicolumn{1}{c}{2.24$\times$1.90} &  &  &  &  &  &  &  \\ 
\multicolumn{1}{c}{UH007} & \multicolumn{1}{c}{2.30} & \multicolumn{1}{c}{2021-07-10} & \multicolumn{1}{c}{414} & \multicolumn{1}{c}{0.29} & \multicolumn{1}{c}{8.14$\times$2.99} &  &  &  &  &  &  &  \\ 
\multicolumn{1}{c}{UH007} & \multicolumn{1}{c}{8.67} & \multicolumn{1}{c}{2021-07-10} & \multicolumn{1}{c}{414} & \multicolumn{1}{c}{0.06} & \multicolumn{1}{c}{13.60$\times$3.02} &  &  &  &  &  &  &  \\ 
\multicolumn{1}{c}{BL292} & \multicolumn{1}{c}{4.87} & \multicolumn{1}{c}{2022-06-21} & \multicolumn{1}{c}{708} & \multicolumn{1}{c}{0.04} & \multicolumn{1}{c}{3.94$\times$1.81} &  &  &  &  &  &  &  \\ 
\multicolumn{1}{c}{UH007} & \multicolumn{1}{c}{2.30} & \multicolumn{1}{c}{2022-10-23} & \multicolumn{1}{c}{438} & \multicolumn{1}{c}{0.25} & \multicolumn{1}{c}{8.29$\times$3.10} &  &  &  &  &  &  &  \\ 
\multicolumn{1}{c}{UH007} & \multicolumn{1}{c}{8.67} & \multicolumn{1}{c}{2022-10-23} & \multicolumn{1}{c}{438} & \multicolumn{1}{c}{0.09} & \multicolumn{1}{c}{4.12$\times$2.22} &  &  &  &  &  &  &  \\ 
\multicolumn{1}{c}{UH007} & \multicolumn{1}{c}{2.30} & \multicolumn{1}{c}{2022-12-12} & \multicolumn{1}{c}{352} & \multicolumn{1}{c}{0.25} & \multicolumn{1}{c}{9.47$\times$3.81} &  &  &  &  &  &  &  \\ \cline{1-6}
\end{tabular}
\end{adjustbox}
\end{table*}

\subsection{Connected-element and single-dish observations}

\begin{table*}[t!]
\caption{Single dish and connected interferometer data used in this study. References: LOFAR \citealt{LOFARDR2}, TGSS \citealt{TGSSDR1}, GLEAM-X, \citealt{GLEAMXDR2}, TEXAS \citealt{TEXASDR1}, RACS \citealt{RACSDR1}, FIRST \citealt{FIRSTDR}, NVSS \citealt{NVSSDR}, VLASS \citealt{VLASSQuickLook}, GB6 \citealt{GB6Catalog}, CRATES \citealt{CRATESDR}, CLASS \citealt{Myers2003}, This work: VLA observations described in Sect.~\ref{sec:integrateddata}.} 
\label{table:usedsingledishdata}
\renewcommand{\arraystretch}{1.2}
\centering
\begin{adjustbox}{max width=1\textwidth}
\begin{tabular}{ccccccccccc}
\cline{1-5} \cline{7-11}
Source     & $\mathbf{\nu}$ [GHz] & $S_{\nu}$ [mJy] & $\sigma_{S_{\nu}}$ [mJy] & Survey                 & \multicolumn{1}{c}{} & Source     & $\nu$ [GHz] & $S_{\nu}$ [mJy] & $\sigma_{S_{\nu}}$ [mJy] & \multicolumn{1}{c}{Survey}                 \\ \cline{1-5} \cline{7-11} 
J0024-4202 & 0.087                                              & 19.739                                        & 1.579                                                  & GLEAM-X       & \multicolumn{1}{c}{} & J1632+3547 & 0.144                                              & 105.2                                         & 1.3                                                    & \multicolumn{1}{c}{LOFAR}        \\
                    & 0.118                                              & 32.233                                        & 2.579                                                  & GLEAM-X       & \multicolumn{1}{c}{} &                     & 0.150                                              & 69.3                                          & 9.4                                                    & \multicolumn{1}{c}{TGSS}        \\
                    & 0.150                                                & 89.00                                         & 10.44                                                  & TGSS         & \multicolumn{1}{c}{} &                     & 0.365                                              & 358                                           & 23                                                     & \multicolumn{1}{c}{TEXAS}       \\
                    & 0.154                                              & 69.911                                        & 5.593                                                  & GLEAM-X       & \multicolumn{1}{c}{} &                     & 1.4                                                & 555.65                                        & 0.152                                                  & \multicolumn{1}{c}{FIRST}        \\
                    & 0.185                                              & 113.272                                       & 9.062                                                  & GLEAM-X       & \multicolumn{1}{c}{} &                     & 1.4                                                & 536.3                                         & 16.1                                                   & \multicolumn{1}{c}{NVSS}         \\
                    & 0.215                                              & 178.440                                       & 14.275                                                 & GLEAM-X       & \multicolumn{1}{c}{} &                     & 3.0                                                & 352.55                                        & 52.88                                                  & \multicolumn{1}{c}{VLASS} \\
                    & 4.739                                              & 1750.10                                       & 90.07                                                  & This work                          & \multicolumn{1}{l}{} &                     & 5.0                                                & 280                                           & 25                                                     & \multicolumn{1}{c}{GB6}     \\  
                    & 4.8                                                & 2036.0                                        & 101.8                                                  & CRATES        & \multicolumn{1}{c}{} &                     & 8.4                                                & 194                                           & 9.7                                                    & \multicolumn{1}{c}{CLASS}      \\ \cline{7-11}
                    & 5.251                                              & 1570.12                                       & 80.68                                                  & This work                          & \multicolumn{1}{l}{} & J1653+3503 & 0.144                                              & 112.43                                        & 11.243                                                 & \multicolumn{1}{c}{LOFAR}       \\
                    & 5.718                                              & 1420.16                                       & 75.09                                                  & This work                          & \multicolumn{1}{l}{} &                     & 0.150                                              & 44.0                                          & 6.9                                                    & \multicolumn{1}{c}{TGSS}        \\
                    & 6.251                                              & 1262.40                                       & 65.86                                                  & This work                          & \multicolumn{1}{l}{} &                     & 1.4                                                & 272.23                                        & 0.134                                                  & \multicolumn{1}{c}{FIRST}        \\
                    & 8.4                                                & 816.80                                        & 40.84                                                  & CRATES        & \multicolumn{1}{c}{} &                     & 1.4                                                & 285.5                                         & 8.6                                                    & \multicolumn{1}{c}{NVSS}         \\
                    & 8.539                                              & 867.62                                        & 43.98                                                  & This work                          & \multicolumn{1}{l}{} & \textbf{}           & 3.0                                                & 178.42                                        & 26.76                                                  & \multicolumn{1}{c}{VLASS} \\
                    & 9.051                                              & 802.35                                        & 40.71                                                  & This work                          & \multicolumn{1}{l}{} &                     & 5.0                                                & 141                                           & 13                                                     & \multicolumn{1}{c}{GB6}     \\ 
                    & 9.539                                              & 744.09                                        & 37.80                                                  & This work                          & \multicolumn{1}{l}{} &                     & 8.4                                                & 91.80                                         & 4.59                                                   & \multicolumn{1}{c}{CLASS}      \\  \cline{7-11}
                    & 10.051                                             & 693.47                                        & 35.27                                                  & This work                          & \multicolumn{1}{l}{} & J1805-0438 & 0.150                                              & 218.9                                         & 24.1                                                   & \multicolumn{1}{c}{TGSS}        \\ \cline{1-5}
J0210-2213 & 0.087                                              & 30.209                                        & 2.417                                                  & GLEAM-X       & \multicolumn{1}{c}{} &                     & 0.365                                              & 735                                           & 35                                                     & \multicolumn{1}{c}{TEXAS}       \\
                    & 0.118                                              & 44.745                                        & 3.580                                                  & GLEAM-X       & \multicolumn{1}{c}{} &                     & 1.4                                                & 516.4                                         & 15.5                                                   & \multicolumn{1}{c}{NVSS}         \\  
                    & 0.150                                                & 85.0                                          & 10.5                                                   & TGSS         & \multicolumn{1}{c}{} &                     & 3.0                                                & 205.69                                        & 30.85                                                  & \multicolumn{1}{c}{VLASS} \\ \cline{7-11}
                    & 0.154                                              & 66.706                                        & 5.336                                                  & GLEAM-X       & \multicolumn{1}{c}{} & J2347-1856 & 0.087                                              & 227.692                                       & 18.215                                                 & \multicolumn{1}{c}{GLEAM-X}      \\
                    & 0.185                                              & 79.463                                        & 6.357                                                  & GLEAM-X       & \multicolumn{1}{c}{} &                     & 0.118                                              & 223.149                                       & 17.852                                                 & \multicolumn{1}{c}{GLEAM-X}      \\
                    & 0.215                                              & 92.435                                        & 7.395                                                  & GLEAM-X       & \multicolumn{1}{c}{} &                     & 0.150                                              & 184.5                                         & 19.3                                                   & \multicolumn{1}{c}{TGSS}        \\
                    & 0.8875                                             & 751.1                                         & 53.654                                                 & RACS         & \multicolumn{1}{c}{} &                     & 0.154                                              & 225.422                                       & 18.034                                                 & \multicolumn{1}{c}{GLEAM-X}      \\
                    & 1.4                                                & 947.1                                         & 28.4                                                   & NVSS          & \multicolumn{1}{c}{} & \textbf{}           & 0.185                                              & 229.393                                       & 18.351                                                 & \multicolumn{1}{c}{GLEAM-X}      \\
                    & 3.0                                                & 765.35                                        & 114.80                                                 & VLASS  & \multicolumn{1}{c}{} &                     & 0.215                                              & 234.798                                       & 18.874                                                 & \multicolumn{1}{c}{GLEAM-X}      \\
                    & 4.8                                                & 618.0                                         & 30.9                                                   & CRATES        & \multicolumn{1}{c}{} &                     & 0.8875                                             & 539.029                                       & 38.239                                                 & \multicolumn{1}{c}{RACS}        \\
                    & 8.4                                                & 265.30                                        & 13.63                                                  & CRATES        & \multicolumn{1}{c}{} &                     & 1.4                                                & 654.9                                         & 19.7                                                   & \multicolumn{1}{c}{NVSS}         \\ \cline{1-5}
J0923-3435 & 0.150                                              & 57.5                                          & 8.7                                                    & TGSS         & \multicolumn{1}{c}{} & \textbf{}           & 3.0                                                & 586.7                                         & 88.0                                                   & \multicolumn{1}{c}{VLASS} \\
                    & 0.8875                                             & 399.20                                        & 28.48                                                  & RACS         & \multicolumn{1}{c}{} &                     & 4.739                                              & 491.51                                        & 24.58                                                  & \multicolumn{1}{c}{This work}                          \\
                    & 1.4                                                & 503.9                                         & 15.1                                                   & NVSS          & \multicolumn{1}{c}{} &                     & 4.8                                                & 523.0                                         & 26.2                                                   & \multicolumn{1}{c}{CRATES}       \\
                    & 4.8                                                & 311.0                                         & 15.6                                                   & CRATES        & \multicolumn{1}{c}{} &                     & 5.251                                              & 460.18                                        & 24.37                                                  & \multicolumn{1}{c}{This work}                          \\
                    & 8.4                                                & 143.0                                         & 7.2                                                    & CRATES        & \multicolumn{1}{c}{} &                     & 5.718                                              & 426.92                                        & 23.26                                                  & \multicolumn{1}{c}{This work}                          \\ \cline{1-5}
J1143+1834 & 1.4                                                & 300.740                                       & 0.173                                                  & FIRST         & \multicolumn{1}{c}{} &                     & 6.251                                              & 401.08                                        & 22.33                                                  & \multicolumn{1}{c}{This work}                          \\
                    & 1.4                                                & 310.5                                         & 9.3                                                    & NVSS          & \multicolumn{1}{c}{} &                     & 8.4                                                & 347.4                                         & 17.4                                                   & \multicolumn{1}{c}{CRATES}       \\
                    & 3.0                                                & 384.14                                        & 57.62                                                  & VLASS  & \multicolumn{1}{c}{} &                     & 8.539                                              & 314.97                                        & 16.14                                                  & \multicolumn{1}{c}{This work}                          \\
                    & 4.8                                                & 315.00                                        & 15.75                                                  & CRATES        & \multicolumn{1}{c}{} &                     & 9.051                                              & 299.03                                        & 15.35                                                  & \multicolumn{1}{c}{This work}                          \\
                    & 5.0                                                & 314                                           & 28                                                     & GB6      & \multicolumn{1}{c}{} &                     & 9.539                                              & 284.50                                        & 14.62                                                  & \multicolumn{1}{c}{This work}                          \\  
                    & 8.4                                                & 244.70                                        & 15.24                                                  & CRATES        &      \multicolumn{1}{c}{}                 &                     &           10.051                                         &                     271.62                          &             13.98                                           &                               \multicolumn{1}{c}{This work}                           \\ \cline{1-5} \cline{7-11}
J1218-2159 & 0.8875                                             & 46.6                                          & 3.8                                                    & RACS         &                       &                     &                                                    &                                               &                                                        &                                                         \\
                    & 3.0                                                & 187.61                                        & 28.14                                                  & VLASS  &                       &                     &                                                    &                                               &                                                        &                                                         \\
                    & 4.8                                                & 187.0                                         & 9.4                                                    & CRATES        &                       &                     &                                                    &                                               &                                                        &                                                         \\
                    & 8.4                                                & 105.7                                         & 5.3                                                    & CRATES        &                       &                     &                                                    &                                               &                                                        &                                                        \\ \cline{1-5}                                                     
\end{tabular}
\end{adjustbox}
\end{table*}

\section{VLBI intensity maps}\label{app:vlbimaps}

\begin{figure*}[h]
    \centering
    \includegraphics[width=0.33\textwidth]{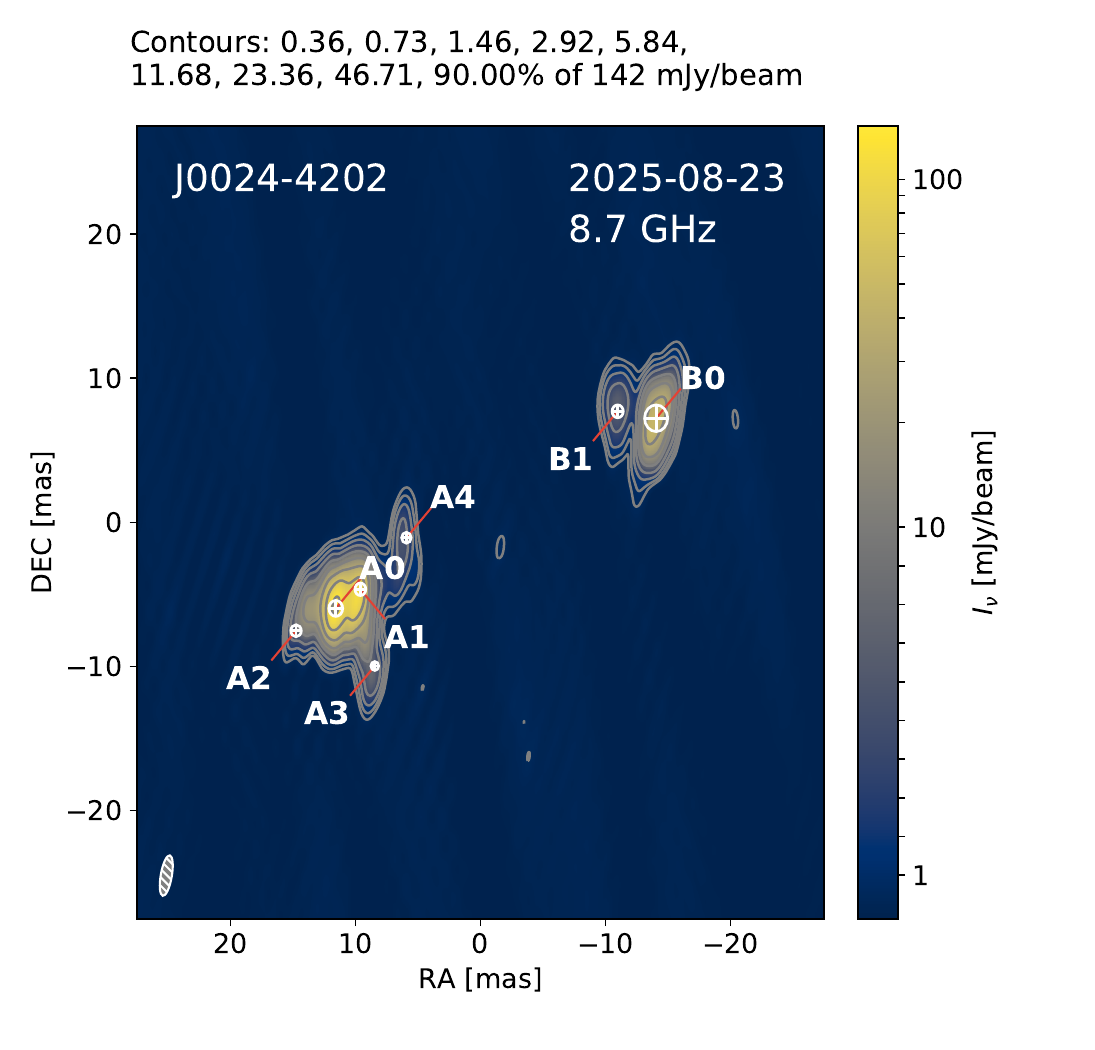}
    \includegraphics[width=0.33\textwidth]{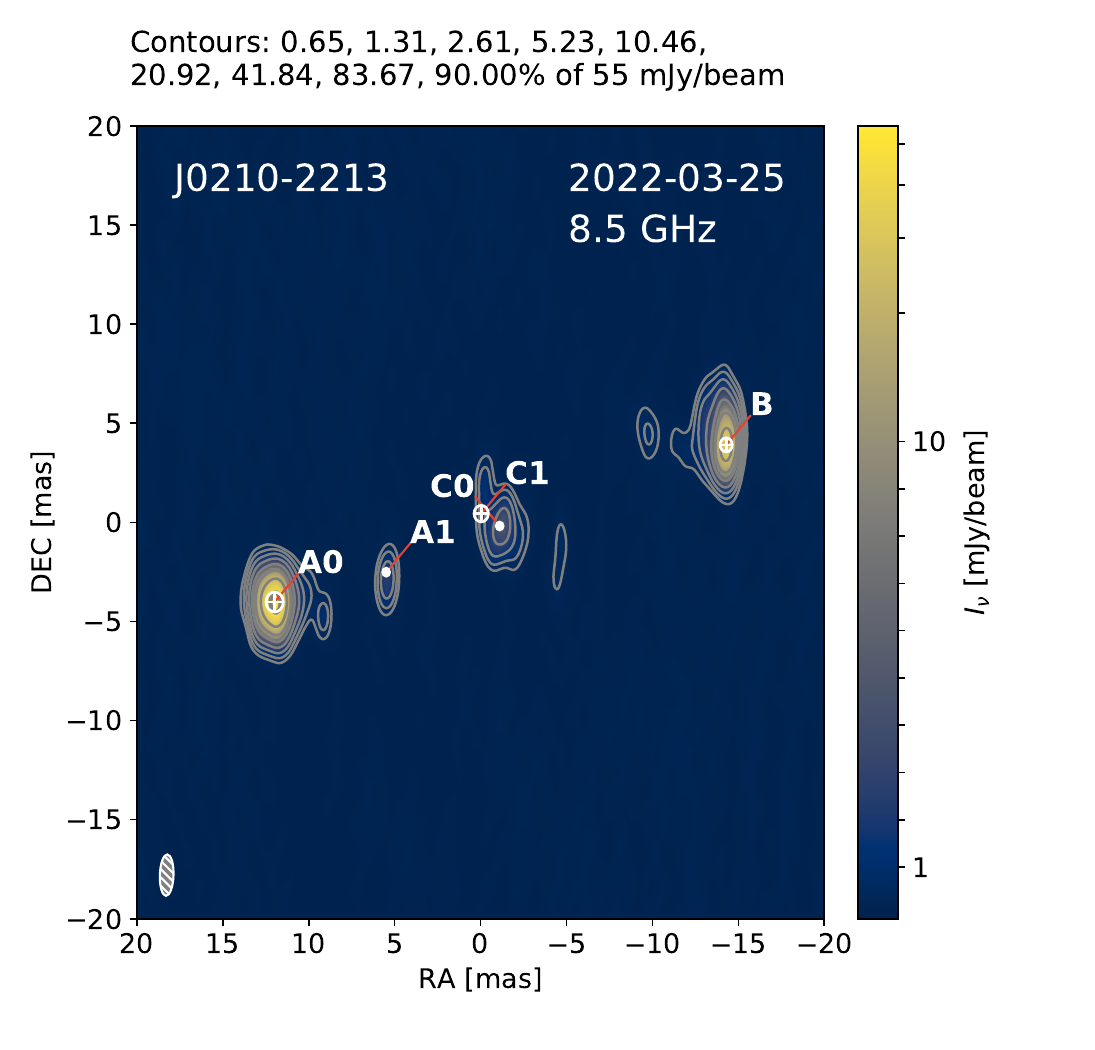}
    \includegraphics[width=0.33\textwidth]{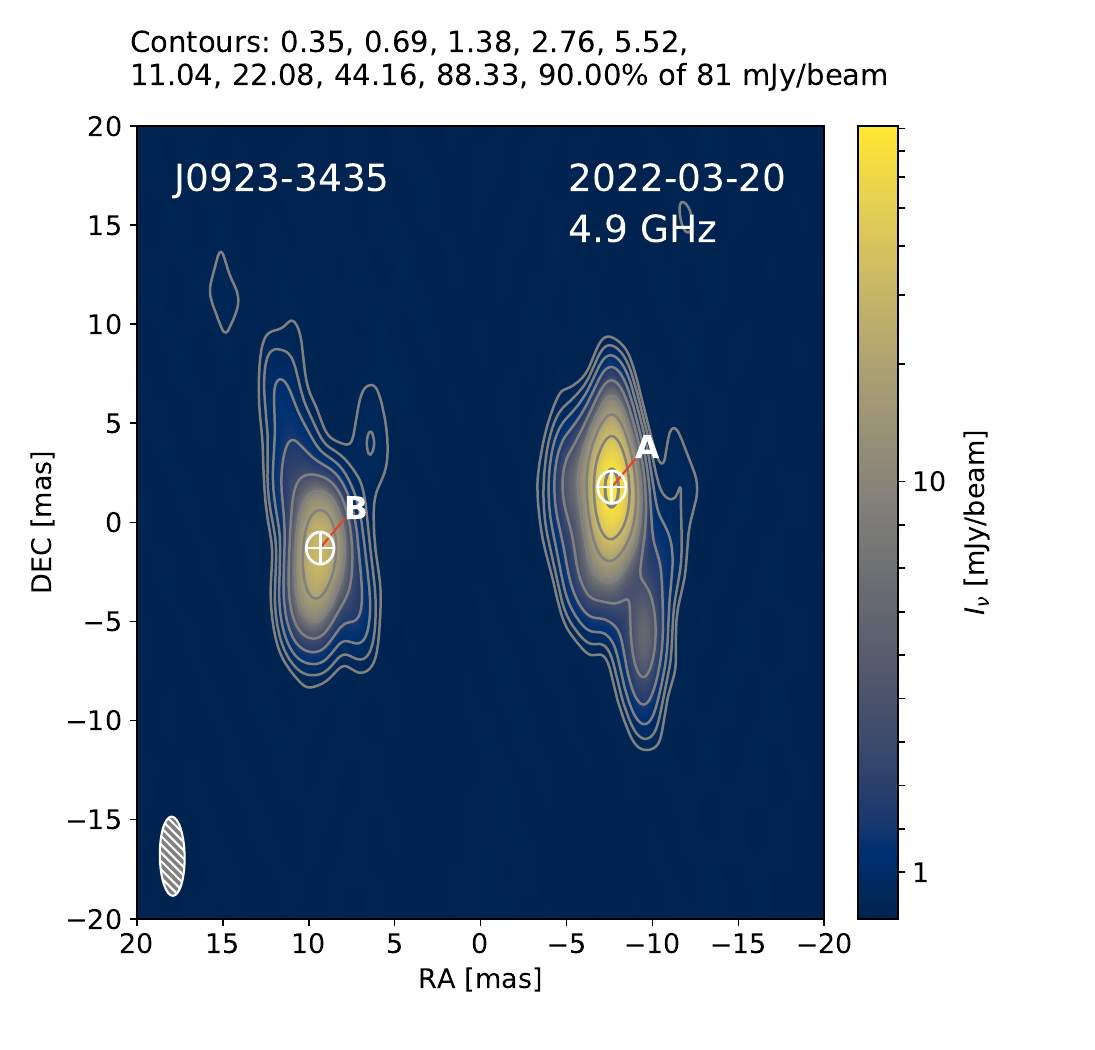}\\
    \includegraphics[width=0.33\textwidth]{./Figures/VLBIMaps/J1143+1834_K_2024_03_13_dao_map_log.pdf}
    \includegraphics[width=0.33\textwidth]{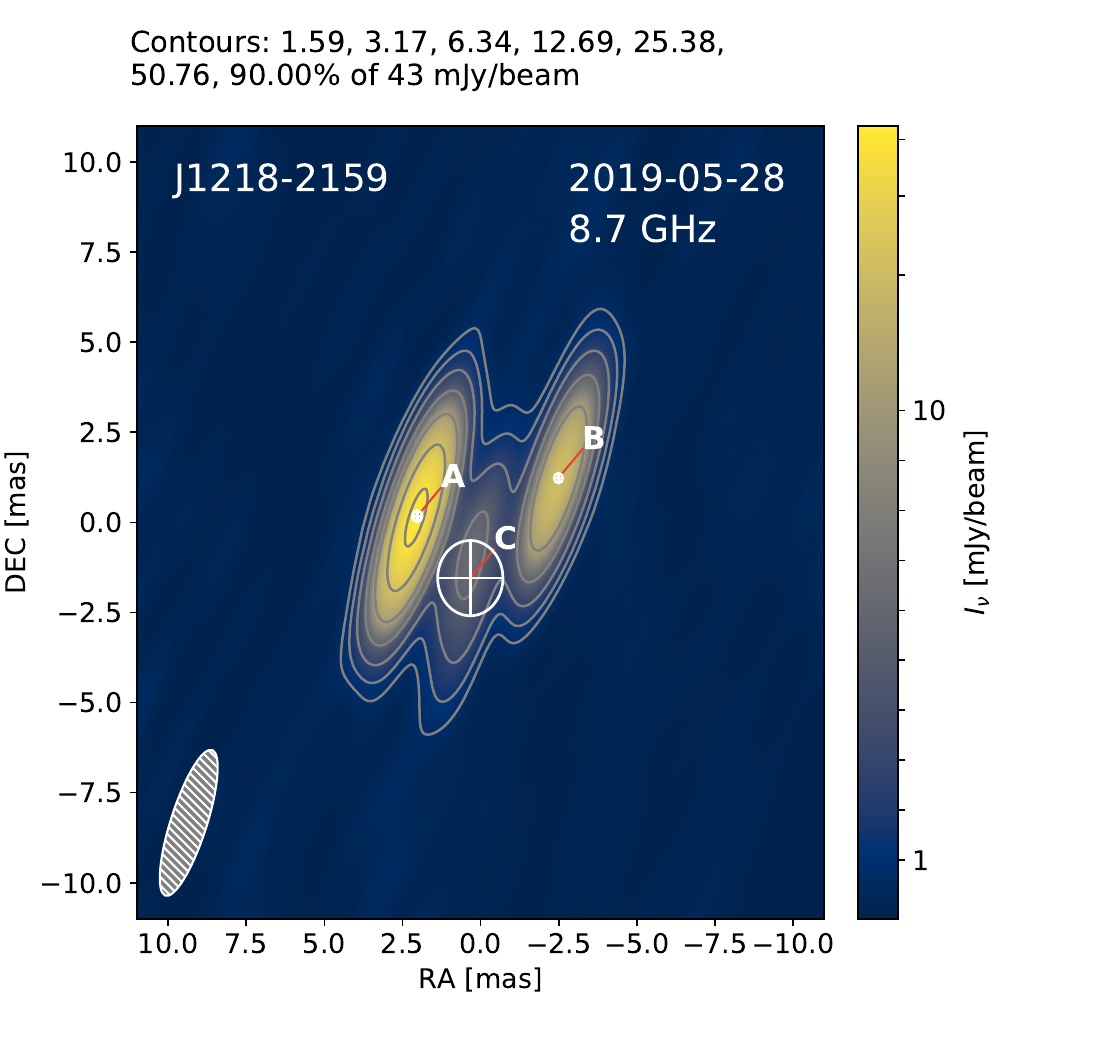}
    \includegraphics[width=0.33\textwidth]{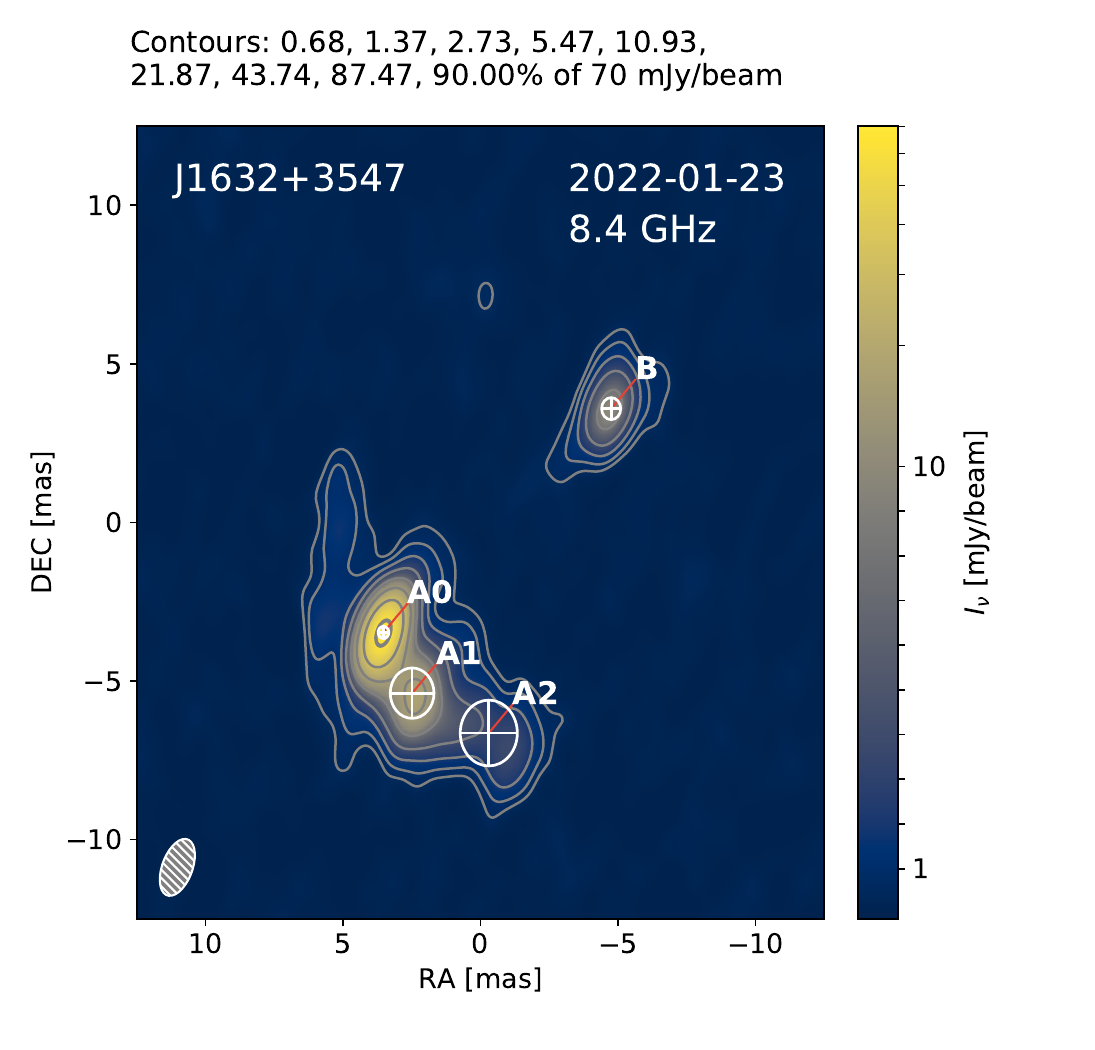}\\
    \includegraphics[width=0.33\textwidth]{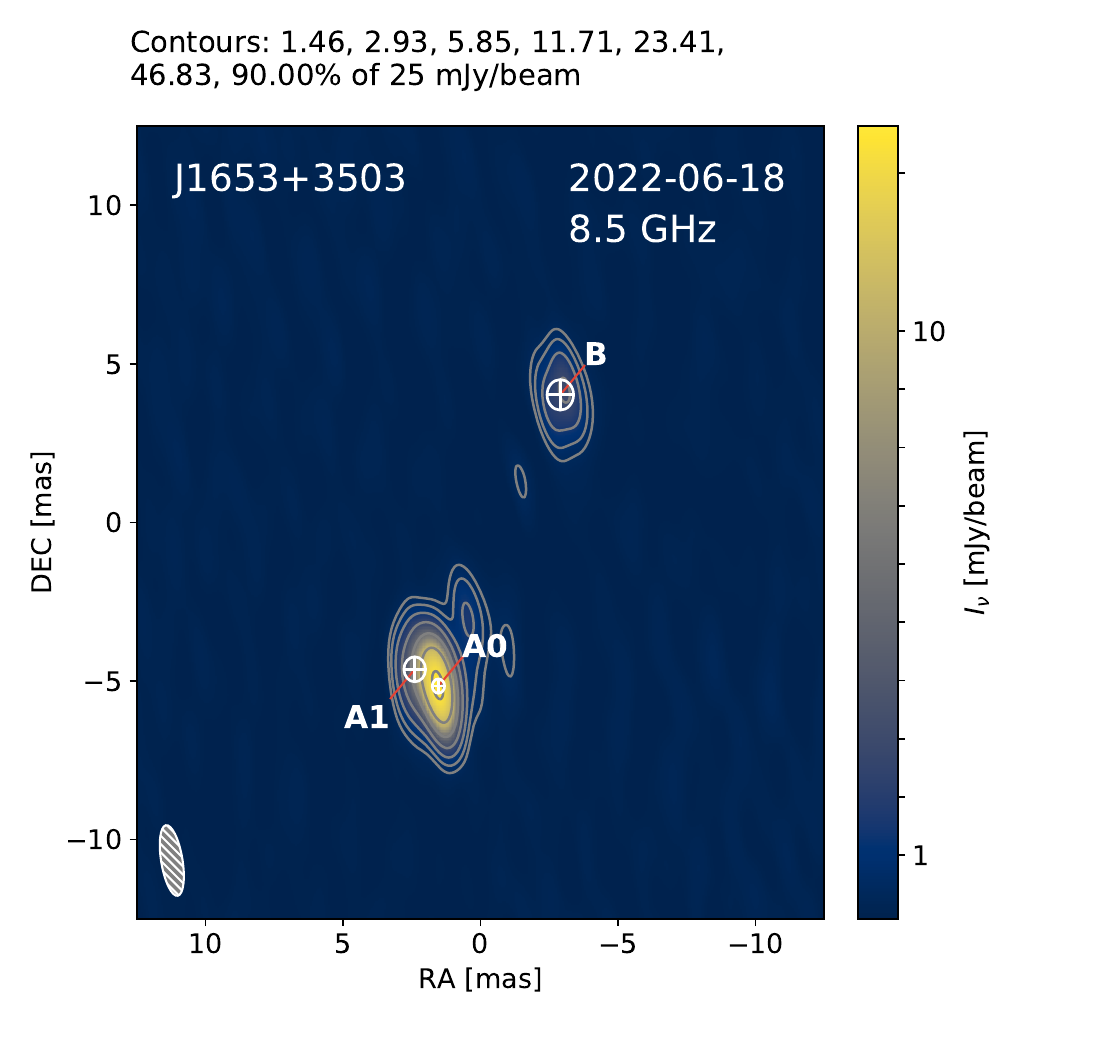}
    \includegraphics[width=0.33\textwidth]{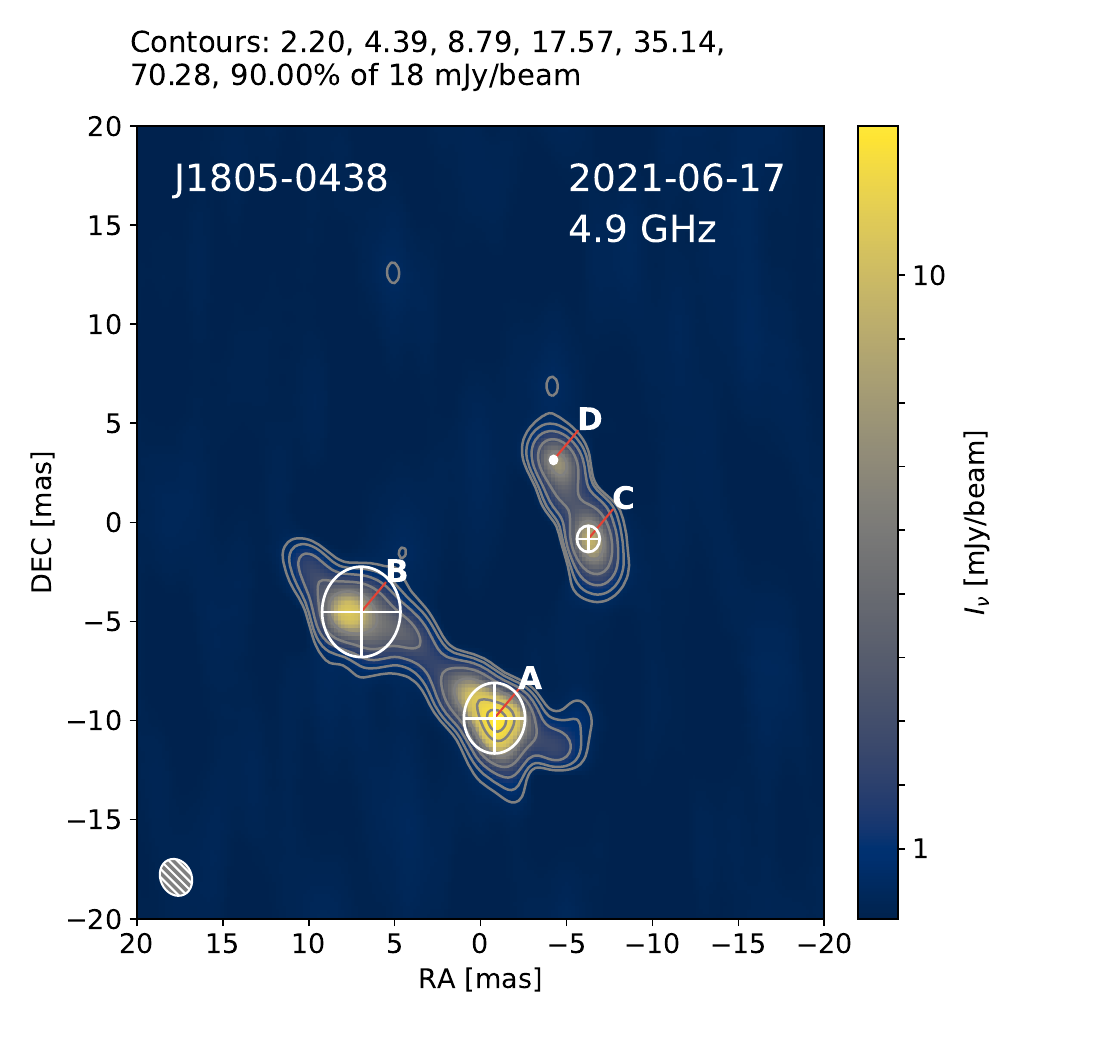}
    \includegraphics[width=0.33\textwidth]{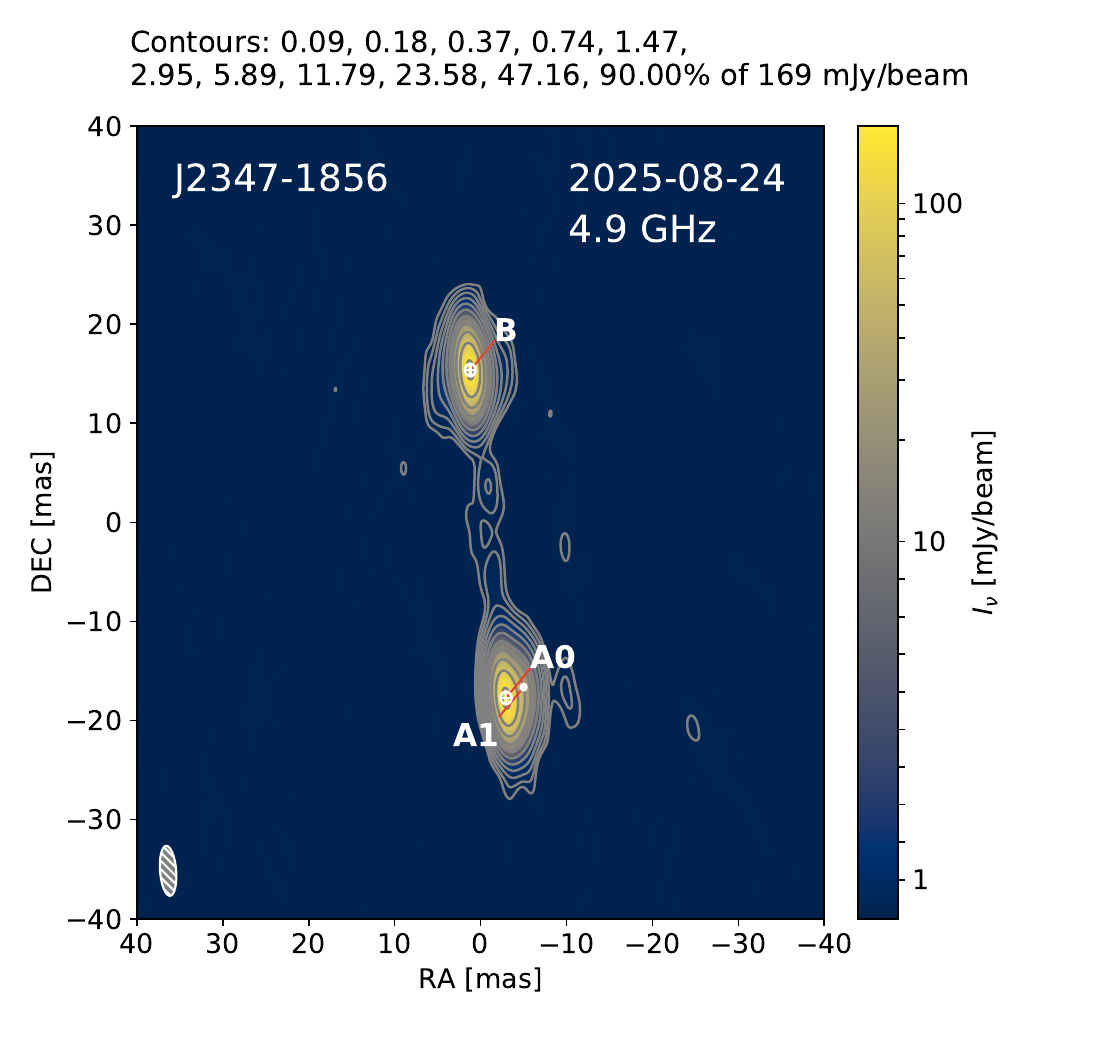}
\caption{Representative VLBI intensity maps of each of the nine milli-lens candidates. Lowest contours are drawn at four times the rms of the residuals from the CLEAN model.}
    \label{fig:vlbimaps}
\end{figure*}

\section{VLBI synchrotron self-absorption spectra}\label{app:ssaspectra}

\begin{figure*}[h]
    \centering
    \includegraphics[width=0.33\textwidth]{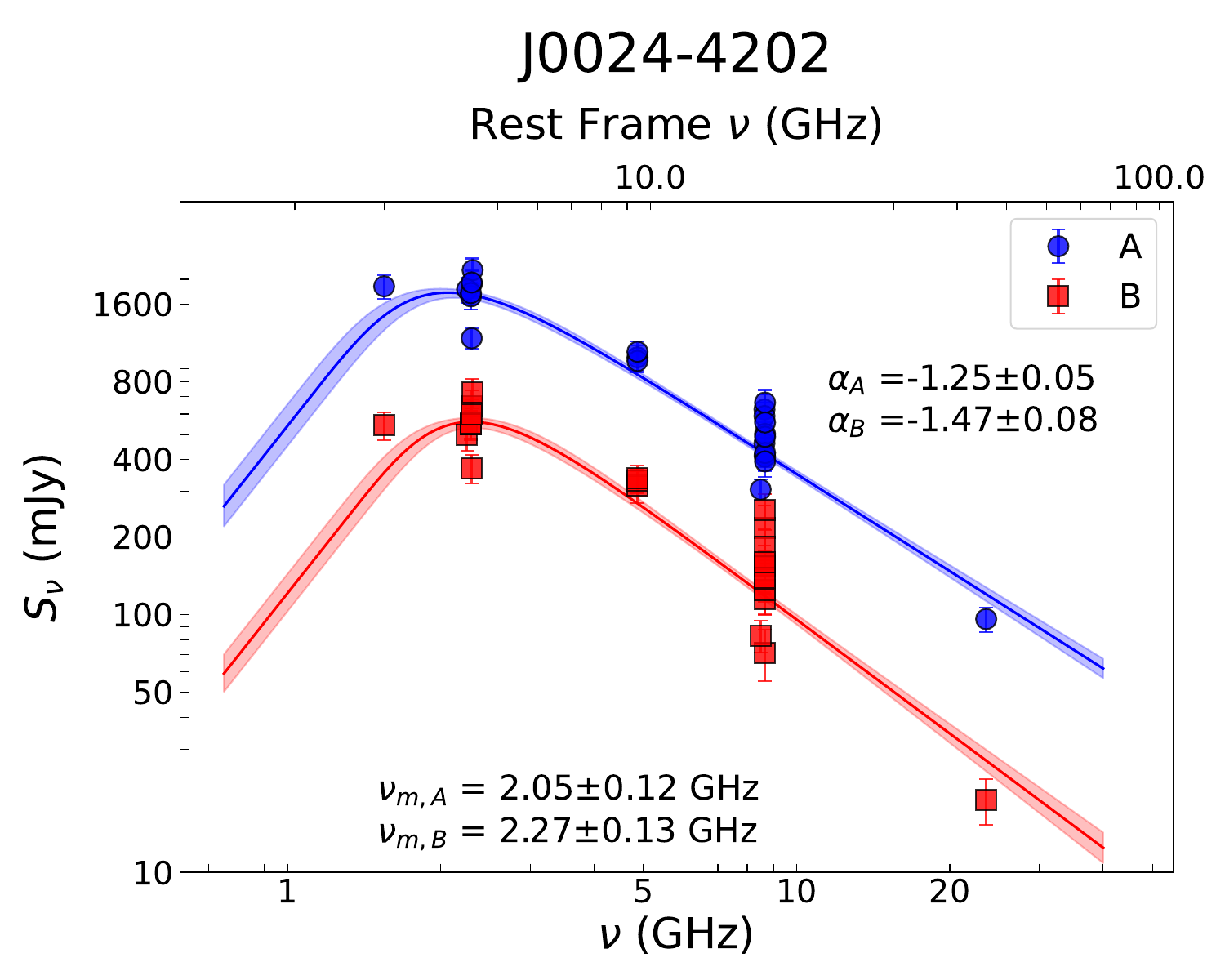}
    \includegraphics[width=0.33\textwidth]{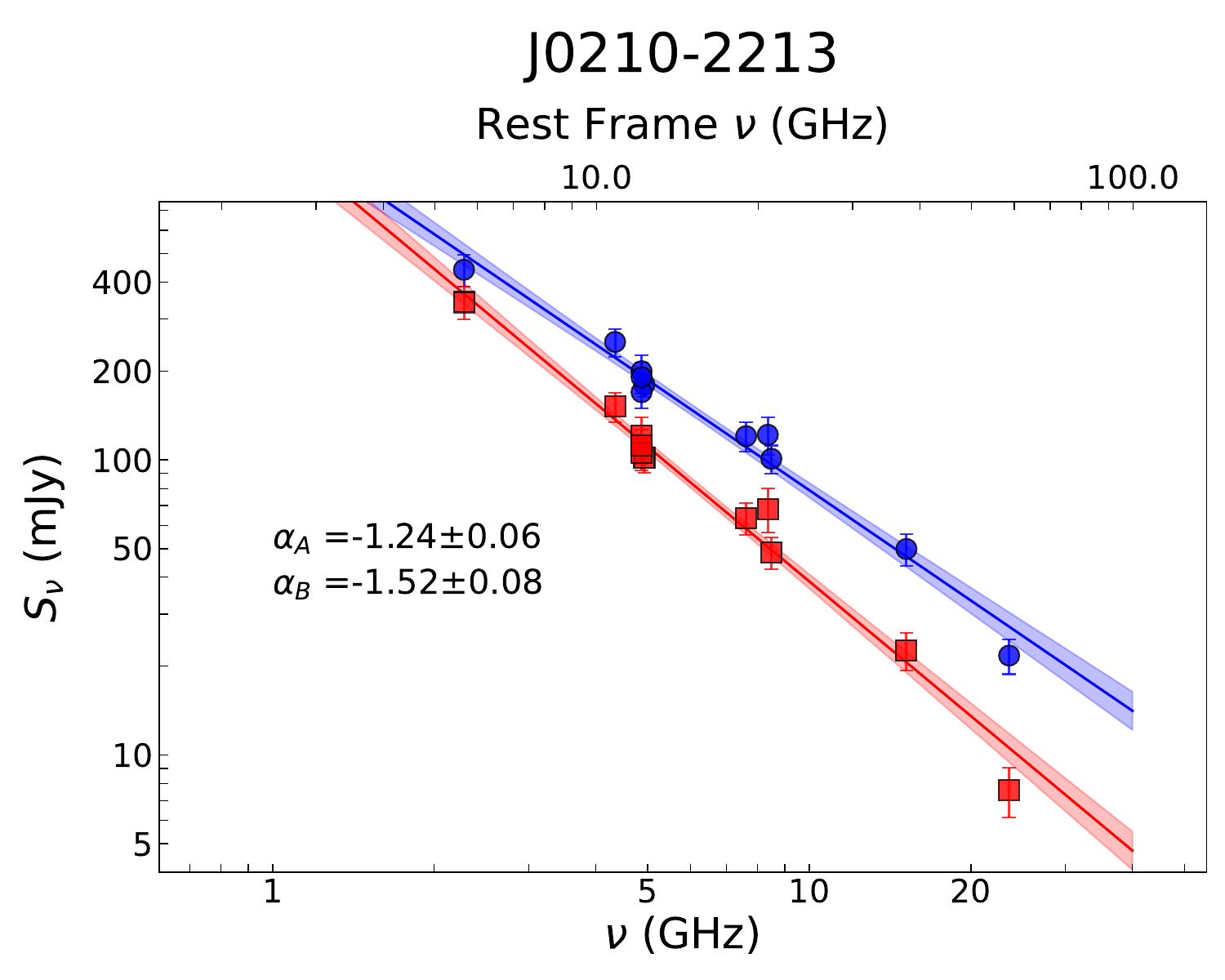}
    \includegraphics[width=0.33\textwidth]{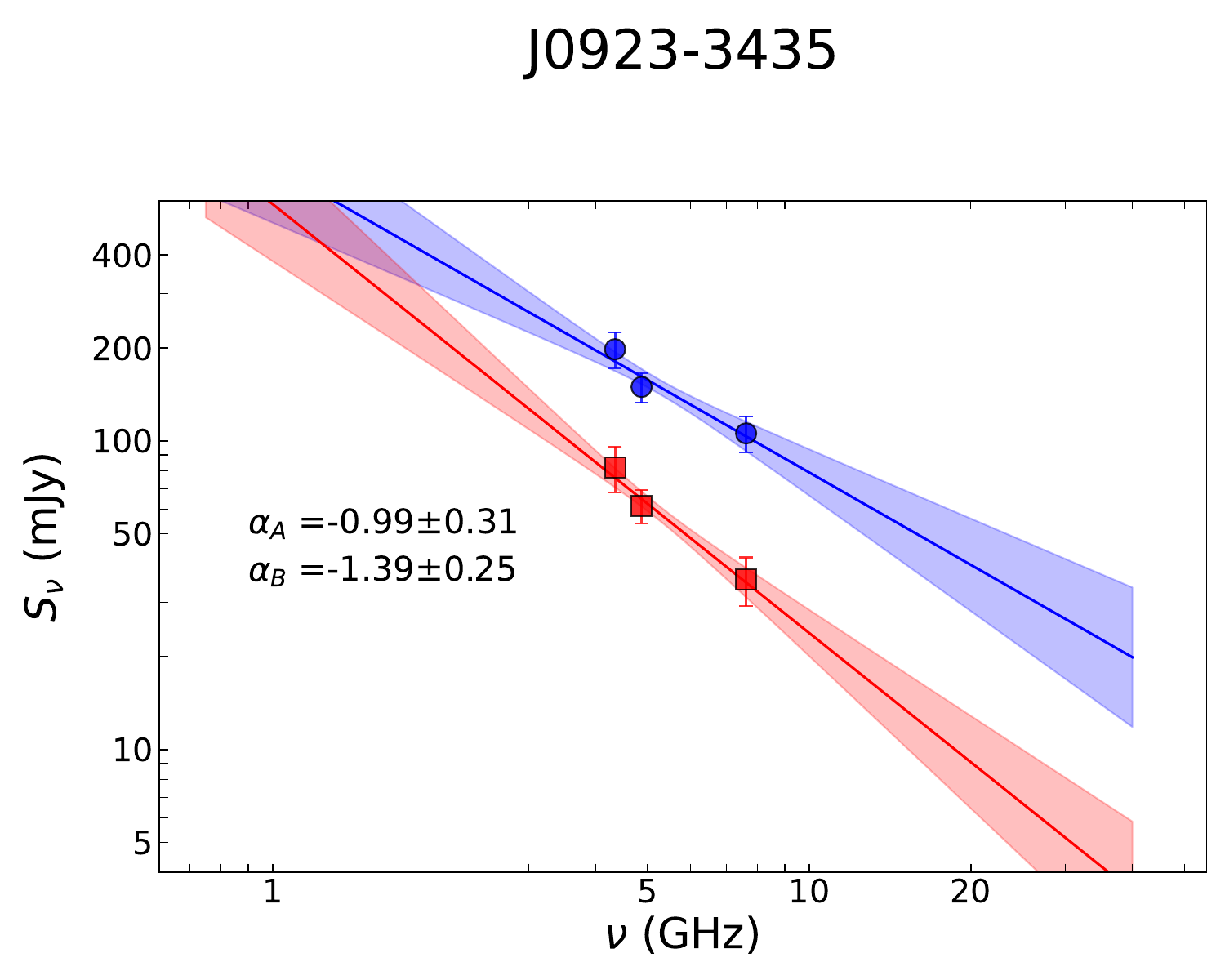}\\
    \includegraphics[width=0.33\textwidth]{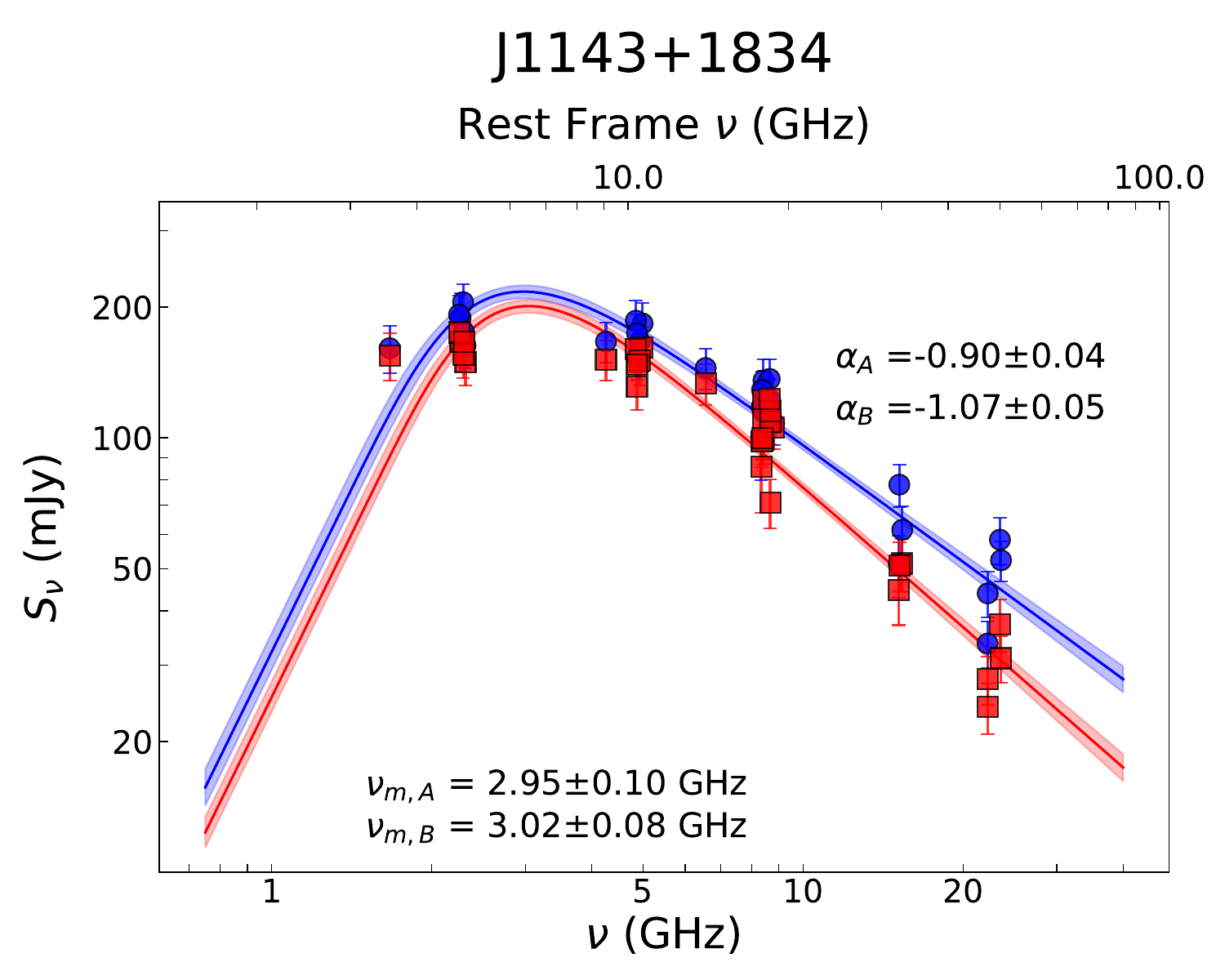}
    \includegraphics[width=0.33\textwidth]{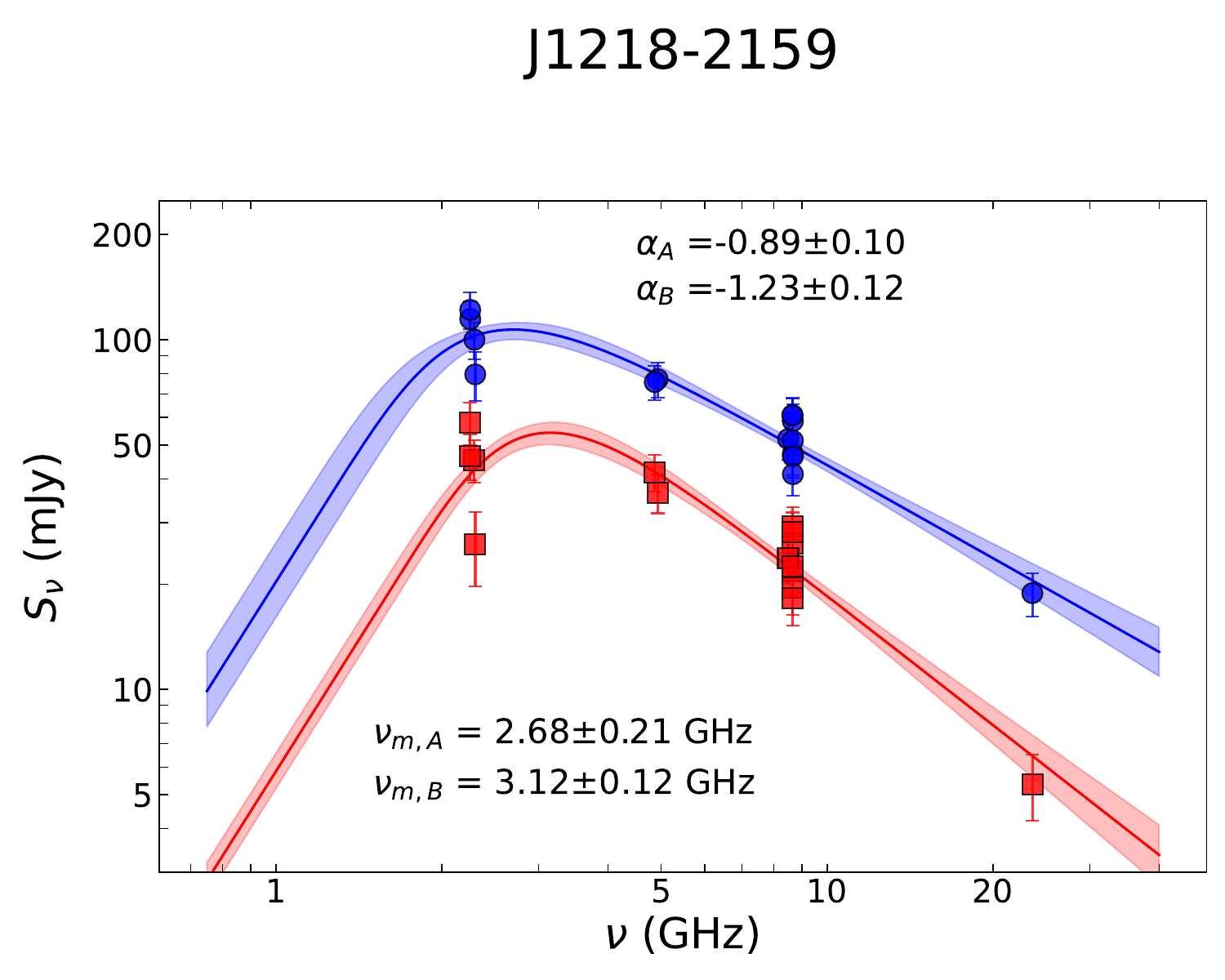}
    \includegraphics[width=0.33\textwidth]{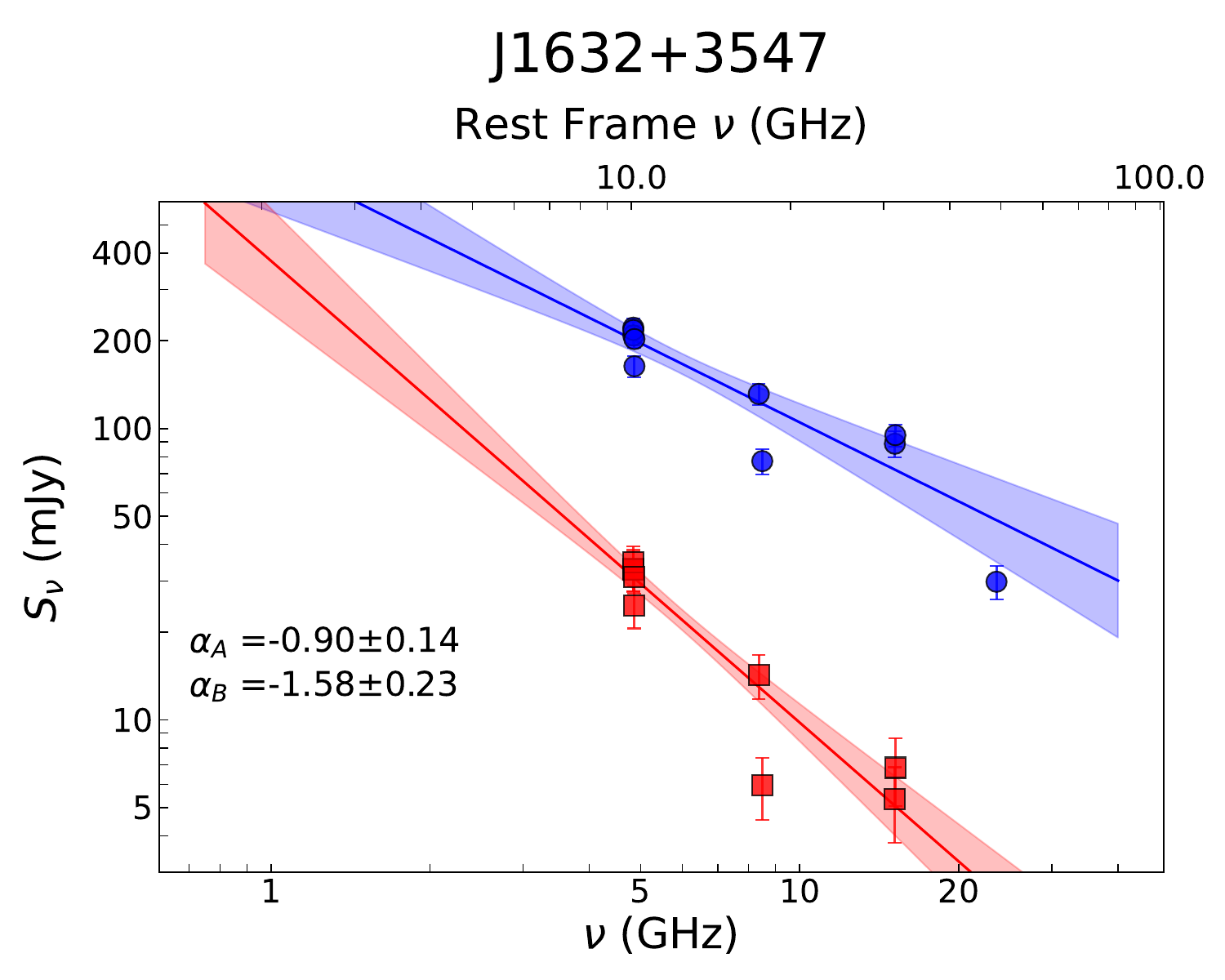}\\
    \includegraphics[width=0.33\textwidth]{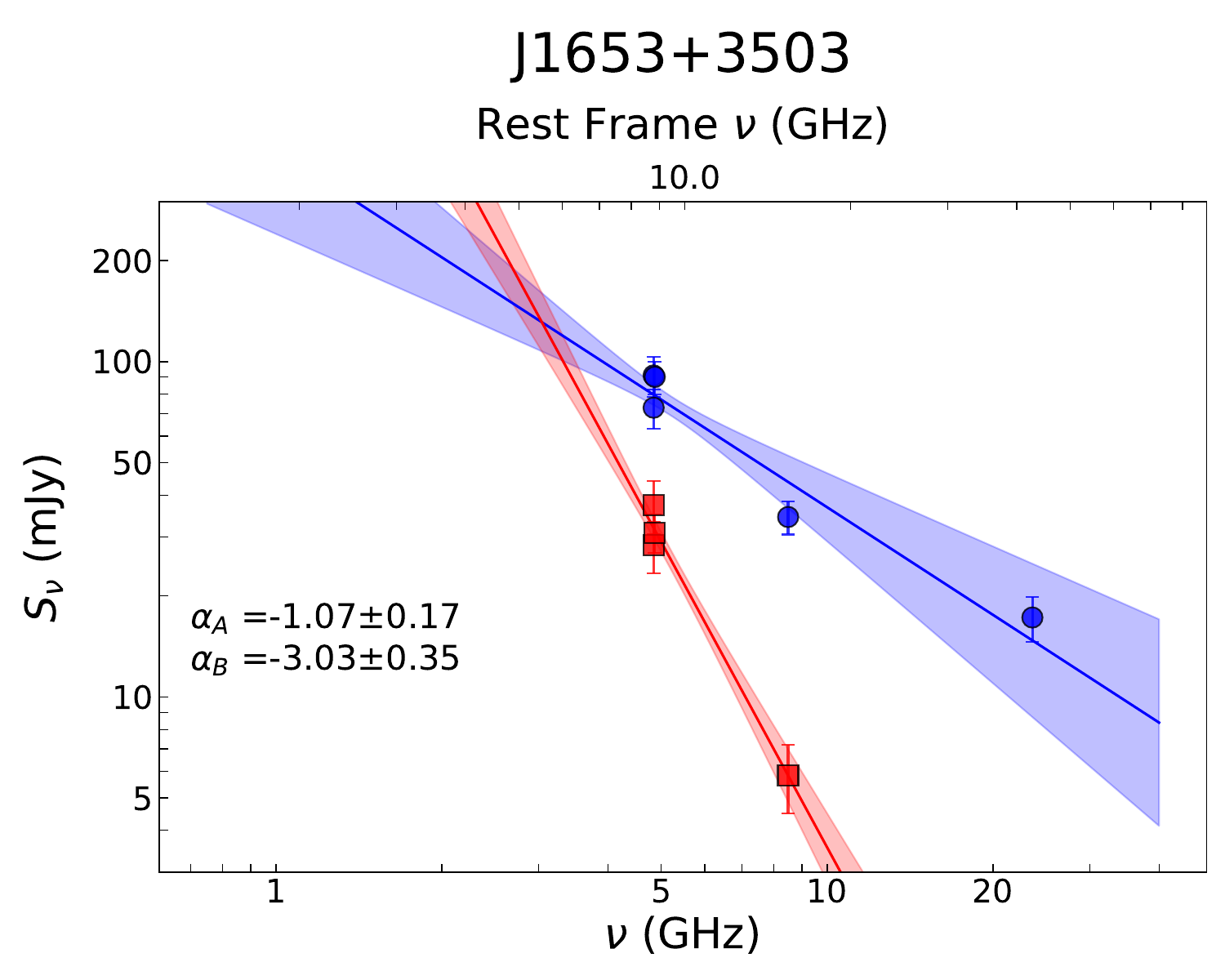}
    \includegraphics[width=0.33\textwidth]{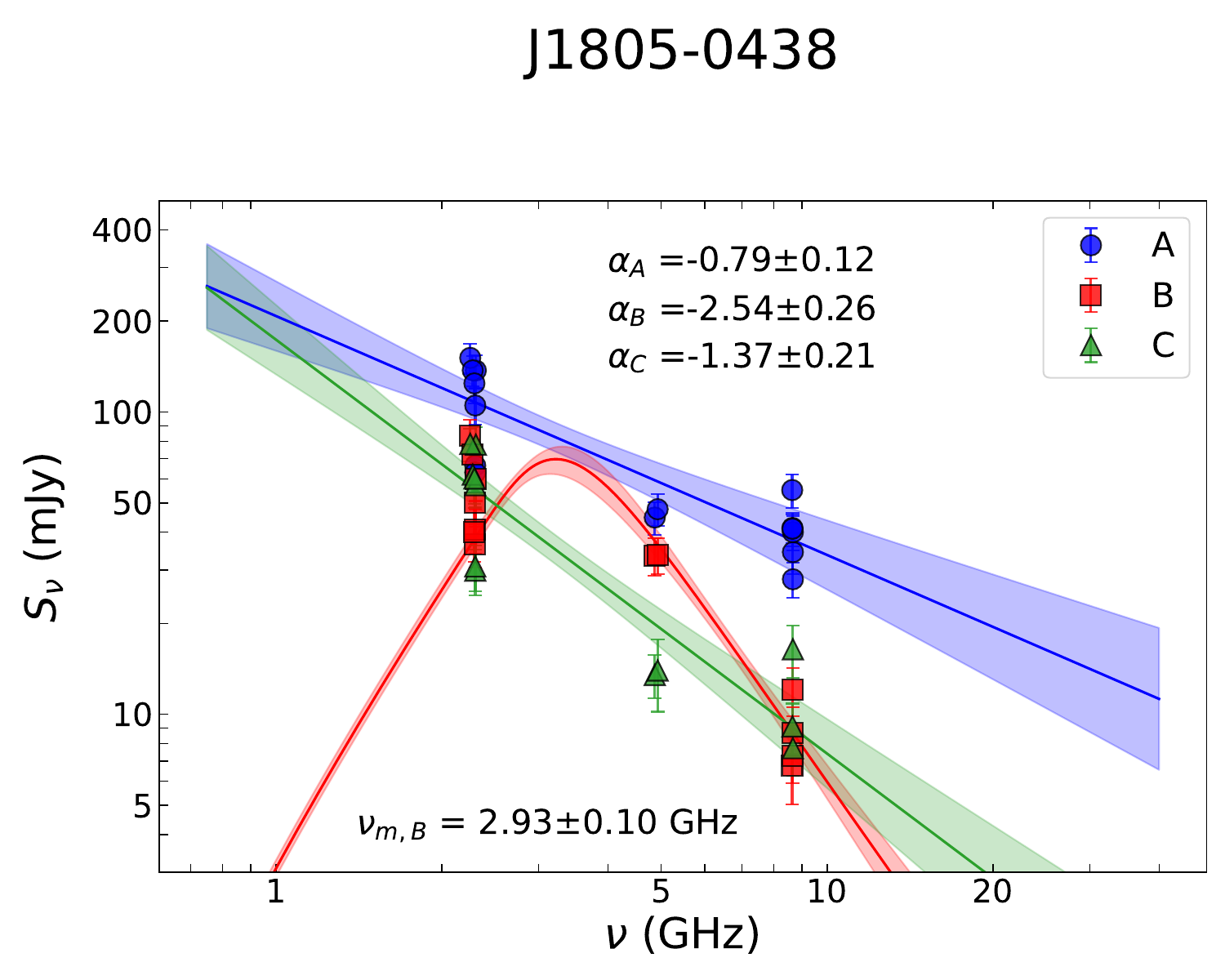}
    \includegraphics[width=0.33\textwidth]{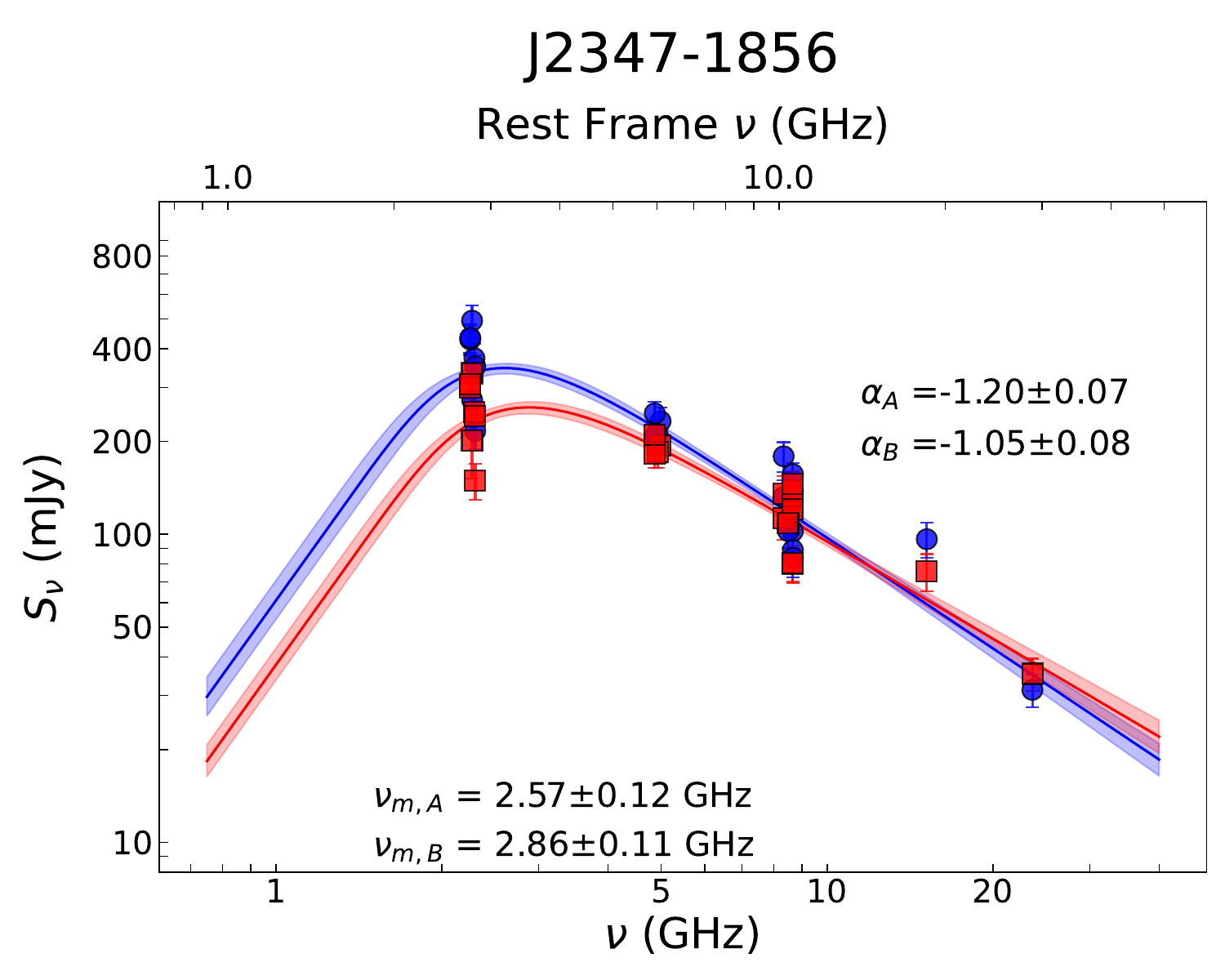}
\caption{VLBI spectra of the different components of each of the milli-lens candidates. Overplotted are the synchrotron self-absorption spectra along with their 1-$\sigma$ confidence envelope. Best fit values of the turnover frequency and the spectral index and their errors are indicated on each plot.}
    \label{fig:vlbispectra}
\end{figure*}

\section{VLBI spectral index maps}\label{app:spixmaps}

\begin{figure*}[h]
    \centering
    \includegraphics[width=0.33\textwidth]{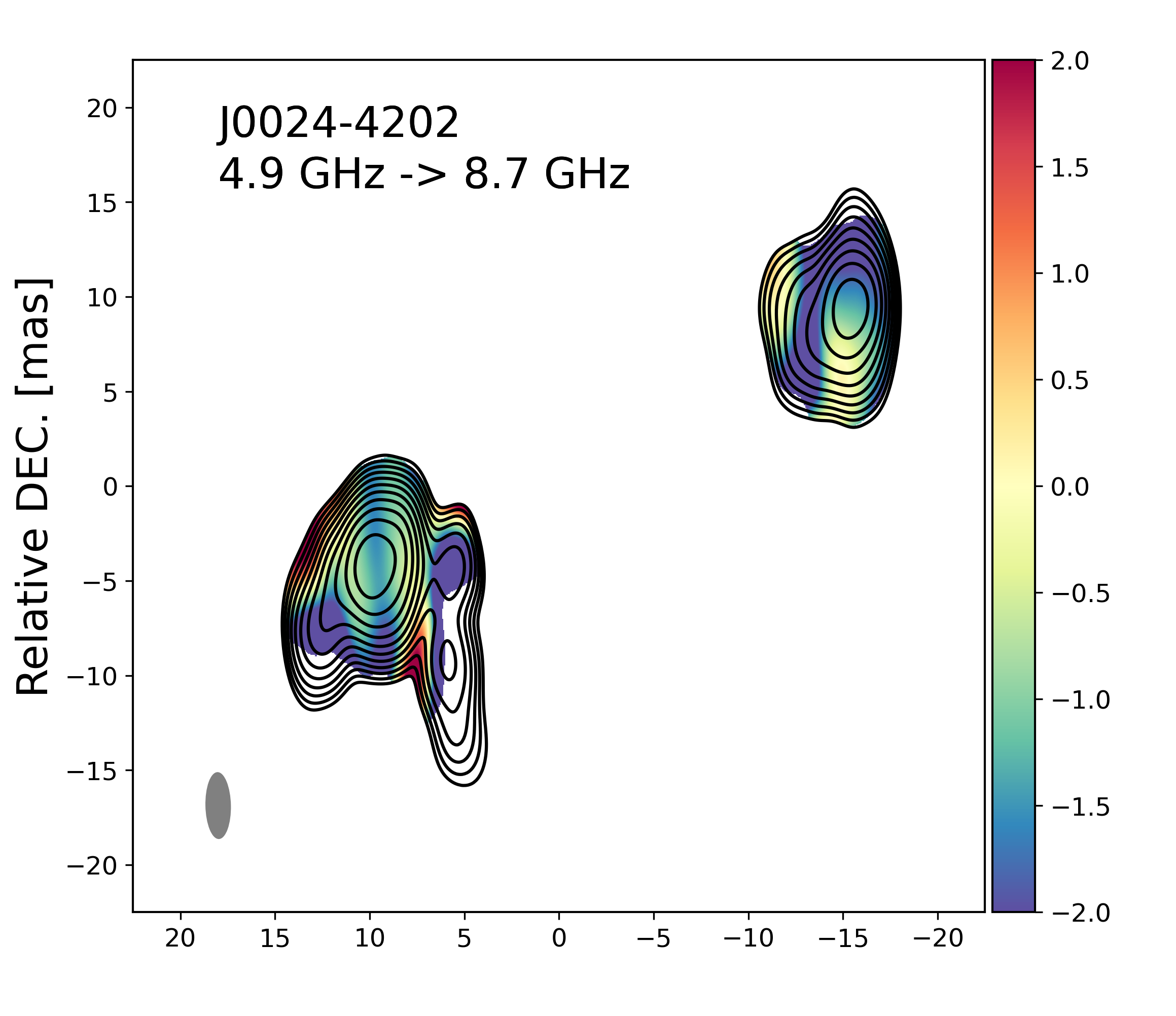}
    \includegraphics[width=0.33\textwidth]{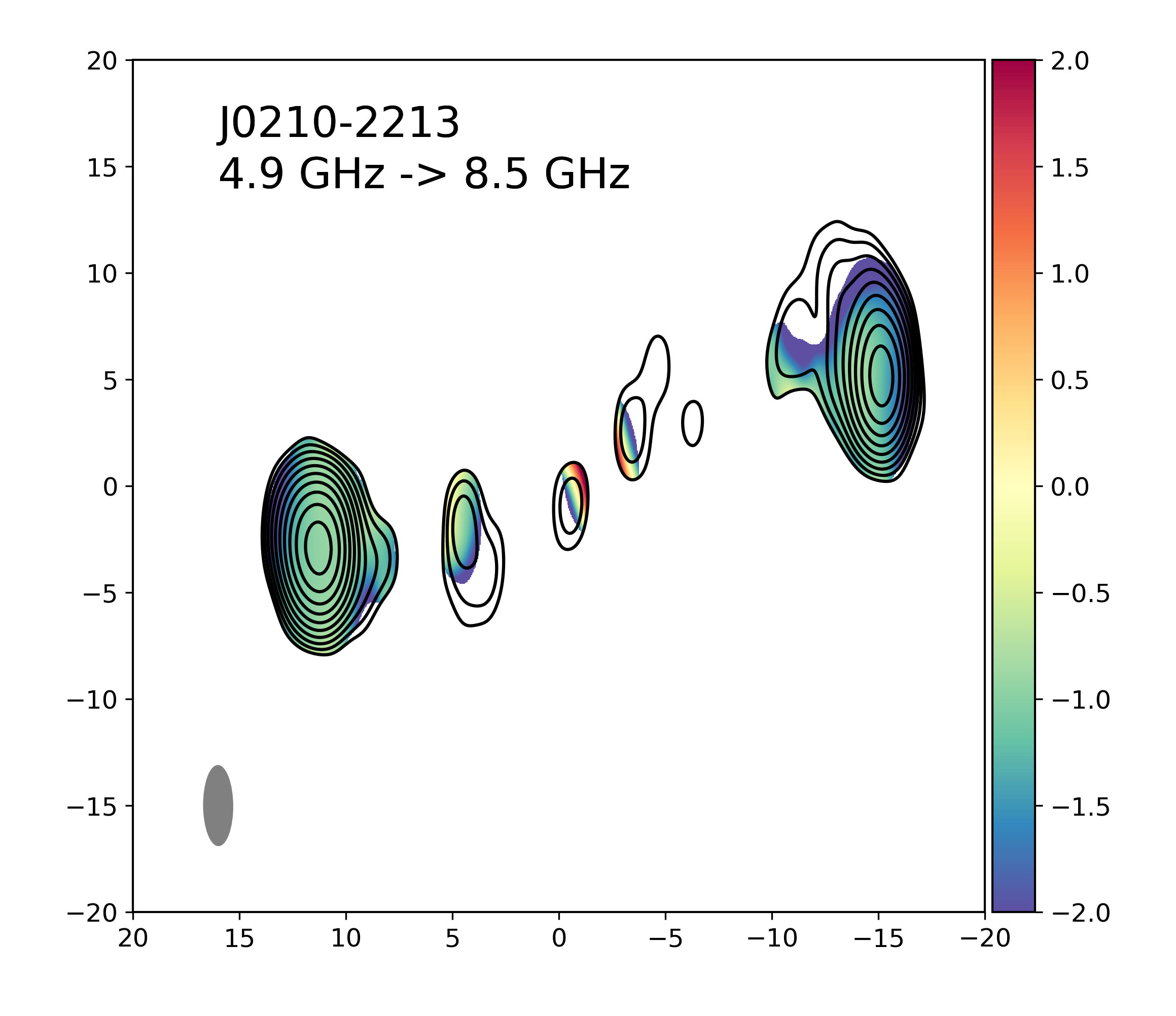}
    \includegraphics[width=0.33\textwidth]{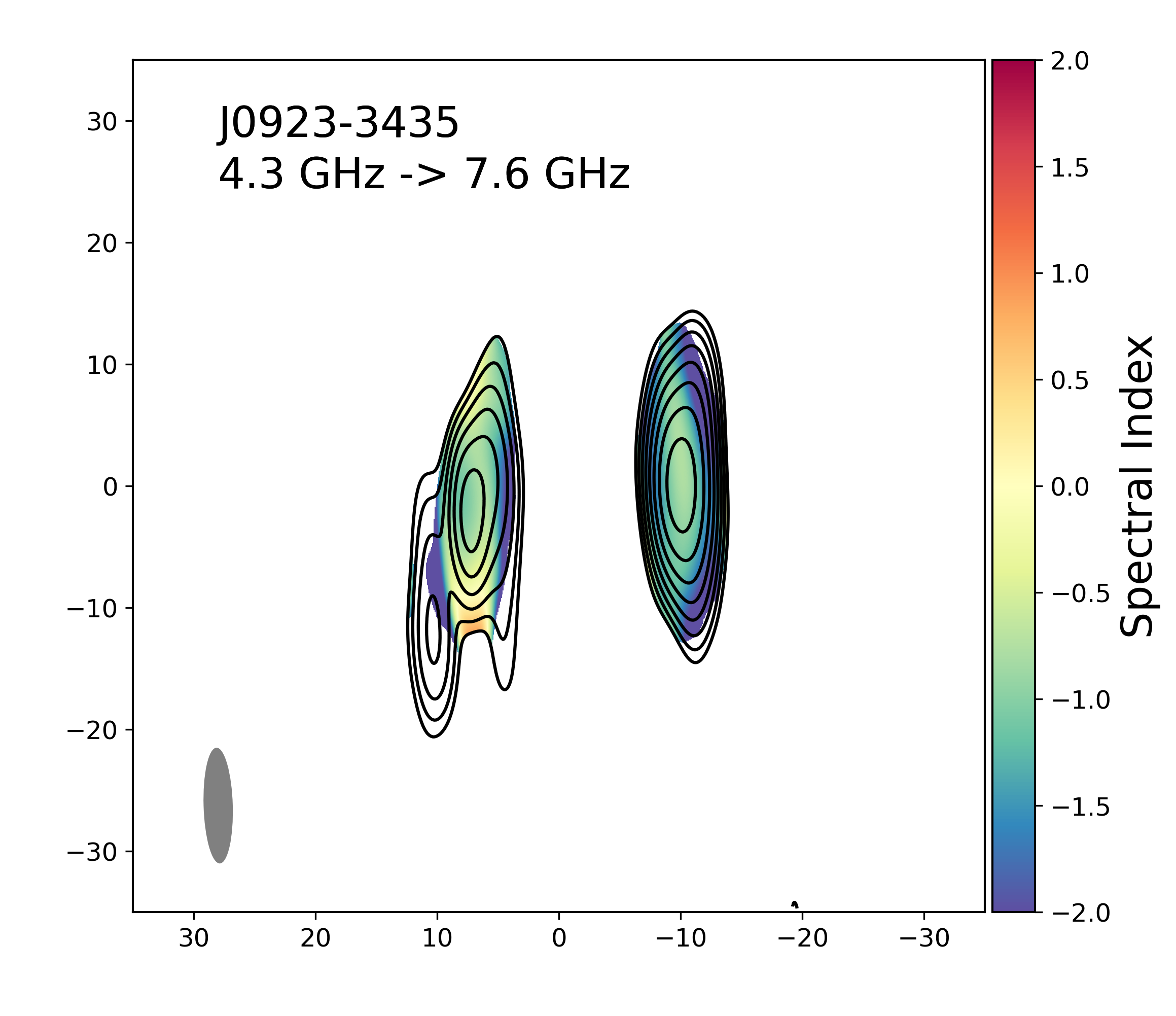}\\
    \includegraphics[width=0.33\textwidth]{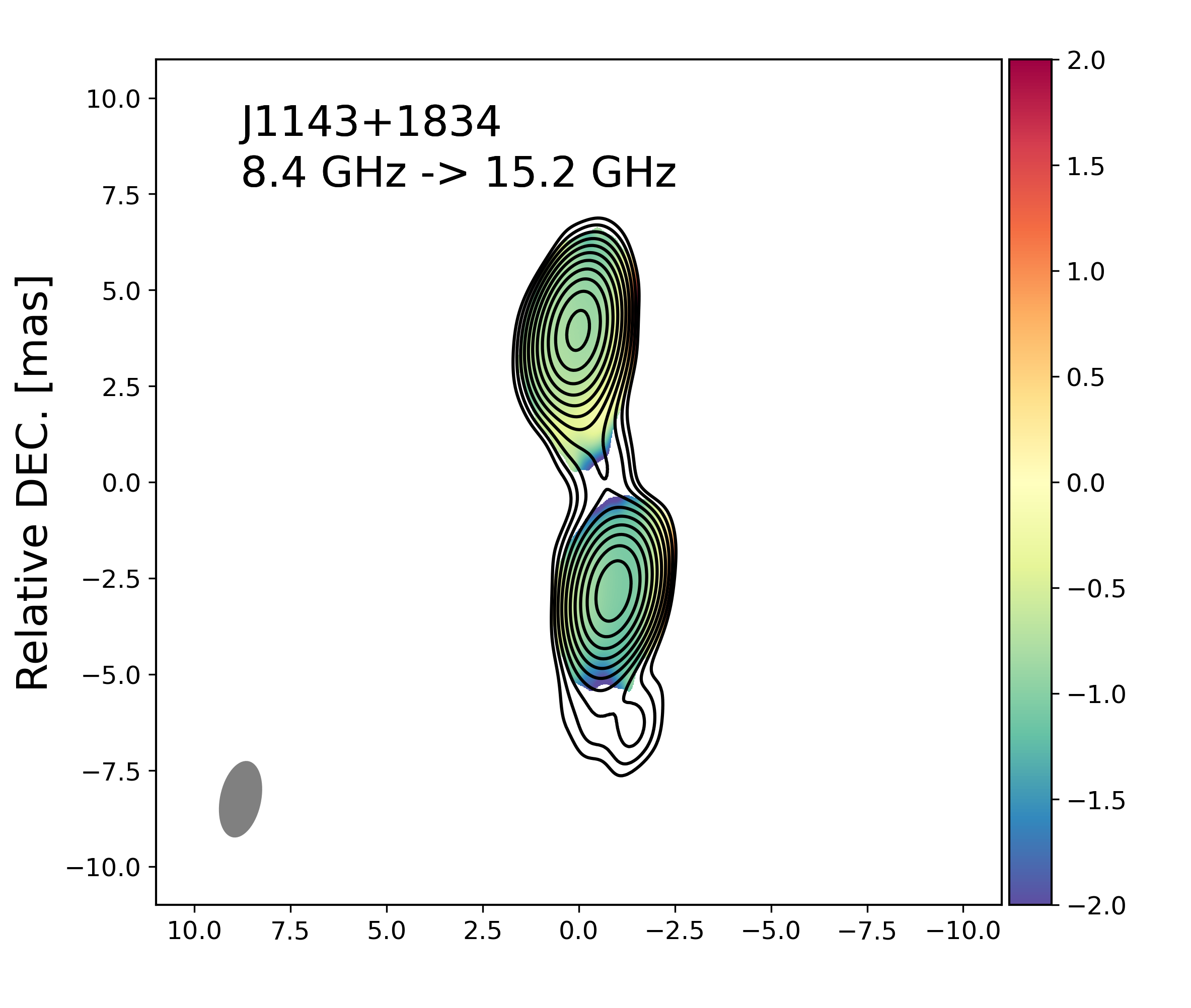}
    \includegraphics[width=0.33\textwidth]{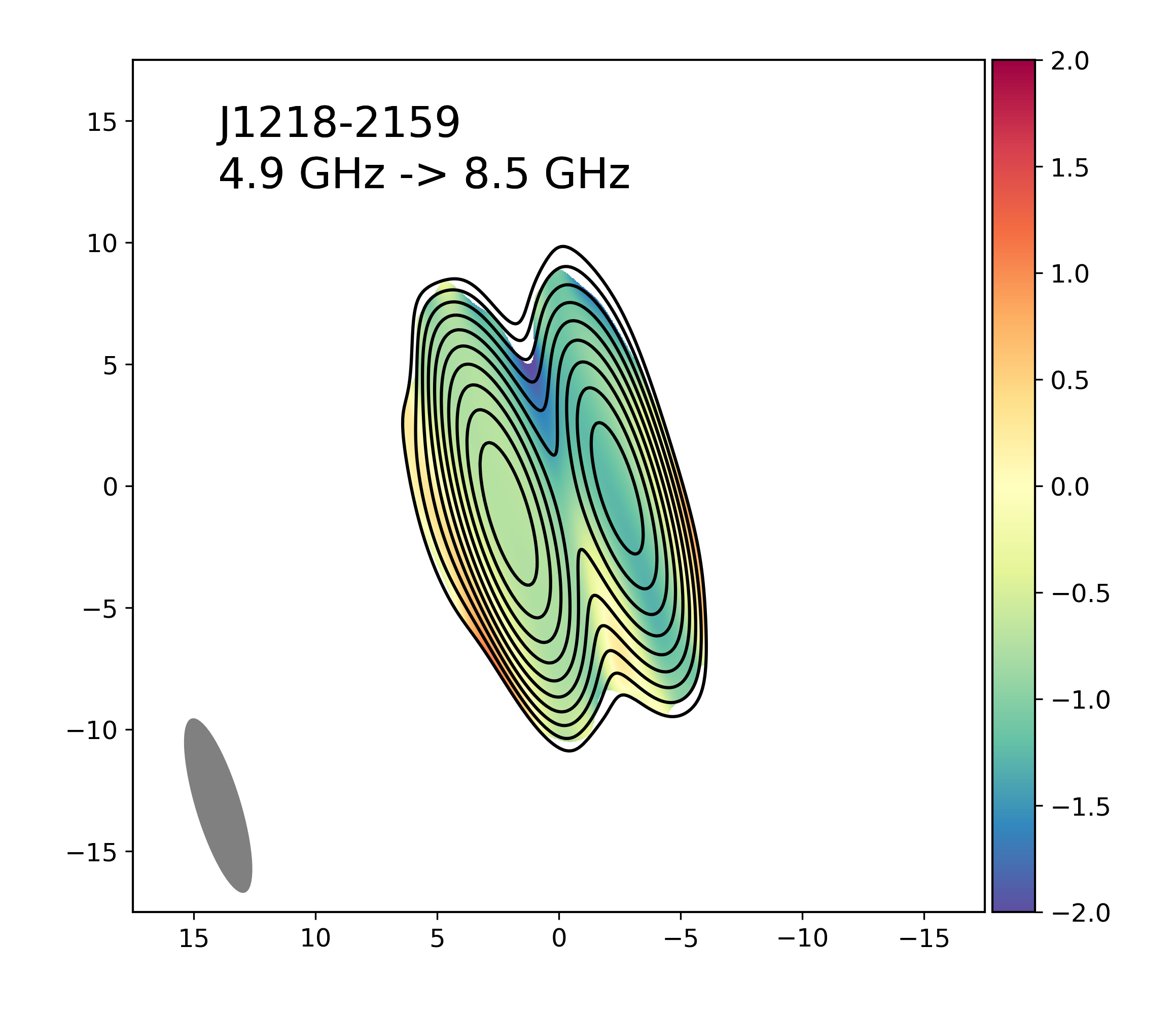}
    \includegraphics[width=0.33\textwidth]{./Figures/SpixMaps/J1632+3547_VIMAP_SPIXMAP}\\
    \includegraphics[width=0.33\textwidth]{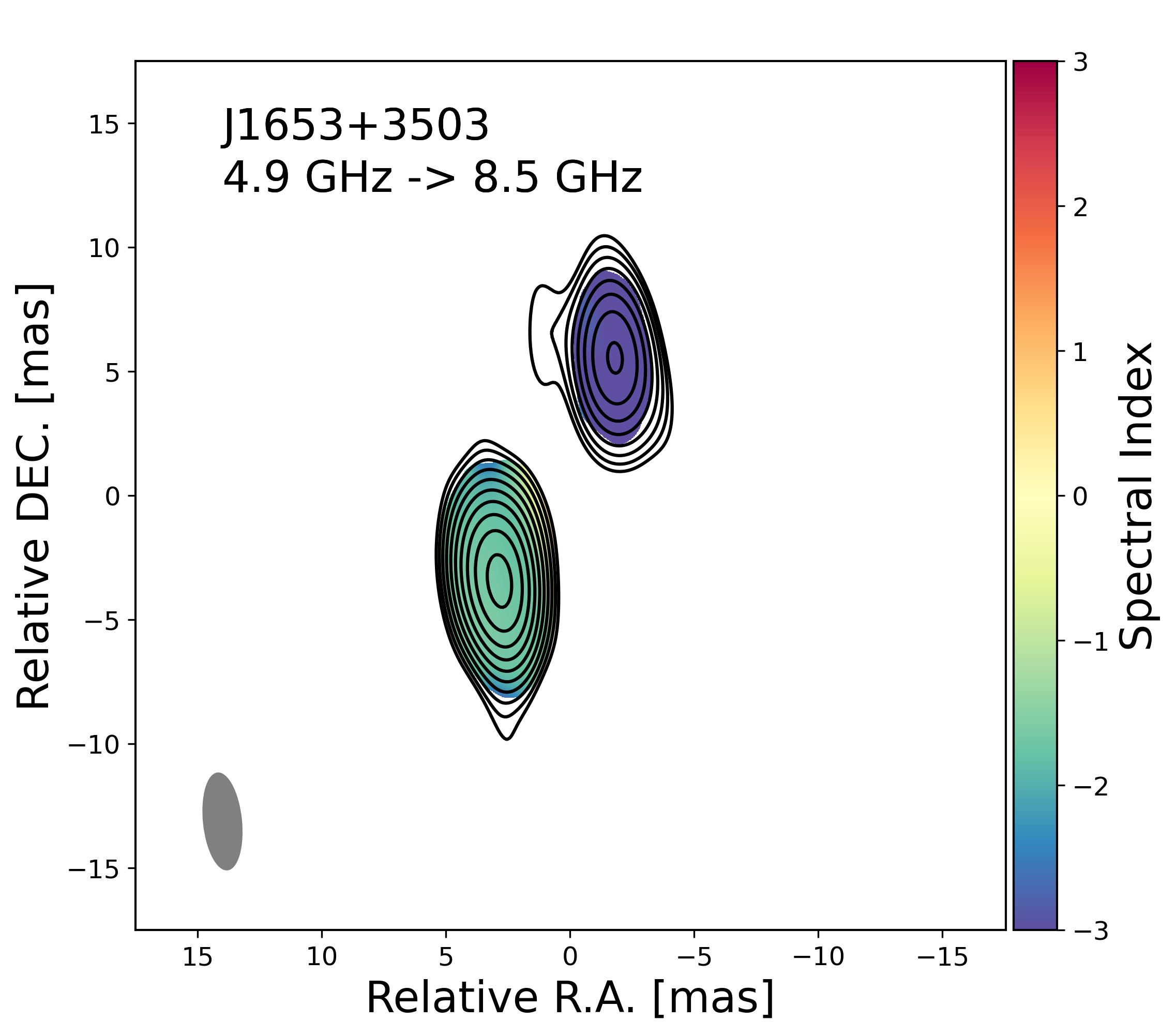}
    \includegraphics[width=0.33\textwidth]{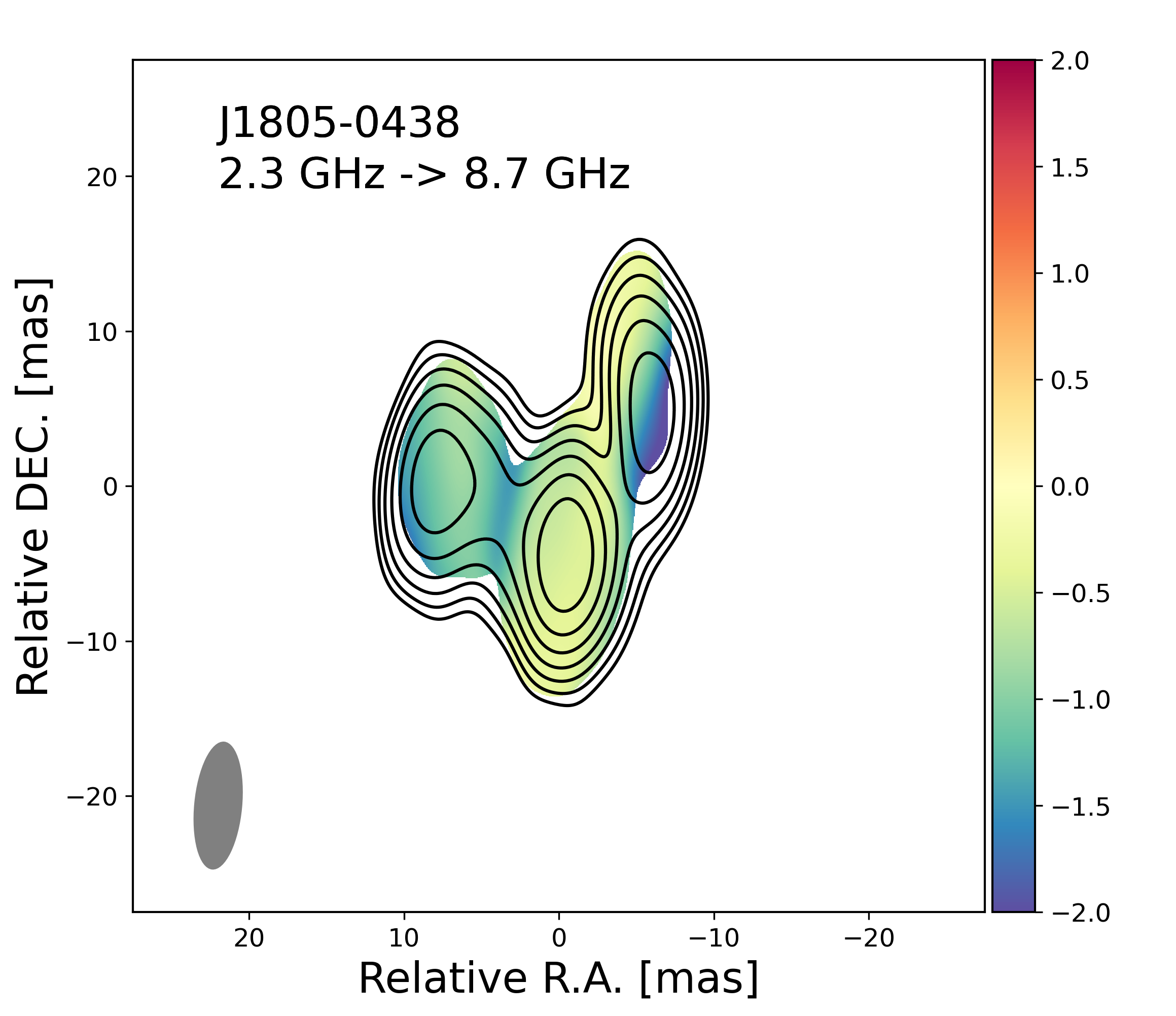}
    \includegraphics[width=0.33\textwidth]{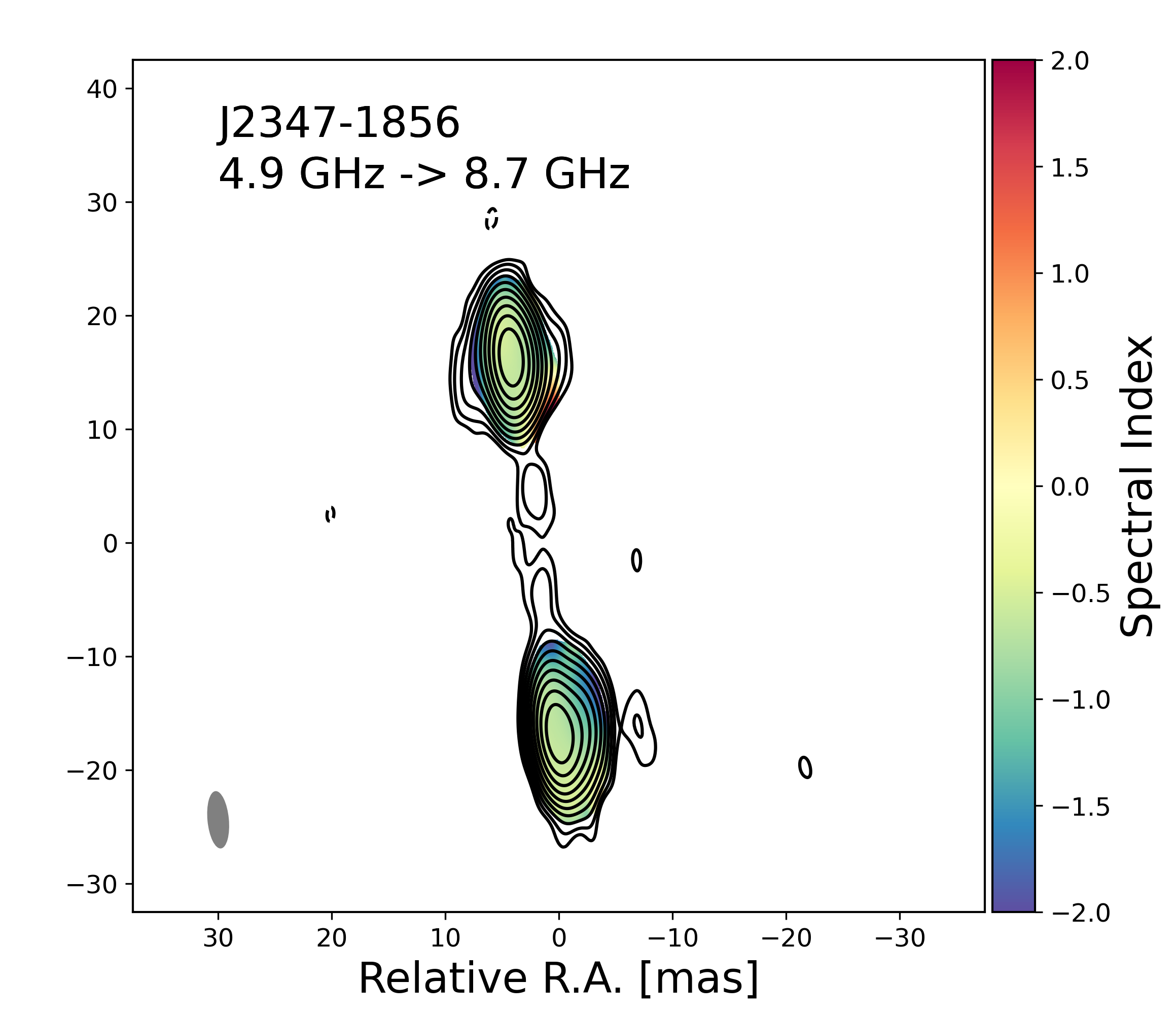}
\caption{Spectral index maps produced with \texttt{VCAT}. }
    \label{fig:spixmaps}
\end{figure*}

\section{Integrated flux spectra}\label{app:sdspectra}

\begin{figure*}[h]
    \centering
    \includegraphics[width=0.33\textwidth]{./Figures/SD_Spectra/J0024_SD_comparison.pdf}
    \includegraphics[width=0.33\textwidth]{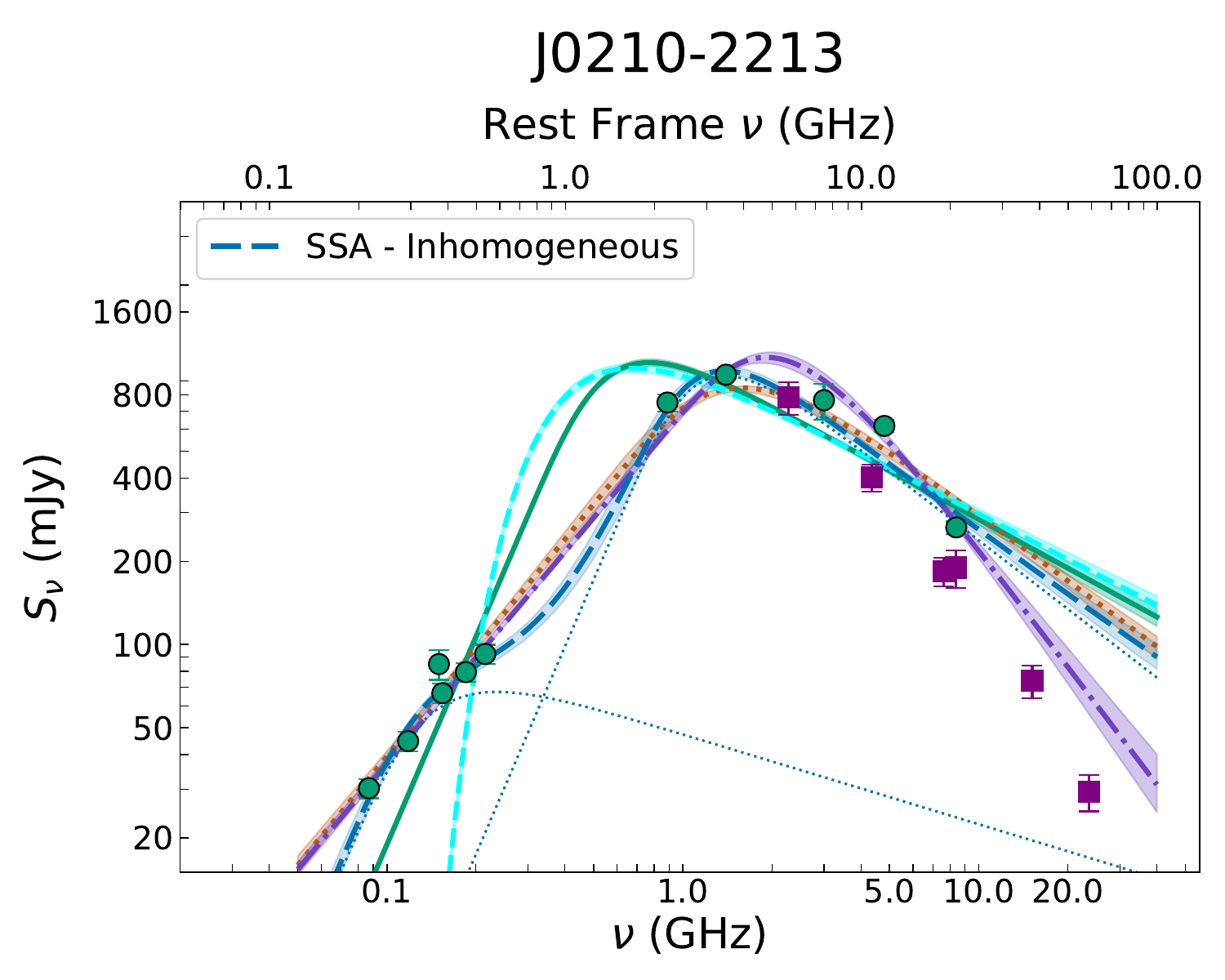}
    \includegraphics[width=0.33\textwidth]{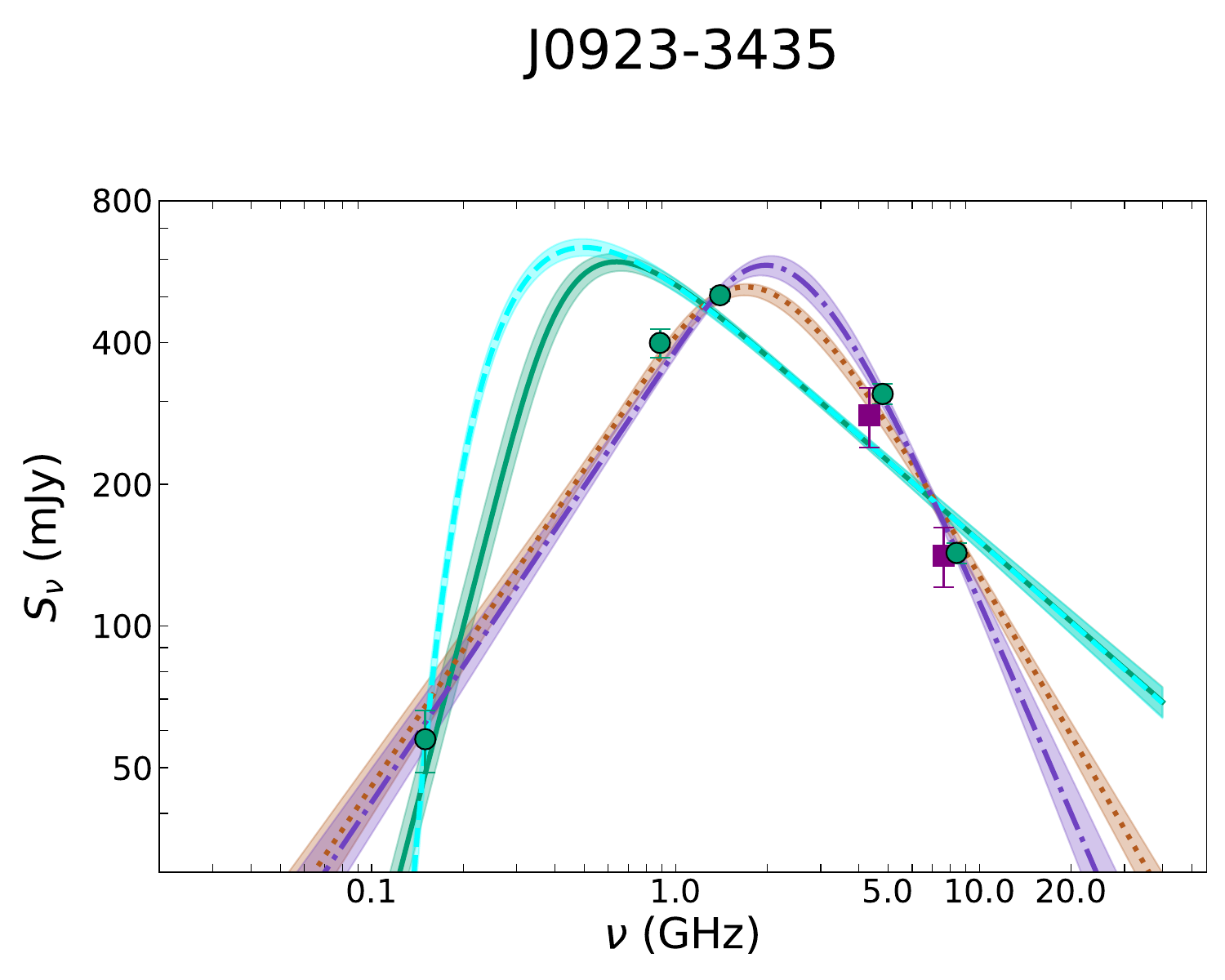}\\\vspace{15pt}
    \includegraphics[width=0.33\textwidth]{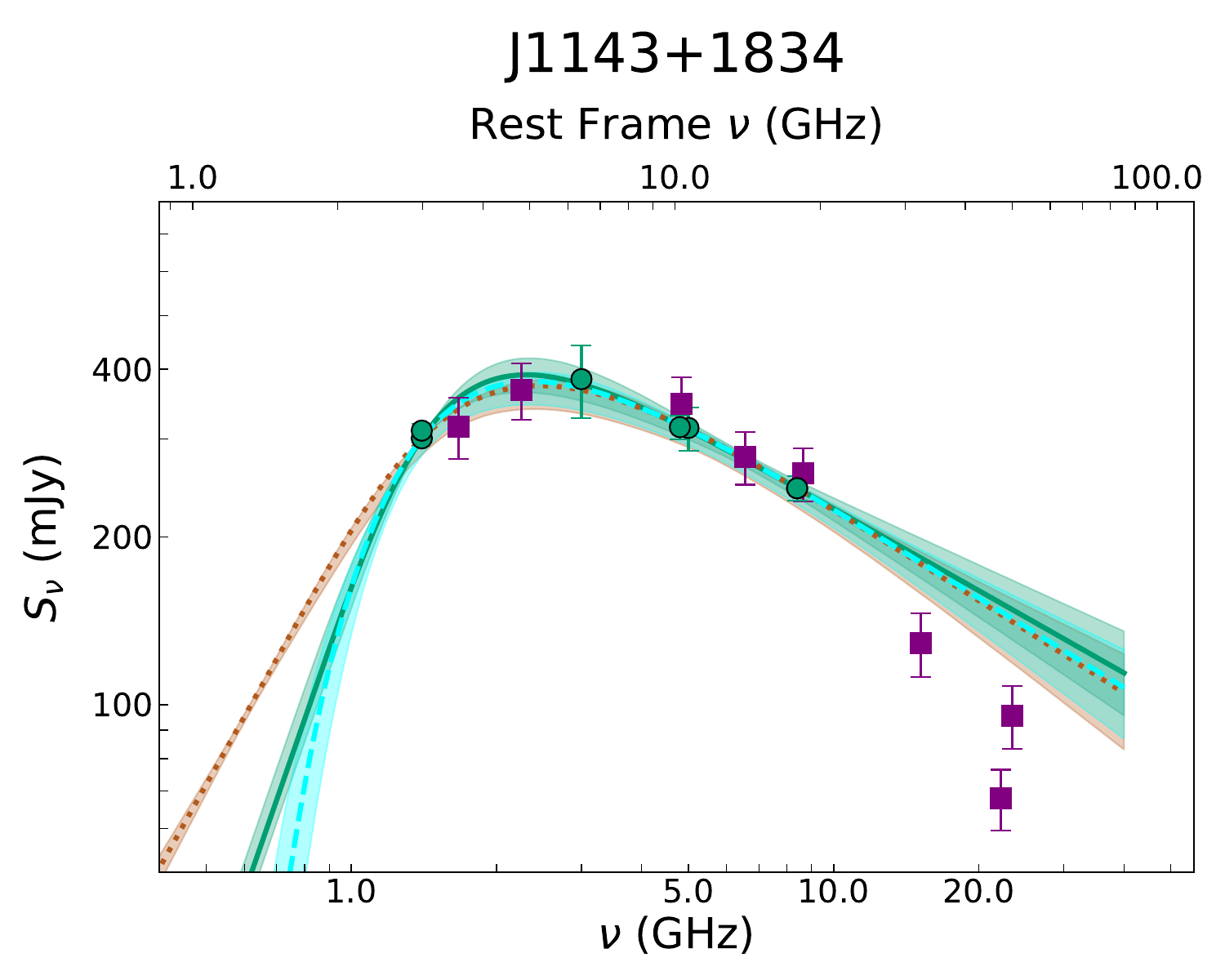}
    \includegraphics[width=0.33\textwidth]{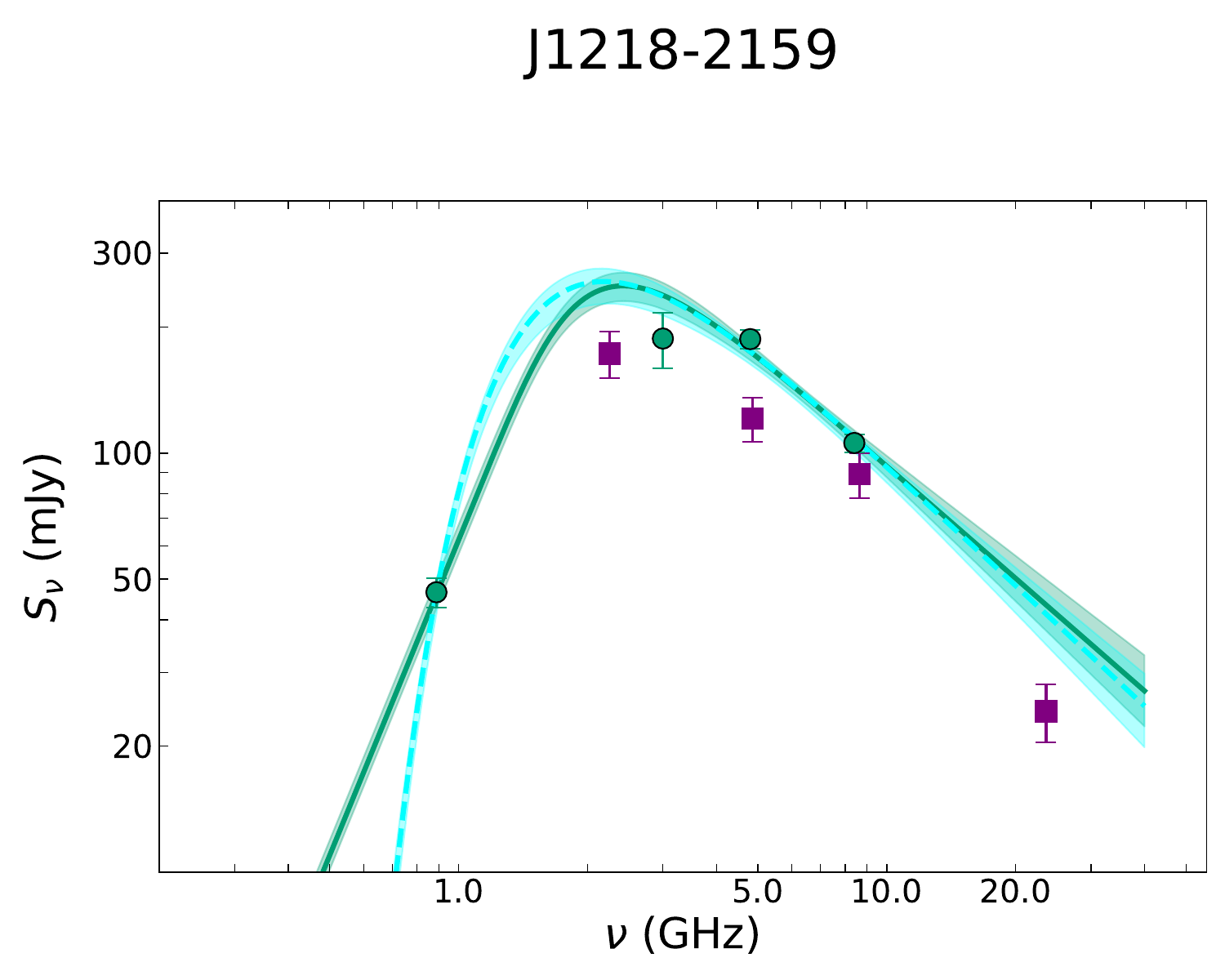}
    \includegraphics[width=0.33\textwidth]{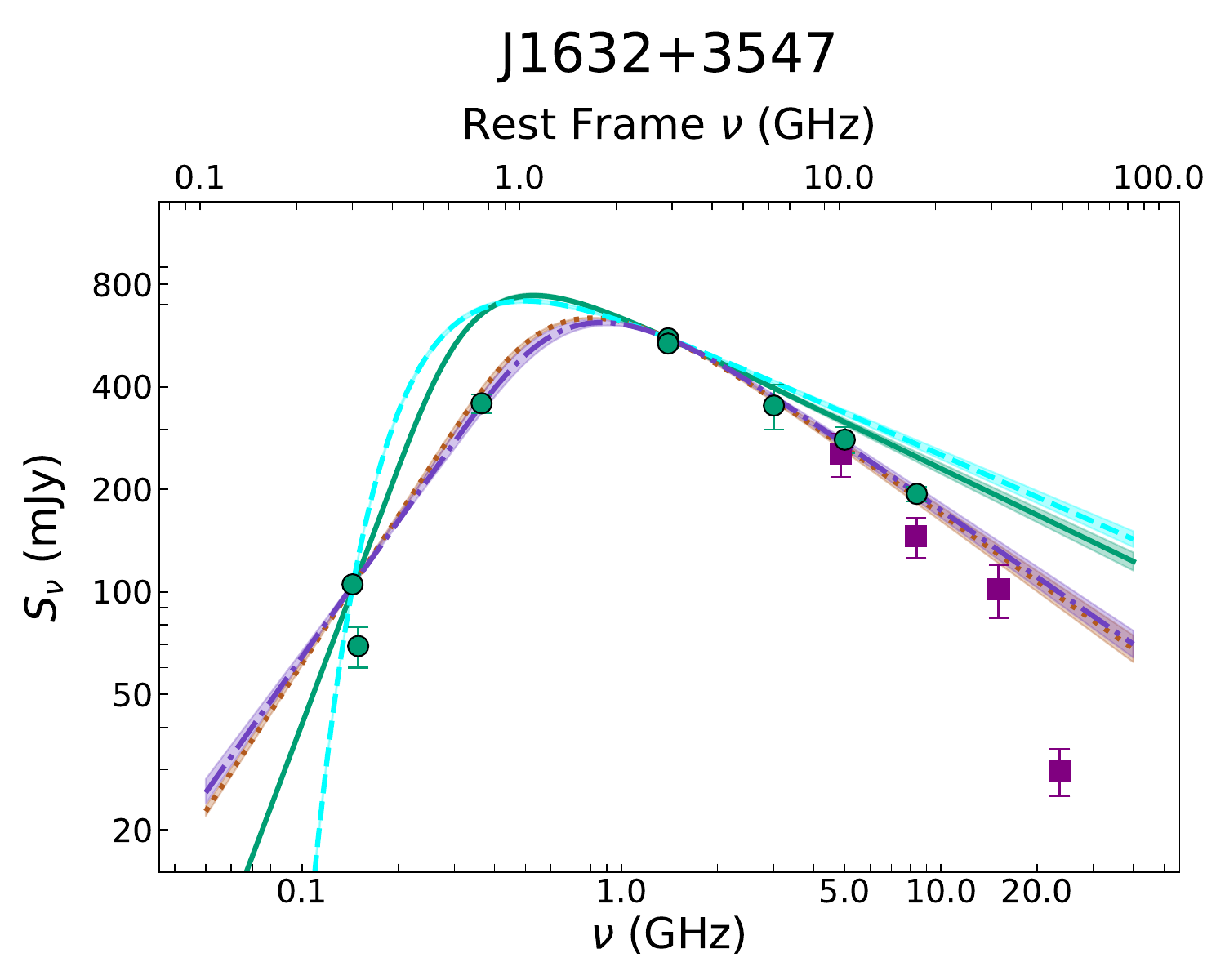}\\\vspace{15pt}
    \includegraphics[width=0.33\textwidth]{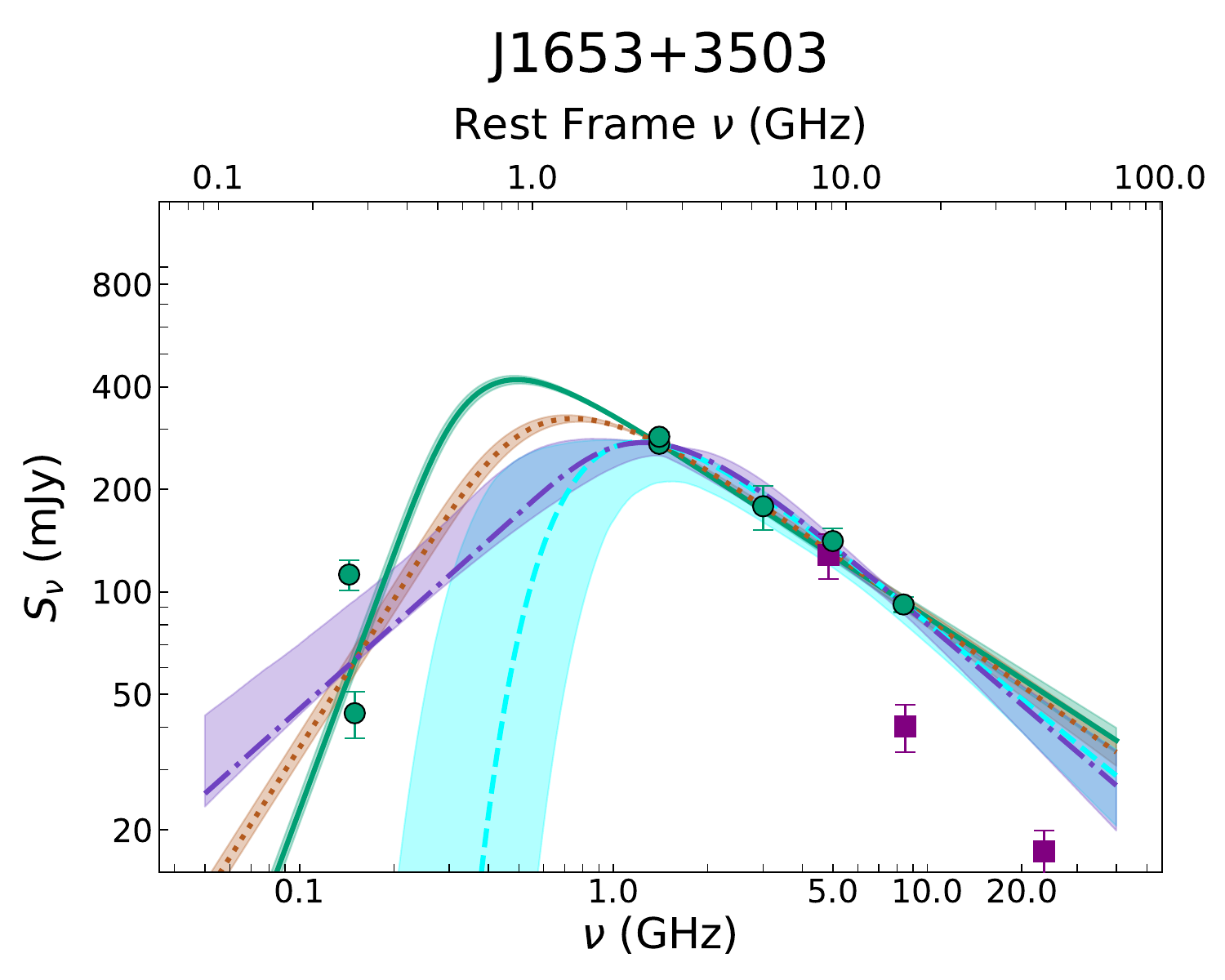}
    \includegraphics[width=0.33\textwidth]{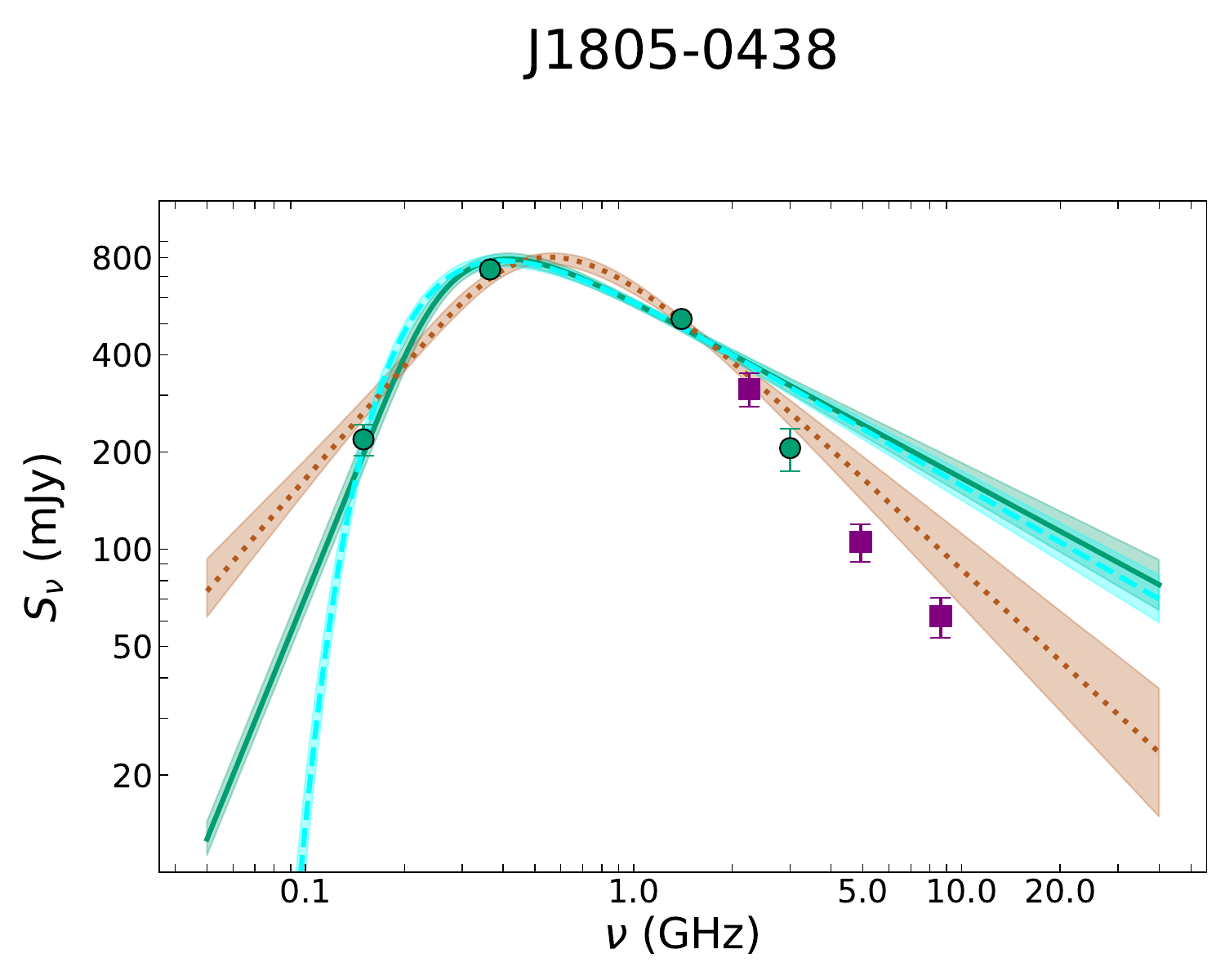}
    \includegraphics[width=0.33\textwidth]{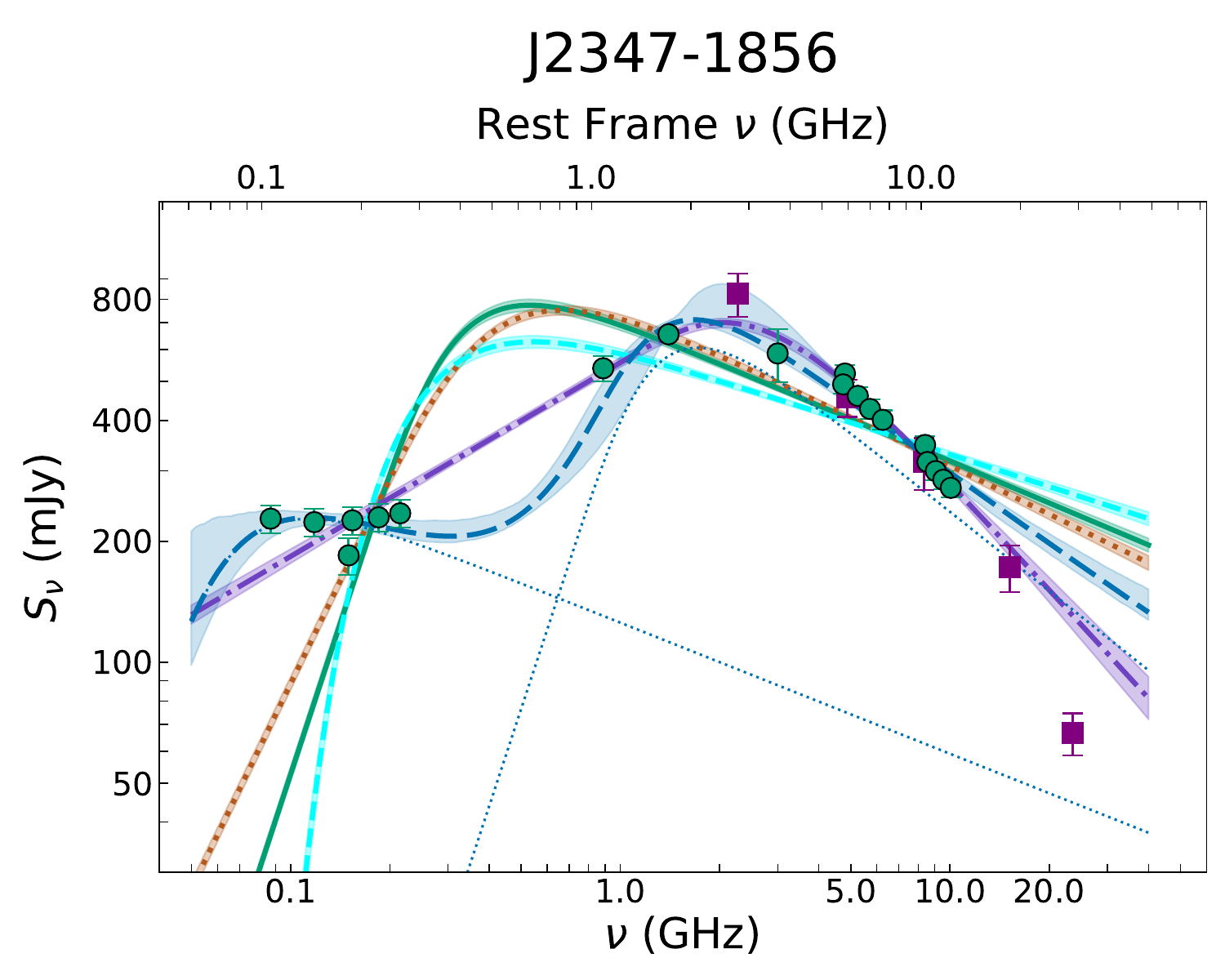}
\caption{Integrated radio spectra of the nine milli-lens candidates. Green circles show the single-dish and connected-element interferometer flux densities, and purple squares the largest sum of the VLBI component flux densities. The lines show the best-fit SSA, internal FFA, homogeneous external FFA, and inhomogeneous external FFA models, and, for J0210-2213 and J2347-1856, the inhomogeneous SSA model (i.e. the sum of two SSA components; its individual components are shown as dotted lines). Shaded regions indicate the 1-$\sigma$ confidence bands. Models that the data cannot constrain are not shown. Best-fit parameters are listed in Tables~\ref{tab:FFAtable} and~\ref{tab:SSA2table}.}
    \label{fig:integratedspectra}
\end{figure*}

\section{VLBI lightcurves}\label{app:vlbilightcurves}

\begin{figure*}[h]
    \centering
    \includegraphics[width=0.98\textwidth]{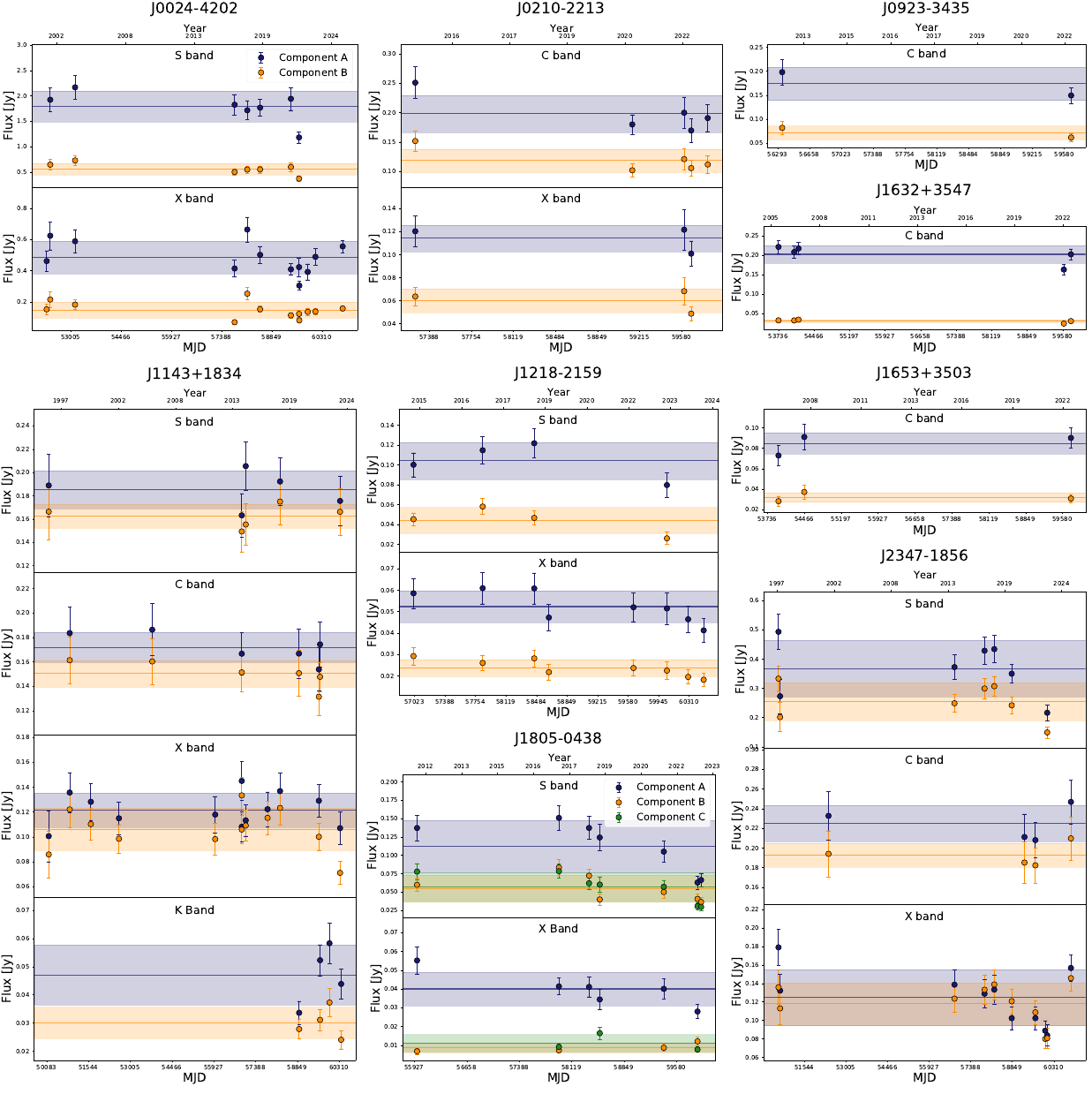}
\caption{VLBI light curves of the individual components for each source. Observing bands were selected based on the number of available data points.}
\label{fig:timevar}
\end{figure*}
\end{appendix}

\end{document}